\documentclass[%
 reprint,
 superscriptaddress,
nofootinbib,
 amsmath,amssymb,
 aps,
]{revtex4-1}

\usepackage{tikz}
\usepackage{graphicx}
\usepackage{dcolumn}
\usepackage{bm}
\usepackage{hyperref}

\usepackage{multirow, array}
\newcommand\Tstrut{\rule{0pt}{3.3ex}}         
\newcommand\Bstrut{\rule[-2ex]{0pt}{0pt}}   
\newcommand{\dittotikz}{%
    \tikz{
        \draw [line width=0.12ex] (-0.2ex,0) -- +(0,0.8ex)
            (0.2ex,0) -- +(0,0.8ex);
        \draw [line width=0.08ex] (-0.6ex,0.4ex) -- +(-0.0em,0)
            (0.6ex,0.4ex) -- +(0.0em,0);
    }%
}

\hypersetup{
    colorlinks = true
}
\usepackage{aas_macros}
\newcommand{\bx}{\boldsymbol{x}}

\newcommand{\B}[1]{\ensuremath{\boldsymbol{#1}}}
\newcommand{\D}[1]{\ensuremath{\text{d}#1}}
\newcommand{\iMpc}{\ensuremath{h/\mathrm{Mpc}}}

\newcommand{\Mpc}{\ensuremath{\mathrm{Mpc}/h}}
\newcommand{\Gpc}{\ensuremath{\mathrm{Gpc}/h}}

\def\fun#1#2{\lower3.6pt\vbox{\baselineskip0pt\lineskip.9pt
        \ialign{$\mathsurround=0pt#1\hfill##\hfil$\crcr#2\crcr\sim\crcr}}}

\newcommand{\beq}{\begin{equation}}
\newcommand{\eeq}{\end{equation}}
\newcommand{\beqa}{\begin{eqnarray}}
\newcommand{\eeqa}{\end{eqnarray}}

\begin{document}


\title{Testing one-loop galaxy bias: joint analysis of power spectrum and bispectrum}

\author{Alexander Eggemeier}
\email{alexander.eggemeier@durham.ac.uk}
\affiliation{%
  Institute for Computational Cosmology, Department of Physics, Durham University, South Road, Durham DH1 3LE,
  United Kingdom
}%

\author{Rom\'an Scoccimarro}
\affiliation{
 Center for Cosmology and Particle Physics, Department of Physics, New York University, NY 10003, New York, USA
}%

\author{Robert E. Smith}
\affiliation{%
 Astronomy Centre, School of Mathematical and Physical Sciences, University of Sussex, Brighton BN1 9QH, United
 Kingdom
}%

\author{Martin Crocce}
\affiliation{%
 Institute of Space Sciences (ICE, CSIC), Campus UAB, Carrer de Can Magrans, s/n, 08193 Barcelona, Spain
}%
\affiliation{%
  Institut d’Estudis Espacials de Catalunya (IEEC), 08034 Barcelona, Spain
}%

\author{Andrea Pezzotta}
\affiliation{%
  Institute of Space Sciences (ICE, CSIC), Campus UAB, Carrer de Can Magrans, s/n, 08193 Barcelona, Spain
}%
\affiliation{%
  Institut d’Estudis Espacials de Catalunya (IEEC), 08034 Barcelona, Spain
}%
\affiliation{%
  Max-Planck-Institut f\"{u}r extraterrestrische Physik, Postfach 1312, Giessenbachstr., 85741 Garching, Germany
}%

\author{Ariel G. S\'anchez}
\affiliation{%
 Max-Planck-Institut f\"{u}r extraterrestrische Physik, Postfach 1312, Giessenbachstr., 85741 Garching, Germany
}%

\date{\today}

\begin{abstract} 
  We present a joint likelihood analysis of the real-space power spectrum and bispectrum measured from a variety
  of halo and galaxy mock catalogs. A novel aspect of this work is the inclusion of nonlinear triangle
  configurations for the bispectrum, made possible by a complete next-to-leading order (``one-loop'')
  description of galaxy bias, as is already common practice for the power spectrum. Based on the goodness-of-fit
  and the unbiasedness of the parameter posteriors, we accomplish a stringent validation of this model compared
  to the leading order (``tree-level'') bispectrum. Using measurement uncertainties that correspond to an effective
  survey volume of $6\,(\mathrm{Gpc}/h)^3$, we determine that the one-loop corrections roughly double the
  applicable range of scales, from $\sim 0.17\,h/\mathrm{Mpc}$ (tree-level) to $\sim 0.3\,h/\mathrm{Mpc}$. This
  converts into a $1.5 - 2$x improvement on constraints of the linear bias parameter at fixed cosmology, and a
  $1.5 - 2.4$x shrinkage of uncertainties on the amplitude of fluctuations $A_s$, which clearly demonstrates the
  benefit of extracting information from nonlinear scales despite having to marginalize over a larger number of
  bias parameters. Besides, our precise measurements of galaxy bias parameters up to fourth order allow for thorough
  comparisons to coevolution relations, showing excellent agreement for all contributions generated by the
  nonlocal action of gravity. Using these relations in the likelihood analysis does not compromise the model
  validity and is crucial for obtaining the quoted improvements on $A_s$. We also analyzed the impact of
  higher-derivative and scale-dependent stochastic terms, finding that for a subset of our tracers the former
  can boost the performance of the tree-level model with constraints on $A_s$ that are only slightly degraded
  compared to the one-loop model.
\end{abstract}

\pacs{Valid PACS appear here}
\maketitle


\section{Introduction}
\label{sec:introduction}

Upcoming surveys of the large-scale structure \cite[e.g.,][]{DESI, Euclid, SKA}, will enable precise measurements
of higher-order statistics, which probe the non-Gaussian nature of the distribution of galaxies and other
luminous tracers in our Universe. These statistics include the three-point correlation function and its Fourier
transform, the bispectrum, whose importance for exploiting the full potential of the new survey generation has
been stressed in several recent studies. In particular, they are expected to provide significant advances in our
understanding of dark energy and gravity
\cite{SefCroPue0607,SonTarOka1508,GagSam1705,ByuEggReg1710,YanPor1902,AgaDesJeo2007}, the determination of
neutrino masses \cite{ChuIva1911,HahVilCas2003,KamSle2011,HahVil2012}, and for a glimpse into the initial inflationary
phase \cite{ScoHuiMan1204,SefCroDes1210,TelRosTas1606,KarLazLig1806,KarFonRoy2010,MorBiaSef2010}. Past applications of the two 
measures to data collected in the Baryon Oscillation Spectroscopic Survey (BOSS) have been mostly limited to
large scales \cite{GilNorVer1507,GilPerVer1702,SleEisBeu1706,SleEisBro1708,PeaSam1808}, while in order to
satisfy the aforementioned expectations it will be critical that their signal can also be reliably extracted
from at least the mildly nonlinear regime.

This requires careful modeling of nonlinearities that arise from the gravitational evolution of the matter
density, redshift space distortions due to peculiar velocities, and the connection between matter and tracer
densities, also known as galaxy bias. Considerable developments on the bispectrum have taken place over the last
years for all three of these modeling challenges
\cite[e.g.,][]{SmiSheSco0807,RamWon1206,BerCroSco1206,Assassi:2014,BalMerMir1505,AngForSch1510,LazGiaSch1604,
  HasRasTar1708,DesJeoSch1812,EggScoSmi1906}, but their implementation in the analysis of real or mock data is
still largely lacking. In this work we intend to take an important step in this direction by conducting a
detailed test of the perturbative galaxy bias model at next-to-leading, or one-loop, order that was presented in
\cite{EggScoSmi1906}, which we apply here to a variety of mock data samples.

The aim of the perturbative galaxy bias expansion (for a comprehensive review, see \cite{Desjacques:2018}) is to
capture all gravitational effects that can influence galaxy formation on scales much larger than their host dark
matter halos, and absorb the uncertainties of the formation process into a set of unknown bias parameters. The
lowest-order term in this expansion is the dark matter density itself, leading to the well-known linear
relationship $\delta_g = b_1\,\delta$ \cite{Kaiser:1984}, where $\delta_g$ and $\delta$ denote the galaxy and
matter density contrasts respectively, and $b_1$ is the linear bias parameter. Nonlinear corrections to this
relation involve powers of the matter field \cite{Coles:1993,FryGaz9308}, as well as the tidal field
\cite{CatLucMat9807,Catelan:2000,McDonald:2009,Matsubara:2011}, which are generated by the anisotropic collapse
of overdensities. Both of their signatures appear at leading order in the bispectrum and have already been
detected at great significance in analyses of simulated data
\cite{GuoJin0909,Pollack:2012,Chan:2012,Baldauf:2012,SheChaSco1304,PolSmiPor1405,SaiBalVla1405,
  BiaDesKeh1408,AngFasSen1503,BelHofGaz1510, LazSch1712,AbiBal1807,OddSefPor2003}, whereas constraints on the
former have also been reported from BOSS \cite{GilNorVer1507,ChiWagSan1509,GilPerVer1702,SleEisBeu1706}. As
shown in \cite{Assassi:2014,EggScoSmi1906}, the bispectrum model at one-loop order receives a number of
additional contributions from the bias expansion. Since each of the associated bias parameters needs to be
marginalized over for making robust inferences on cosmology, one of the leading questions we want to address is
whether this increased parameter space strongly penalizes the amount of information that can be extracted from
the nonlinear regime.

An important question related to this problem is whether the full set of galaxy bias parameters is truly
required to describe the measurements of the bispectrum, or whether one can propose universal simplifications
that do not compromise the model's validity. Principal candidates for that are the so-called coevolution
relations between galaxy bias parameters \cite{Fry9604,Chan:2012,Baldauf:2012,EggScoSmi1906}, which arise from
making simplifying assumptions about the matter-tracer connection at some time in the far past, and allow one to
fix a subset of the parameters at later times. Previous data analyses have relied heavily upon such relations,
fixing varying numbers of bias parameters in the power spectrum or two-point function
\cite[e.g.,][]{SanScoCro1701,GilPerVer1606,BeuSeoSai1704,GriSanSal1705}. This motivated various studies to check
their validity by making precise measurements of bias parameters from dark matter halo catalogs using either the
combination of the power spectrum and bispectrum \cite{SaiBalVla1405}, or more recently, using the
cross-correlation between the halo density and multiple higher-order fields \cite{LazSch1712,AbiBal1807}. These
studies have shown that the coevolution relations generally provide a good description of the measurements,
although deviations have been reported in particular for the tidal bias parameter. Due to parameter space
degeneracies this is likely not a concern for analyses of the power spectrum alone \cite{EggScoCro2006}, but in
combination with the bispectrum, which is affected by tidal bias even on large scales, application of the
coevolution relation can lead to significant errors. 

Our strategy for tackling these questions closely follows the approach presented in the other two installments
of this series of papers \cite{EggScoCro2006,PezCroSco2101}. We perform full likelihood fits to the measured
power spectra and bispectra from a diverse pool of tracers, including samples that mimic the SDSS Main, as well
as the BOSS LOWZ and CMASS galaxies, in addition to four halo catalogs with different mass cuts and
redshifts. Subsequently, we assess the quality of these fits based on the unbiasedness of the recovered
parameters and the goodness-of-fit, in order to quantitatively determine the range of scales over which the
model can be considered to be valid. Contrasting the constraining power of various modeling options at the
respective maximum scale of validity gives a straight-forward and robust means of comparison of the adopted
assumptions. As our focus in this paper lies on identifying a potentially optimal model for galaxy bias, we
ignore redshift-space distortions and keep cosmological parameters fixed, with the exception of the amplitude of
fluctuations $A_s$. We also note that our set of samples does not cover the main targets of the upcoming Dark
Energy Spectroscopic Instrument \cite{DESI} or the Euclid satellite mission \cite{Euclid}, but its diversity
still allows for a strong test of the universality of the one-loop bias model and for shedding new light on the
validity of the coevolution relations.

The paper is organized as follows: in Sec.~\ref{sec:model} we give a brief review of the theoretical
background including all relevant expressions that enter the models; in Sec.~\ref{sec:data} we provide
details on our samples, measurements and fitting methodology. Sec.~\ref{sec:results} presents the main
analysis of this paper with an estimation of the model validity ranges and a detailed test of the coevolution
relations under fixed cosmology, while in Sec.~\ref{sec:joint-analysis-with-As} we check the impact of
varying $A_s$. Our conclusions are given in Sec.~\ref{sec:conclusions}.

\section{Statistics of biased tracers in perturbation theory}
\label{sec:model}

In order to describe the clustering of galaxies, the bias expansion must account for all properties of the
matter field that affect the formation and evolution of the galaxies. Based on the equivalence principle and
symmetry considerations these properties --- in the following also called \emph{operators} --- have recently
been systematically identified and organized into a basis of terms at each order of perturbation theory, where
increasingly higher orders become relevant at smaller and smaller scales
\cite{Sen1406,Mirbabayi:2015,Desjacques:2018,EggScoSmi1906}. In this section we present the operator basis up to
fourth order as required by one-loop corrections to the bispectrum, starting from effects that are purely
induced by gravity. We review the multi-point propagator formalism for galaxy bias, which simplifies the
computation of the correlation functions, and finally discuss contributions to the bias expansion from two
additional types of effects: ``higher-derivative'' operators and stochasticity.

\subsection{Galaxy bias expansion}
\label{sec:bias-expansion}

In the basis presented in \cite{EggScoSmi1906} the galaxy density contrast is written as
\begin{equation}
  \label{eq:model.expansion}
  \begin{split}
    \delta_g(\bx) = \; &\bar{b}_1\,\delta(\bx) + \frac{\bar{b}_2}{2}\,\delta^2(\bx) + \bar{\gamma}_2\,{\cal
      G}_2(\Phi_v|\,\bx) + \frac{\bar{b}_3}{6} \delta^3(\bx) \\ &\hspace{-3em}+ 
    \bar{\gamma}_2^{\times}\,{\cal G}_2(\Phi_v|\,\bx)\,\delta(\bx) + \bar{\gamma}_3\,{\cal G}_3(\Phi_v|\,\bx) + \bar{\gamma}_{21}\,{\cal G}_2(\varphi_2,\varphi_1|\,\bx) \\
    &\hspace{-3em}+ \bar{\gamma}_{21}^{\times}\,{\cal G}_2(\varphi_2,\varphi_1|\,\bx)\,\delta(\bx) + \bar{\gamma}_{211}\,{\cal
      G}_3(\varphi_2,\varphi_1,\varphi_1|\,\bx) \\ &\hspace{-3em}+ \bar{\gamma}_{22}\,{\cal G}_2(\varphi_2,\varphi_2|\,\bx) + \bar{\gamma}_{31}\,{\cal
      G}_2(\varphi_3,\varphi_1|\,\bx) + \ldots\,,
  \end{split}
\end{equation}
where we have suppressed all time dependencies as well as ignored terms that do not contribute to the one-loop
power spectrum or bispectrum. Each term in Eq.~(\ref{eq:model.expansion}) is a combination of an operator (to be
described in the following) and a bias coefficient denoted by the symbol $b_X$ or $\gamma_X$, while $\Phi_v$ is
the (scaled) velocity potential,\footnote{We scale the peculiar matter velocity field $\B{u}$ by the growth rate
  and conformal Hubble rate, such that $\B{v} \equiv -\B{u}/(f {\cal H})$. The corresponding scaled velocity
  potential is defined by $\nabla^2\,\Phi_v \equiv \theta \equiv \nabla \cdot \B{v}$.} and $\varphi_n$ are the
Lagrangian perturbation theory (LPT) potentials.

In this particular representation of the bias expansion we distinguish between two different groups of effects,
designated either as local or nonlocal evolution operators (LE and NLE operators, respectively). The former are
constructed exclusively out of products of the matter density perturbations, $\delta$, and the two Galileons,
${\cal G}_2(\Phi_v)$ and ${\cal G}_3(\Phi_v)$, which are defined as follows (repeated indices are summed over):
\begin{align}
  {\cal G}_2(\Phi_v) \equiv \,&\left(\nabla_{ij}\Phi_v\right)^2 -
  \left(\nabla^2\Phi_v\right)^2\,, \label{eq:model.G2} \\ 
  {\cal G}_3(\Phi_v) \equiv \,&\left(\nabla^2\Phi_v\right)^3 +
  2\nabla_{ij}\Phi_v\,\nabla_{jk}\Phi_v\,\nabla_{ki}\Phi_v \nonumber \\ &-
  3\left(\nabla_{ij}\Phi_v\right)^2\nabla^2\Phi_v\,. \label{eq:model.G3}
\end{align}
Apart from $\nabla^2\Phi_v$ these are the only other two scalar invariants of the tensor $\nabla_{ij}\Phi_v$ in
three dimensions, and the usual tidal field enters the bias relation at second order through the combination
${\cal G}_2(\Phi_v) + 2/3\,\delta^2$. Furthermore, at leading order each of the LE operators is a local function
of the linear velocity potential, and if gravitational instability was a local process, this would remain true
at all orders, such that we could expect the LE operators to be sufficient for the description of galaxy bias
\cite{Chan:2012} (barring higher-derivative and stochasticity effects to be discussed in
Sec.s~\ref{sec:high-deriv-bias} and \ref{sec:stochasticity}).

Gravity, however, acts over long distances and the meaning of the second group of operators is to account for
the new dependencies that emerge as a consequence of this nonlocality. In LPT the nonlocal nature of gravity is
manifest in all terms beyond the Zel'dovich approximation \cite{KofPog9503,Ber9501}, for instance the
second-order LPT potential $\varphi_2$, given by $\nabla^2\,\varphi_2 = {\cal G}_2(\varphi_1)$, is a nonlocal
function of the linear (Zel'dovich) potential $\varphi_1 = - \nabla^{-2}\,\delta$. Using this fact it was
demonstrated in \cite{EggScoSmi1906} that the new dependencies can also be captured by Galileons, but involving
the higher-order LPT potentials, i.e.,
\begin{equation}
  {\cal G}_2(\varphi_n,\varphi_m) \equiv \nabla_{ij}\varphi_n\,\nabla_{ij}\varphi_m - \nabla^2\varphi_n\,\nabla^2\varphi_m\,,
\end{equation}
and analogously for ${\cal G}_3(\varphi_n,\varphi_m,\varphi_k)$. At leading order we have $\Phi_v = -\varphi_1$,
so the first time an NLE operator can appear in the bias expansion is at third order, with the only possibility
being ${\cal G}_2(\varphi_2,\varphi_1)$. At fourth order four additional terms must be taken into account whose
expressions are summarized in Appendix~\ref{sec:fourth-order-bias}. 

The set of LE and NLE operators in Eq.~(\ref{eq:model.expansion}) covers all non-redundant (linear independent)
combinations that can be constructed out of second derivatives of the gravitational and velocity
potentials. According to the equivalence principle and the Galilean invariance of the equations of motion these
are precisely the leading gravitational effects that impact the formation of galaxies. Based on the same
principles, another complete bias basis up to fourth order was previously presented in \cite{Desjacques:2018}
following the work of \cite{Mirbabayi:2015}, and is equivalent to the one adopted above by means of the
relations provided in Appendix A.2 in~\cite{EggScoSmi1906}.

\subsection{Multi-point propagator formalism}
\label{sec:MPformalism}

The coefficients --- or bias parameters --- of the expansion in Eq.~(\ref{eq:model.expansion}) depend on the
order at which this series is truncated, which implies that they are not immediately comparable with direct
measurements of the bias parameters, for instance through cross-correlations with the matter field
\cite{LazSch1712,AbiBal1807}. This was first pointed out in \cite{McDonald:2006} and can be remedied by
appropriate redefinitions (``renormalization'') of the original parameters that preserves the overall form of
the bias expansion.

An alternative (but equivalent) way of thinking about this complication is to guarantee that the expansion
coefficients are strictly defined as observable quantities. As shown in \cite{EggScoSmi1906} this can be
achieved by expanding the galaxy density contrast in terms of Wiener-Hermite functionals ${\cal H}_n$, such that
\begin{equation}
  \label{eq:model.MPexpansion}
  \delta_g = \Gamma_g^{(1)} \otimes {\cal H}_1 + \frac{1}{2} \Gamma_g^{(2)} \otimes {\cal H}_2
  + \frac{1}{3!} \Gamma_g^{(3)} \otimes {\cal H}_3 + \ldots\,.
\end{equation}
The role of the bias parameters is here taken by the scale-dependent functions $\Gamma_g^{(n)}$, which we call
galaxy \emph{multi-point propagators} in analogy to similar quantities appearing in the context of renormalized
perturbation theory \cite{CroSco0603a}, and they are defined as ensemble averaged derivatives of $\delta_g$ with
respect to the linear matter perturbations $\delta_L$. At $n$th order and written in Fourier space\footnote{We
  use the following Fourier space convention
  $\delta(\B{x}) = \int_{\B{k}} \exp{(-i \B{k}\cdot\B{x})}\,\delta(\B{k})\,,$ and write all $\B{k}$-space
  integrals with the short-hand notation
  $\int_{\B{k}_1,\ldots,\B{k}_n} \equiv \int \D{^3k_1}/(2\pi)^3 \cdots \D{^3k_n}/(2\pi)^3$.} we have
\begin{equation}
  \label{eq:eq.model.MPdef}
  \begin{split}
    \left<\frac{\partial^n\,\delta_g(\B{k})}{\partial\,\delta_L(\B{k}_1) \cdots
        \partial\,\delta_L(\B{k}_n)}\right> \equiv
    \;&(2\pi)^3\,\Gamma_g^{(n)}(\B{k}_1,\ldots,\B{k}_n) \\ &\times\,\delta_D(\B{k}-\B{k}_{1\cdots n})\,,
  \end{split}
\end{equation}
where $\delta_D$ denotes the Dirac delta function and $\B{k}_{1 \cdots n} \equiv \B{k}_1 + \cdots +
\B{k}_n$. The Wiener-Hermite functionals are given in terms of the linear matter perturbations and the
first three read as follows ($*$ stands for complex conjugation)
\begin{alignat}{2}
  \label{eq:model.hermite1-3}
  \begin{aligned}
    {\cal H}_1(\B{k}) &= &&\hspace*{-0.5em}\delta_L^*(\B{k})\,, \\[0.25em]
    {\cal H}_2(\B{k}_1,\B{k}_2) &= &&\hspace*{-0.5em}\delta_L^*(\B{k}_1)\,\delta_L^*(\B{k}_2) -
    \langle\delta_L(\B{k}_1)\,\delta_L(\B{k}_2)\rangle\,, \\[0.25em]
    {\cal H}_3(\B{k}_1,\B{k}_2,\B{k}_3) &=
    &&\hspace*{-0.5em}\delta_L^*(\B{k}_1)\,\delta_L^*(\B{k}_2)\,\delta_L^*(\B{k}_3) \\ & &&\hspace*{-0.5em}- \Big[
    \langle\delta_L(\B{k}_1)\,\delta_L(\B{k}_2)\rangle\,\delta_L^*(\B{k}_3) + \text{cyc.}\Big]\,,
  \end{aligned}
\end{alignat}
while the product $\otimes$ in Eq.~(\ref{eq:model.MPexpansion}) is defined as
\begin{equation}
  \begin{split}
    \left[\Gamma_g^{(n)} \otimes {\cal H}_n\right](\B{k}) \equiv \; &(2\pi)^3 \int_{\B{k}_1,\ldots,\B{k}_n}
    \delta_D(\B{k}-\B{k}_{1\cdots n}) \\ &\times\,\Gamma_g^{(n)}(\B{k}_1,\ldots,\B{k}_n)\,{\cal H}_n(\B{k}_1,\ldots,\B{k}_n)\,.
  \end{split}
\end{equation}
One can show that the multi-point propagators are related to cross-correlations between the galaxy and matter
fields \cite{CroSco0603b}, for example
$\left<\partial\delta_g/\partial\delta\right> =
\left<\delta_g\,\delta\right>/\left<\delta\,\delta\right>$.\footnote{Note that this property provides a
  connection to the bias renormalization procedure outlined in \cite{Assassi:2014}.} This highlights that these
quantities are indeed directly measurable and in the particular case of $\Gamma_g^{(1)}$ we see that the low-$k$
limit matches the linear bias parameter typically extracted from simulations or mock catalogs. The
scale-dependence of the multi-point propagators is determined by the functional form of the various operators
that appear in the bias expansion and for the basis presented in Sec.~\ref{sec:bias-expansion} they have a
particularly simple structure in Lagrangian space (i.e., at an initial time where nonlinearities in the matter
density perturbations are negligible). More precisely, \cite{EggScoSmi1906} demonstrated that the $n$-point
propagator is given by all contributions that enter Eq.~(\ref{eq:model.expansion}) at $n$th order, in addition
to loop corrections (of order $n+2$ at one-loop level) that can only involve NLE operators. Given this, and
using only operators up to fourth order, the first three Lagrangian (indicated by the subscript ${\cal L}$)
multi-point propagators are:
\begin{equation}
  \label{eq:model.MP1}
    \Gamma_{g,{\cal L}}^{(1)}(\B{k}) = b_{1,{\cal L}} + 2\gamma_{21,{\cal L}} \int_{\B{q}}
    K(\B{k}-\B{q},\B{q})\,K(\B{k},\B{q})\,P_L(q)\,,
\end{equation}
\begin{equation}
  \label{eq:model.MP2}
  \begin{split}
    \Gamma_{g,{\cal L}}^{(2)}(\B{k}_1,\B{k}_2) = \; & b_{2,{\cal L}} + 2 \gamma_{2,{\cal L}}\,K(\B{k}_1,\B{k}_2) \\
    &\hspace{-3em}+ 12 \int_{\B{q}}\Big[\gamma_{21,{\cal L}}^{\times}\,{\cal K}_{\delta{\cal
        G}_2(\varphi_2,\varphi_1)}^{(4,\text{F})} + \gamma_{211,{\cal L}}\,{\cal K}_{{\cal
        G}_3(\varphi_2,\varphi_1,\varphi_1)}^{(4)}\Big. \\ &\hspace{-3em}\Big.+ \gamma_{22,{\cal L}}\,{\cal
      K}_{{\cal G}_2(\varphi_2,\varphi_2)}^{(4)} + \gamma_{31,{\cal L}}\,{\cal K}_{{\cal
        G}_2(\varphi_3,\varphi_1)}^{(4)}\Big]\,P_L(q)\,,
  \end{split}
\end{equation}
\begin{equation}
  \label{eq:model.MP3}
  \begin{split}
    \Gamma_{g,{\cal L}}^{(3)}(\B{k}_1,\B{k}_2,\B{k}_3) = \; & b_{3,{\cal L}} + 2\gamma_{2,{\cal L}}^{\times}\left[K(\B{k}_1,\B{k}_2) +
      \text{cyc.}\right] \\
    &+2\gamma_{21,{\cal L}}\left[K(\B{k}_1,\B{k}_2)\,K(\B{k}_{12},\B{k}_3) + \text{cyc.}\right] \\
    &+6\gamma_{3,{\cal L}}\,L(\B{k}_1,\B{k}_2,\B{k}_3)\,,
  \end{split}
\end{equation}
where $P_L$ is the linear matter power spectrum, and $K$ and $L$ are the Fourier space
kernels of the two Galileons ${\cal G}_2(\Phi_v)$ and ${\cal G}_3(\Phi_v)$,
\begin{align}
  K(\B{k}_1,\B{k}_2) &\equiv \mu_{12}^2 - 1\,, \label{eq:model.K}\\
  L(\B{k}_1,\B{k}_2,\B{k}_3) &\equiv 2\,\mu_{12}\,\mu_{23}\,\mu_{31} - \mu_{12}^2 - \mu_{23}^2 - \mu_{31}^2 +
  1\,, \label{eq:model.L}
\end{align}
with $\mu_{ij} \equiv \B{k}_i \cdot \B{k}_j / k_i\,k_j$. The square bracket in the integral appearing in
Eq.~(\ref{eq:model.MP2}) is evaluated for the argument $(\B{k}_1,\B{k}_2,\B{q},-\B{q})$ and the kernel functions
associated to the fourth order NLE operators are collected in Appendix~\ref{sec:fourth-order-bias}. We stress
that the parameters in Eqs.~(\ref{eq:model.MP1})-(\ref{eq:model.MP3}) are automatically ``renormalized'' and no
longer depend on the order of the bias expansion, which we have indicated by the lack of an overbar. In this way, 
we bypass the cumbersome process of renormalization altogether.

The transition from Lagrangian to Eulerian space, that is, to the observed late-time galaxy density
fluctuations, induces corrections to the above propagators because of nonlinear evolution. This leads to lower
order ($< n$) bias contributions entering a given $n$-point propagator, as well as additional loop corrections
that are no longer constrained to NLE operators alone. Assuming the conservation of tracers\footnote{This is not
  a strong assumption, all it requires is that for each tracer of interest identified at redshift $z$, one
  follows back the center of mass of its constituents to the time of the initial conditions. This defines the tracers in Lagrangian space and ensures conservation by construction.}, all these terms can
be computed by the nonlinear evolution of the multi-point propagators themselves, which are determined by a set
of recursion relations \cite{EggScoSmi1906}. If we combine $\Gamma_g^{(n)}$ with the multi-point propagators for
the matter field and velocity divergence (which are defined in analogy with Eq.~\ref{eq:eq.model.MPdef}) into
the three-vector $\Gamma_a^{(n)} \equiv \left[\Gamma_m^{(n)},\,\Gamma_{\theta}^{(n)},\,\Gamma_g^{(n)}\right]$, and
using the logarithm of the growth factor as the time variable $\eta \equiv \ln{D}$, we can write the recursion
relations as
\begin{equation}
  \label{eq:model.MPrecursion}
  \begin{split}
    \Gamma_a^{(n)}(\B{k}_1,\ldots,\B{k}_n,\eta) = \; & g_{ab}(\eta)\,\Gamma_{b,{\cal
        L}}^{(n)}(\B{k}_1,\ldots,\B{k}_n) \\ &\hspace{-2.em}+ \int_0^\eta \text{d}\eta'\,g_{ab}(\eta-\eta')
    \left[\Gamma_{b,\mathrm{tree}}^{(n)} + \Gamma_{b,\mathrm{1L}}^{(n)}\right]\,,
  \end{split}
\end{equation}
where $\Gamma_{m,{\cal L}}^{(n)} = \Gamma_{\theta,{\cal L}}^{(n)} = 1$ for $n=1$ and otherwise zero, and the last two
terms are given by
\begin{equation}
  \begin{split}
    \Gamma_{b,\mathrm{tree}}^{(n)} = \; & \sum_{m=1}^{n-1} \Big[\gamma_{bcd}(\B{k}_{1
        \cdots m},\B{k}_{m+1 \cdots n})\,\Gamma_c^{(m)}(\B{k}_1,\ldots,\B{k}_m,\eta')\Big. \\ &\Big.\times\,\Gamma_d^{(n-m)}(\B{k}_{m+1},\ldots,\B{k}_n,\eta')
    + \mathrm{sym.}\Big]\,,
  \end{split}
\end{equation}
and 
\begin{equation}
  \label{eq:model.MPloop}
  \begin{split}
    \Gamma_{b,\mathrm{1L}}^{(n)} = \; & \sum_{m=1}^{n+1} \int_{\B{q}} \Big[\gamma_{bcd}(\B{k}_{1
        \cdots m-1} + \B{q},\B{k}_{m \cdots n} - \B{q}) \Big.  \\
     & \times\,\Gamma_c^{(m)}(\B{k}_1,\ldots,\B{k}_{m-1},\B{q},\eta') \\ &\Big.\times\,\Gamma_d^{(n-m+2)}(\B{k}_{m},\ldots,\B{k}_n,-\B{q},\eta')
    + \mathrm{sym.}\Big]\,.
  \end{split}
\end{equation}
The expressions are symmetrized over all participating $\B{k}$-modes and definitions for the linear time
propagator $g_{ab}(\eta)$ as well as the vertices $\gamma_{abc}$ are provided in
Appendix~\ref{sec:evolotion_tracers}. The first term in Eq.~(\ref{eq:model.MPrecursion}) is thus a linear
extrapolation of the initial conditions, while the remaining two terms give rise to the nonlinear evolution
corrections at either tree-level or one-loop order. Although these relations are straightforward to evaluate,
the resulting expressions are cumbersome to reproduce, so we instead direct the reader to a \texttt{Mathematica}
notebook accompanying this paper\footnote{\url{https://doi.org/10.5281/zenodo.4529886}}, which implements
Eq.~(\ref{eq:model.MPrecursion}) and computes all relevant quantities.

\subsection{Coevolution and peak-background split relations}
\label{sec:coevolution}

A subset of the contributions generated by the nonlinear evolution of the multipoint propagators has the same
scale-dependence as those terms already present at initial time. We can group those terms together and
define them as the Eulerian bias parameters, which yields the following coevolution relations for the
parameters appearing in Eqs.~(\ref{eq:model.MP1})-(\ref{eq:model.MP3}) \cite{EggScoSmi1906}:
\begin{align}
  \gamma_2 &= -\frac{2}{7}\left(b_1-1\right) + \gamma_{2,{\cal L}}\,, \label{eq:model.g2evo} \\
  \gamma_2^{\times} &= -\frac{2}{7}b_2 + \gamma_{2,{\cal L}}^{\times}\,, \\
  \gamma_3 &= -\frac{1}{9}\left(b_1-1\right) - \gamma_2 + \gamma_{3,{\cal L}}\,, \\
  \gamma_{21} &= \frac{2}{21}\left(b_1-1\right) + \frac{6}{7}\gamma_2 + \gamma_{21,{\cal L}}\,, \label{eq:model.g21evo} \\
  \gamma_{21}^{\times} &= \frac{2}{21}b_{2} + \frac{6}{7}\gamma_2^{\times} + \gamma_{21,{\cal
                         L}}^{\times}\,, \label{eq:model.g21xevo} \\
  \gamma_{211} &= {5\over 77}\left(b_1-1\right)+{15\over 14}\gamma_2-{9\over 7}\gamma_3+\gamma_{21} +
                 \gamma_{211,{\cal L}}\,, \\
  \gamma_{22} &= -{6\over 539}\left(b_1-1\right) - {9\over 49} \gamma_2 + \gamma_{22,{\cal L}}\,, \\
  \gamma_{31} &= -{4\over 11} \left(b_1-1\right) - 6 \gamma_2 +\gamma_{31,{\cal L}}\,, \label{eq:model.g31evo}
\end{align}
where $b_1 = 1 + b_{1,{\cal L}}$ and $b_2 = b_{2,{\cal L}}$. We see, in particular, that bias operators that
might have been absent at initial times (vanishing Lagrangian bias parameter) are sourced by nonlinear
evolution, in which case the corresponding parameters are fixed in terms of the remaining ones. A special case,
the so-called local Lagrangian approximation, arises if the initial galaxy bias relation only involves
powers of the matter perturbations, such that at late times all bias parameters can be expressed as functions of
$b_1$, $b_2$ etc. This has already been invalidated by detailed measurements in \cite{LazSch1712,AbiBal1807},
which have found that $\gamma_{2,{\cal L}} < 0$ and thus demonstrated an impact of at least the tidal field at
initial time, although in practice Eq.~(\ref{eq:model.g2evo}) with $\gamma_{2,{\cal L}} = 0$ can still be a
reasonable assumption, depending on how sensitive a given observable is to effects from the tidal field
\cite{EggScoCro2006}. Given that all terms deriving from the nonlocality of gravity (our NLE operators) are
inherently linked to nonlinear evolution, a potential simplification is to assume that these operators are not 
needed for characterizing the distribution of proto-halos --- a property that is also manifest in peak bias
models \cite[e.g.][]{MoJinWhi97,ParShe1211}. We will study this assumption in
Sec.~\ref{sec:galaxy-bias-constraints}, but note that it would provide us with a useful simplification as it
significantly reduces the overall number of free parameters in the one-loop bispectrum.

Based on the separate universe approach \cite{WagSchChi1503}, the authors of \cite{LazWagBal1602} measured the
response of the halo population to changes in a long wavelength mode and thus to a local modulation of the
density threshold that triggers halo formation. This corresponds to an exact implementation of the
peak-background split (PBS) and yields precise measurements of the set of local bias parameters $b_n$, which in
turn revealed tight relationships between the higher-order parameters $b_2$ and $b_3$, and the linear bias
parameter. As the PBS is sensitive to the total overdensities, these measurements match the spherically averaged
parameters of our bias basis (see Appendix C.3 of \cite{EggScoSmi1906}), so that
\begin{align}
  b_{2,\mathrm{PBS}} &= b_{2,\mathrm{PBS}}^{\mathrm{sph}}(b_1) + \frac{4}{3}\gamma_2 \label{eq:model.b2PBS} \\
  b_{3,\mathrm{PBS}} &= b_{3,\mathrm{PBS}}^{\mathrm{sph}}(b_1) + 4\gamma_2^{\times} - \frac{4}{3}\gamma_3 -
                       \frac{8}{3}\gamma_{21} - \frac{32}{21}\gamma_2\,, \label{eq:model.b3PBS}
\end{align}
where we denoted the relations found in \cite{LazWagBal1602}, which were fitted by a third order polynomial in
$b_1$, as $b_{2,\mathrm{PBS}}^{\mathrm{sph}}$ and $b_{3,\mathrm{PBS}}^{\mathrm{sph}}$.

\subsection{Power spectrum and bispectrum from multi-point propagators}
\label{sec:PB}

Besides being the (scale-dependent) physical bias parameters, we now show that the multi-point propagators serve
a second important role: they are also the main building blocks of the moments or correlation functions of the
galaxy density fluctuations. In this paper we are interested in the two lowest order correlation functions in
Fourier space, the power spectrum and bispectrum, which are given by
\begin{align}
  \langle \delta_g(\B{k}_1)\,\delta_g(\B{k}_2)\rangle &\equiv
  (2\pi)^3\,P_{gg}(k_1)\,\delta_D(\B{k}_{12})\,,   \label{eq:model.Pdef} \\
  \langle \delta_g(\B{k}_1)\,\delta_g(\B{k}_2)\,\delta_g(\B{k}_3)\rangle &\equiv
  (2\pi)^3\,B_{ggg}(k_1,k_2,k_3)\,\delta_D(\B{k}_{123})\,,   \label{eq:model.Bdef}
\end{align}
with analogous definitions for the matter perturbations. In particular, the linear power spectrum $P_L$
corresponds to the correlation of two linear Fourier modes $\delta_L(\B{k})$. Note that in this work we ignore
redshift space distortions, which means that statistical isotropy holds and the power spectrum and bispectrum
are determined by either a single or three $k_i$'s, respectively.

In order to compute these statistics we have to relate them to the linear matter spectrum by plugging in the
bias expansion. This step is greatly simplified when the bias expansion is written in the form of
Eq.~(\ref{eq:model.MPexpansion}), as we can exploit the orthogonality relations for the Wiener-Hermite
functionals (see e.g. \cite{Matsubara:1995,EggScoSmi1906}), which ensure that many terms have to vanish when taking
products of galaxy densities. It is then easy to see that in case of the power spectrum there is only one term
at each loop order, such that
\begin{align}
  \label{eq:model.Pg1loop}
  P_{gg}(k) = \; & \left[\Gamma_g^{(1)}(k)\right]^2 P_L(k) + \frac{1}{2}\int_{\B{q}}\,
  \left[\Gamma_g^{(2)}(\B{k}-\B{q},\B{q})\right]^2 \nonumber \\ &\times\,P_L(|\B{k}-\B{q}|)\,P_L(q) + \ldots
\end{align}
To compute $P_{gg}$ strictly at the one-loop level we only need to keep those terms that are of order
${\cal O}\left(P_L^2\right)$, which means we require $\Gamma_g^{(1)}$ at next-to-leading order (this includes
the usual $P_{13}$ contribution), but it is sufficient to evaluate $\Gamma_g^{(2)}$ at tree-level. Proceeding to
the bispectrum and using the product formula for three Wiener-Hermite functionals given in \cite{EggScoSmi1906}
(see also \cite{BerCroSco0811} for a direct evaluation), we obtain:
\begin{widetext}
  \begin{align}
    \label{eq:model.Bg1loop}
    B_{ggg}(k_1,k_2,k_3) = \; & \Gamma_g^{(2)}(\B{k}_1,\B{k}_2)\, \Gamma_g^{(1)}(k_1)\, \Gamma_g^{(1)}(k_2)\, P_L(k_1)P_L(k_2)+ {\rm cyc.} \nonumber \\
    &+ \Bigg[\int_{\B{q}} \Gamma_g^{(2)}(\B{k}_1-\B{q},\B{q})\, \Gamma_g^{(2)}(\B{k}_2+\B{q},-\B{q}) \,
    \Gamma_g^{(2)}(\B{k}_1-\B{q},\B{k}_2+\B{q})\, P_L(|\B{k}_1-\B{q}|)\, P_L(|\B{k}_2+\B{q}|)\, P_L(q)  \nonumber \\
    &+ \frac{1}{2} \int_{\B{q}} \Gamma_g^{(3)}(\B{k}_3,\B{k}_2-\B{q},\B{q})\, \Gamma_g^{(2)}(\B{k}_2-\B{q},\B{q})\,
    \Gamma_g^{(1)}(k_3)\, P_L(|\B{k}_2-\B{q}|)\,P_L(q)\,P_L(k_3)+ {\rm cyc.}  \Bigg] + \ldots\,,
  \end{align}
\end{widetext}
which is (up to next-to-leading order) fully determined by the first three multi-point propagators, but opposed
to the power spectrum each term can involve the combination of propagators from different orders. A consistent
computation at one-loop order further limits the expression to ${\cal O}\left(P_L\right)^3$, so we need to take
into account loop corrections of $\Gamma_g^{(1)}$ and $\Gamma_g^{(2)}$, but no more than $\Gamma_g^{(3)}$ at
tree-level. Note that, apart from the tree-level bispectrum, the first line of Eq.~(\ref{eq:model.Bg1loop}) also
contains the terms that are commonly denoted as $B_{321}^{II}$ and $B_{411}$ \cite{Sco97}. From these
expressions we also see clearly that $\Gamma_g^{(2)}$ constitutes a higher-order contribution for the power
spectrum, but enters the bispectrum at leading order. As we will verify in Sec.~\ref{sec:consistency}, this is
one of the reasons why the bispectrum is so helpful in reducing the uncertainties on the second-order parameters
$b_2$ and $\gamma_2$, as is well known.

\subsection{Higher-derivative effects}
\label{sec:high-deriv-bias}

The formation of dark matter halos and galaxies occurs through the gravitational collapse of material from an
extended region of space, whose size can be roughly identified with the Lagrangian radius $R$ of the halos (or
host halos in case of galaxies). This process implies a spatial nonlocality between the matter and tracer
densities that is not accounted for by the galaxy bias expansion in Eq.~(\ref{eq:model.expansion}), because it
implicitly assumes that $\delta_g$ is locally related to each of the bias operators: at position $\B{x}$ the
galaxy density only depends on the value of the bias operators at the \emph{same} position $\B{x}$. Therefore,
we should instead consider each term on the right-hand side of Eq.~(\ref{eq:model.expansion}) as integrated over
a patch of size $R$, in which case the spatially local assumption becomes valid in the limit that we consider
correlations on scales $r \gg R$, or equivalently for Fourier modes $k \ll 1/R$. When approaching smaller scales
(but still larger than $R$) it is possible to capture the resulting effects perturbatively, which leads to the
occurrence of higher-derivative terms, starting from $R^2\,\nabla^2\delta$
\cite{Des0811,McDonald:2009,DesCroSco1011}. We assume here that the scale $1/R$ is of the same order as the
nonlinearity scale $k_{\mathrm{nl}}$, defined as $k_{\mathrm{nl}}^3\,P(k_{\mathrm{nl}})/(2\pi^2) \equiv 1$, such
that $R^2\,\nabla^2\delta$ is the only relevant higher-derivative effect for the one-loop power spectrum. For
the bispectrum, though, the four additional terms,
\begin{equation}
  \nabla^2\delta^2\,, \quad (\B{\nabla}\delta)^2\,, \quad \nabla^2{\cal G}_2(\Phi_v)\,, \quad {\cal
    G}_2(\nabla_i\Phi_v,\nabla_i\Phi_v)\,,
\end{equation}
need to be taken into account, and together they give rise to the following higher-derivative corrections to
Eqs.~(\ref{eq:model.Pg1loop}) and (\ref{eq:model.Bg1loop}), respectively:
\begin{equation}
  \label{eq:model.Phd}
  P_{\nabla}(k) = - \beta_P\,k^2\,P_L(k)\,,
\end{equation}
\begin{equation}
  \label{eq:model.Bhd}
  \begin{split}
    B_{\nabla,123} = - \Big\{&\Big[\beta_{B,a}\,\left(k_1^2 + k_2^2\right) +
    \beta_{B,b}\,k_3^2\Big]\,F_2(\B{k}_1,\B{k}_2) \\ + &\Big[\beta_{B,c}\,\left(k_1^2 + k_2^2\right) +
    \beta_{B,d}\,k_3^2\Big]\,K(\B{k}_1,\B{k}_2) \\ + \hspace{0.005em} &\hspace{0.4em}\beta_{B,e}\,\B{k}_1 \cdot
    \B{k}_2\Big\}\,P_L(k_1)\,P_L(k_2) + \text{cyc.}
  \end{split}
\end{equation}
Here, $F_2$ denotes the second-order SPT kernel \cite{Bernardeau:2002}, and we have introduced $\beta_P$ and
$\beta_{B,a/\cdots/e}$ as the higher-derivative bias parameters, which are not fully independent as one can show
that $\beta_P = \left(\beta_{B,a}+\beta_{B,b}\right)/2$. We have absorbed the dependence on the scale $R$ into the
parameters, which consequently have units of $\left[\mathrm{Length}\right]^2$. Recently, \cite{LazSch1911} have
reported the first measurements of $\beta_P$ for various halo masses and showed that they follow the expected
scaling with Lagrangian radius.

As discussed in \cite{EggScoSmi1906}, corrections that result from a breakdown of the perfect, pressureless
fluid assumption in the nonlinear regime \cite{PueSco0908} are
completely degenerate with higher-derivative galaxy bias. However, if both effects are considered
simultaneously, the relation between $\beta_P$ and $\beta_{B,a/b}$ no longer holds and all six parameters must
enter the model as freely adjustable values. On the other hand, if we can ignore higher-derivative bias while
keeping the stress-tensor effects, we can set $\beta_{B,e} = 0$.

\subsection{Stochasticity}
\label{sec:stochasticity}

Apart from the various deterministic terms discussed thus far, the galaxy bias relation is also subject to
stochasticity, which can be thought of as the impact of deeply nonlinear modes on the formation of halos and
galaxies. Since these are uncorrelated with the large-scale fields (provided that no significant primordial
non-Gaussianities generate such correlations), their contribution appears stochastic in the perturbation theory
regime of validity \cite{Dekel:1999,TarSod9909,Mat9911}. To account for this, we write the galaxy density as a
sum of deterministic and stochastic contributions, $\delta_g(\B{x}) = \delta_g^{\mathrm{det}}(\B{x}) +
\varepsilon_g(\B{x})$, where the stochastic galaxy field,
\begin{equation}
  \label{eq:model.eps_g}
  \varepsilon_g(\B{x}) = \varepsilon(\B{x}) + \varepsilon_{\delta}(\B{x})\,\delta(\B{x}) +
  \varepsilon_{\nabla^2\delta}(\B{x})\,\nabla^2\delta(\B{x}) + \ldots\,,
\end{equation}
can be decomposed into the first-order term $\varepsilon$, and a series of composite terms, such as
$\varepsilon_{\delta}\,\delta$, which are induced by gravitational evolution\footnote{In principle every
  operator in the bias expansion will appear in Eq.~(\ref{eq:model.eps_g}) with an associated stochastic field,
  but for simplicity we have ignored terms of order ${\cal O}(\delta_L^2)$ or higher. }
\cite{Desjacques:2018}. Each of the stochastic fields has vanishing ensemble average and is uncorrelated with
the large-scale density $\delta_L$, which ensures that $\left<\varepsilon_g(\B{x})\right> = 0$ and implies that
correlations among each other must be highly localized in configuration space. In Fourier space we can therefore
express the power spectrum of two stochastic fields as an effective low-$k$ expansion \cite{Desjacques:2018}:
\begin{equation}
  P_{\varepsilon_a\varepsilon_b}(k) \equiv \left<\varepsilon_a(\B{k})\,\varepsilon_b(\B{k}')\right>' =
  P_{\varepsilon_a\varepsilon_b,0} +  P_{\varepsilon_a\varepsilon_b,2}\,k^2 + \ldots\,,
\end{equation}
with the constants $P_{\varepsilon_a\varepsilon_b,0}$ and $P_{\varepsilon_a\varepsilon_b,2}$, and using the
primed ensemble average to indicate that we have dropped a factor of $(2\pi)^3$ as well as the momentum
conserving Dirac delta function. Analogous expressions hold for all higher $N$-point functions.

Based on these considerations, the galaxy stochasticity power spectrum in the large-scale limit is given by
\begin{equation}
  \label{eq:model.Pstoch}
  \begin{split}
    C_{gg}(k) &\equiv \left<\varepsilon_g(\B{k})\,\varepsilon_g(\B{k}')\right>' =
    \left<\varepsilon(\B{k})\,\varepsilon(\B{k}')\right>' + \ldots \\ &=\; N_{P,0} + N_{P,2}\,k^2 + \ldots\,,
  \end{split}
\end{equation}
where contributions from stochastic fields not written down are absorbed by the constants $N_{P,0}$, $N_{P,2}$
etc. Physically, the galaxy stochasticity power spectrum represents deviations from purely Poissonian shot
noise, which can either lead to less large-scale power, $N_{P,0} < 0$ (sub-Poisson), due to reduced small-scale
clustering from the halo exclusion effect, or enhanced power, $N_{P,0} > 0$ (super-Poisson) due to subhalo or
satellite galaxy clustering \cite{MoWhi9609,SheLem99,SmiScoShe0703,BalSelSmi1310}. In the limit $k \to \infty$
we expect the shot noise to approach the Poisson limit and so $\lim_{k\to\infty} C_{gg}(k) = 0$ \cite{Sch1603},
which suggests an anti-correlation between $N_{P,0}$ and the scale-dependent noise parameter $N_{P,2}$. This was
empirically confirmed in \cite{EggScoCro2006}, which presented the first detailed measurements of $N_{P,2}$, and
showed further that its contribution is important for making consistent predictions of the galaxy auto power
spectrum and the galaxy-matter cross spectrum.

For the bispectrum we also have to account for three-point correlations between $\delta_g^{\mathrm{det}}$ and
$\varepsilon_g$, which leads to the following stochasticity bispectrum:
\begin{align}
  C_{ggg}(k_1,k_2,k_3) \equiv \; &\left<\varepsilon_g(\B{k}_1)\,\varepsilon_g(\B{k}_2)\,\varepsilon_g(\B{k}_3)\right>' \nonumber
  \\ &+
       \left[\left<\delta_g^{\mathrm{det}}(\B{k}_1)\,\varepsilon_g(\B{k}_2)\,\varepsilon_g(\B{k}_3)\right>' + \mathrm{cyc.}\right]\,.
\end{align}
In the large-scale limit the first term can be expanded as for the power spectrum, yielding
\begin{equation}
  \left<\varepsilon_g(\B{k}_1)\,\varepsilon_g(\B{k}_2)\,\varepsilon_g(\B{k}_3)\right>' = N_{B,0} +
  N_{B,2}\sum_{n=1}^3k_n^2 + \ldots\,,
\end{equation}
and we make use of Wick's theorem to evaluate the second term
\begin{align}
  \label{eq:model.delta_eps_eps}
    &\left<\delta_g^{\mathrm{det}}(\B{k}_1)\,\varepsilon_g(\B{k}_2)\,\varepsilon_g(\B{k}_3)\right>' \nonumber \\
    &\hspace{3.5em}= b_1 \left[P_{\varepsilon\varepsilon_{\delta}}(k_2) +k_1^2\,
      P_{\varepsilon\varepsilon_{\nabla^2\delta}}(k_2)\right] P_{mm}(k_1) + \ldots\,,                             
\end{align}
where the right-hand side has to be symmetrized over $k_2$ and $k_3$. Expanding the stochasticity power spectra
in Eq.~(\ref{eq:model.delta_eps_eps}) up to order ${\cal O}(k^2)$, and introducing the three new parameters
$M_{B,0}$ and $M_{B,2a/b}$, we finally obtain
\begin{equation}
  \label{eq:model.Bstoch}
  \begin{split}
    C_{ggg}(k_1,k_2,k_3) &= N_{B,0} + \Big\{N_{B,2}\,k_1^2 + \Big[M_{B,0} + M_{B,2a}\,k_1^2 \Big.\Big. \\
    &\hspace{-1em}\Big.\Big. + M_{B,2b}\,\left(k_2^2+k_3^2\right)\Big]\,P_{mm}(k_1) + \text{cyc.}\Big\} + \ldots
  \end{split}
\end{equation}
Again we note that contributions from higher-order stochastic fields can be absorbed by the already included
noise parameters, but we see that the addition of $\varepsilon_{\nabla^2\delta}\,\nabla^2\delta$ was important
as it generates a separate scale-dependence. While the parameters $N_{B,0}$ and $M_{B,0}$ are routinely taken
into account for analyses involving the bispectrum (though not necessarily as independent parameters)
\cite{GilNorVer1407,GilPerVer1702,OddSefPor2003}, the relevance of the three scale-dependent parameters is so
far unexplored. The fact that $N_{P,2}$ was found to be crucial for tracers with strong deviations from Poisson
shot noise (see \cite{EggScoCro2006}) motivates the inclusion of $N_{B,2}$ and $M_{B,2a/b}$, as we are going to
do in Sec.~\ref{sec:detailed-model-test}.

As was shown in \cite{McDonald:2006,EggScoSmi1906}, a subset of the loop corrections to the galaxy power
spectrum and bispectrum have a non-vanishing large-scale limit, which means that the models retain a sensitivity
to the nonlinear regime even on linear scales and thus strongly depend on the order at which we truncate the
perturbative expansion. However, these terms are fully absorbed by the noise parameters, and so we can subtract
them from the power spectrum and bispectrum. In our bias model, the large-scale limit of the loop corrections are explicitly given by
\cite{EggScoSmi1906}
\begin{align}
  \lim_{k \to 0} P_{gg}(k) &= \frac{b_2^2}{2}\,\int_{\B{q}} P_L(q)^2\,,\\
  \lim_{k_1,\,k_2 \to 0} B_{ggg}(k_1,k_2,k_3) &= b_1\,b_2\,\left[\frac{115}{42}b_2 + b_3 -
                                                \frac{8}{3}\gamma_{2}^{\times}\right] \nonumber \\
                           &\hspace{-5em}\times \left(P_1 + P_2 + P_3\right)\,\int_{\B{q}} P_L(q)^2 +  b_2^3\,\int_{\B{q}} P_L(q)^3\,,
\end{align}
where $P_i \equiv P_L(k_i)$.

\section{Measurements and methodology}
\label{sec:data}

\subsection{Galaxy and halo catalogs}
\label{sec:galaxy-halo-catalogs}

The joint fits of the power spectrum and bispectrum in this work will be performed on measurements from the same
set of tracers as those described in \cite{EggScoCro2006}. We give a brief overview in the following, but for
full details on the underlying simulations we direct the reader to \cite{EggScoCro2006} and references therein.

We consider a total of seven different catalogs that were generated from dark-matter only simulations: three
galaxy samples based on a halo occupation distribution (HOD) approach, and four halo samples with different mass
cuts. The galaxy samples have redshifts $z = 0.132$, $0.342$ and $0.57$, and are designed to match the number
densities and clustering properties of the SDSS Main Galaxy Sample, BOSS LOWZ and BOSS CMASS, respectively
(referred to as MGS, LOWZ and CMASS for the remainder of the paper), but do not account for the survey geometry
or any systematic effects. The volumes of a single simulation box in these three cases are $(1000\,\Mpc)^3$,
$(2400\,\Mpc)^3$ and $(1500\,\Mpc)^3$, and we make use of 40 independent realizations for MGS and LOWZ, and 100
for CMASS. Our halo samples HALO1 and HALO2 cover the mass ranges $[1,10] \times 10^{13}\,M_{\odot}$ and
$[10,\infty] \times 10^{13}\,M_{\odot}$ at $z = 0$, while HALO3 and HALO4 have $z = 0.974$ and contain halo
masses in the intervals $[1.3,2] \times 10^{13}\,M_{\odot}$ and $[2,\infty] \times 10^{13}\,M_{\odot}$. In all
of these cases we have 40 realizations, each with a volume of $(2400\,\Mpc)^3$.

\vspace*{-0.5em}
\subsection{Measurements of the power spectrum and bispectrum}
\label{sec:measurements}

We measure the power spectrum on scales ranging from $k_{\mathrm{min}} = \Delta k_P$ to
$k_{\mathrm{max}} = 0.3\,\iMpc$, where the bin width is chosen to be $\Delta k_P = k_f$ for the galaxy samples
and $\Delta k_P = 2 k_f$ for the halo samples ($k_f \equiv 2\pi/L_{\mathrm{box}}$ denotes the fundamental
frequency of the simulation box). The measurements are corrected for the Poisson shot noise contribution
$P_{\mathrm{Poisson}} = 1/\bar{n}$, depending on the number density $\bar{n}$ of the tracers.

In order to estimate the bispectra, we use the fast algorithm presented in \cite{Sco1510,SefCroSco1512} and for
a given bin width we determine all triangle configurations that satisfy the conditions: 1)
$k_1 \geq k_2 \geq k_3$, and 2) $k_1 \leq k_2 + k_3$. The bin width for the bispectrum measurements does not
have to coincide with that of the power spectrum, and we adopt the values $\Delta k_B = k_f$ for MGS and
$\Delta k_B = 2 k_f$ in all other cases. The maximum scale is kept fixed at $0.3\,\iMpc$, while
$k_{\mathrm{min}} = k_f$ for MGS, $k_{\mathrm{min}} = 2k_f$ for CMASS and $k_{\mathrm{min}} = 4k_f$ for the
remaining samples. In that way we obtain a total of 9959 distinct triangle configurations for MGS, 4353 for
CMASS and 17374 for LOWZ and the halo catalogs. As for the power spectrum we subtract the Poisson shot noise,
which in case of the bispectrum is given by \cite{Pee80}
\begin{equation}
  \label{eq:data.BPoisson}
  B_{\mathrm{Poisson}}(k_1,k_2,k_3) = \frac{1}{\bar{n}^2} + \frac{1}{\bar{n}}\left[\hat{P}(k_1) + \hat{P}(k_2) +
  \hat{P}(k_3)\right]\,,
\end{equation}
where $\hat{P}(k)$ denotes the (shot noise corrected) power spectrum estimate at scale $k$\footnote{The power
  spectrum estimates used for the shot noise correction are averaged over fundamental triangles defined by
  shells with the same bin width as the respective bispectrum measurements.}.

\begin{figure}
  \centering
  \includegraphics{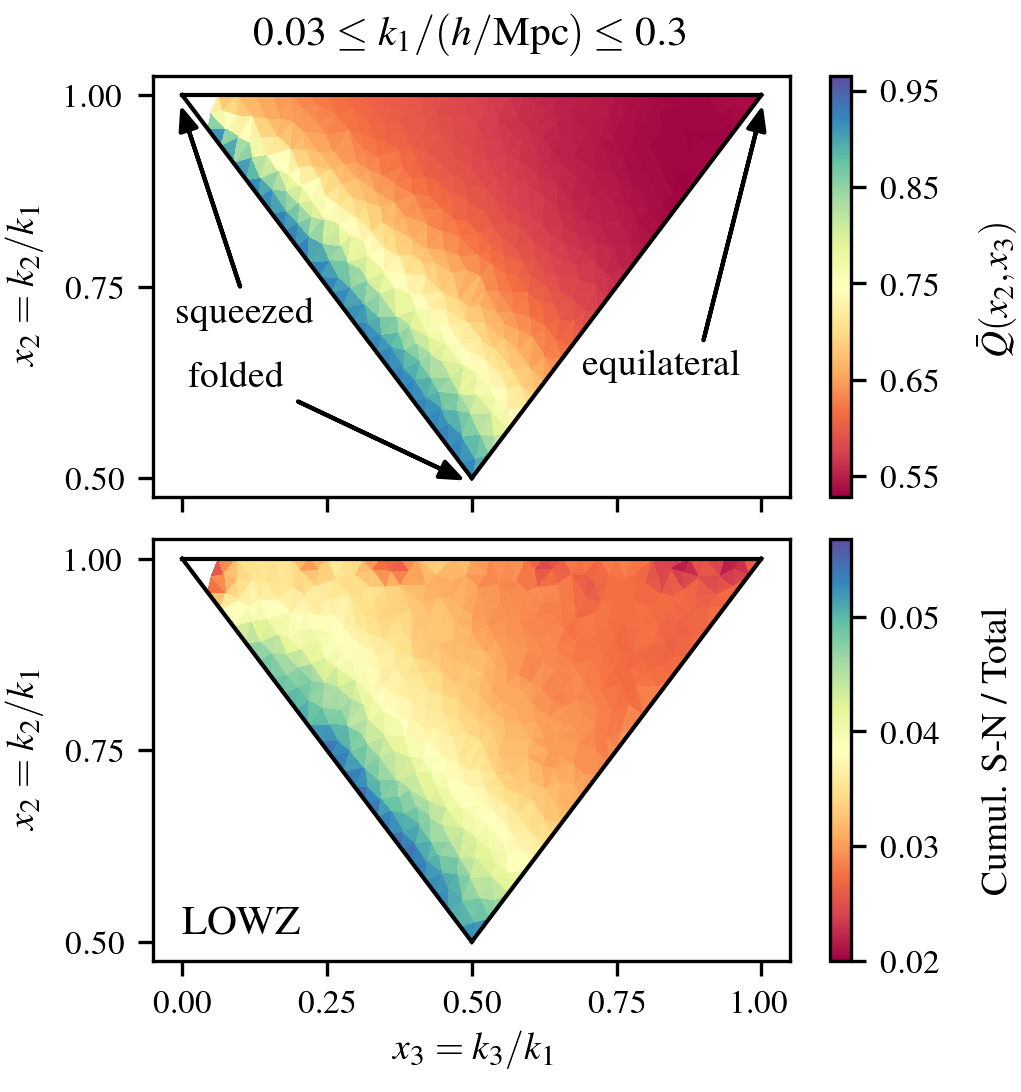}
  \caption{Upper panel: reduced bispectrum measurements from the LOWZ catalog, averaged over $k_1$ between
    $0.03\,\iMpc$ and $0.3\,\iMpc$, and plotted as a function of the triangle side ratios $x_2$ and $x_3$. Lower
    panel: cumulative signal-to-noise per $x_2$-$x_3$ bin for the same measurements, relative to the total
    signal-to-noise.}
  \label{fig:Q_SN}
\end{figure}

\begin{figure*}
  \centering
  \includegraphics{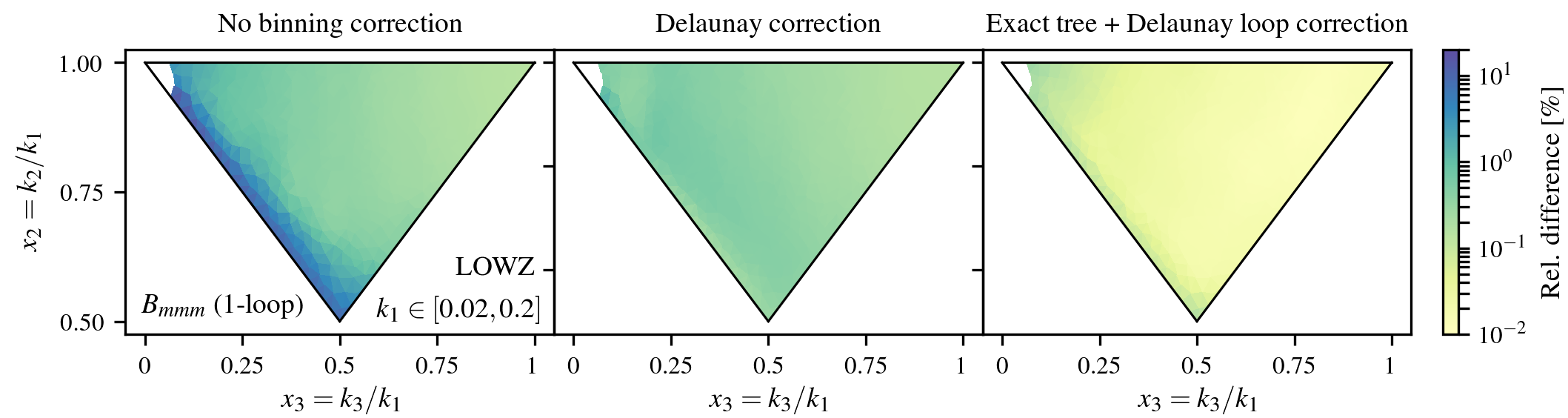}
  \caption{Relative difference between the exact bin-averaged (LOWZ) one-loop matter bispectrum according to
    Eq.~(\ref{eq:data.binavgB}) and two approximate solutions, averaged over $k_1$ in the interval
    $[0.02,0.2]\,\iMpc$: no correction (left panel), correction of tree-level and loop contribution using
    Delaunay interpolation (middle panel), exact bin-average of tree-level contribution and Delaunay correction
    of the one-loop part (right panel).}
  \label{fig:triangle_binning_correction}
\end{figure*}

It is interesting to visualize the configuration dependence of the bispectrum by computing the reduced
bispectrum $\hat{Q}_{123} \equiv \hat{B}_{123}/(\hat{P}_1\,\hat{P}_2+\hat{P}_2\,\hat{P}_3+\hat{P}_3\,\hat{P}_1)$
and averaging over an interval in $k_1$ while keeping the ratios $x_2 = k_2/k_1$ and $x_3 = k_3/k_1$ fixed, such
that
\begin{equation}
  \label{eq:data.Qavg}
  \bar{Q}(x_2,x_3) \equiv 
  \frac{1}{k_{u}-k_{l}}\int_{k_{l}}^{k_{u}} \mathrm{d}k \, \hat{Q}(k,k\,x_2,k\,x_3)\,.
\end{equation}
As an example, we take the LOWZ measurements between $k_l = 0.03\,\iMpc$ and $k_u = 0.3\,\iMpc$, which results
in the plot in the upper panel of Fig.~\ref{fig:Q_SN}, where valid configurations (according to the two
conditions above) are constrained to the triangular plotting area. Note that due to the finite value of
$k_{\mathrm{min}}$, it is not possible to measure arbitrarily ``squeezed'' configurations. The figure displays
the typical shape of the gravitationally induced bispectrum that is already well known from previous studies
\cite[e.g.][]{ScoColFry9803,ScoCouFri9906,SmiSheSco0807}: an enhanced signal for elongated triangle configurations
(along the left side) that continually decreases as the configurations become more equilateral, which reflects
the filamentary nature of the large-scale distribution of matter and galaxies. The three tree-level terms
involving $b_1$, $b_2$ and $\gamma_2$ contribute differently to this behavior of $\bar{Q}$: the dominating
effect stems from the matter contribution (proportional to $b_1^3$), which peaks for elongated triangle shapes,
whereas it is easy to see that the nonlinear bias is independent of configuration and the $\gamma_2$ term
vanishes for elongated triangles. The configuration dependence of the bispectrum therefore enables us to clearly
disentangle their differing effects, which presents a great advantage over the power spectrum, where these can
be largely degenerate (particularly from $b_2$ and $\gamma_2$), as we will see in Sec.~\ref{sec:consistency}.

The covariance matrices used in this work, both for the power spectrum and bispectrum, are assumed to be
diagonal, but not strictly Gaussian. We evaluate the variance in each bin from the independent set of
realizations for each sample, and to reduce noise we compare these estimates with the respective Gaussian
prediction (see e.g. \cite{SefCroPue0607}) on a bin-by-bin basis and retain whichever values are larger. The low
number densities of our tracers work in favor of this approximation as the high degree of shot noise enhances
the variance compared to correlations between different bins and thus drives the covariance matrices to become
more diagonal. More importantly, we do not expect this approximation to impinge on our test of the galaxy bias
modeling --- the main motivation for this work --- though one should bear in mind that the non-zero correlations
in a realistic covariance matrix will likely lead to less stringent constraints than those reported below. We
also ignore any cross-correlation between the power spectrum and bispectrum, which again is not true in
practice. However, this approximation should rather be considered a conservative one, as it has been shown in
\cite{SefCroPue0607,ByuEggReg1710} that inclusion of the cross-covariance helps shrinking parameter
uncertainties (in particular for amplitude-like parameters, such as galaxy bias and $\sigma_8$). Finally, like
in \cite{EggScoCro2006} we rescale each covariance matrix by an appropriate factor $\eta$,
\begin{equation}
  C_{ij} \rightarrow C_{ij}/\eta\,,
  \label{eq:eta}
\end{equation}
so that they match the effective volume \cite{FelKaiPea9405,Teg9711} of our LOWZ catalog at $k = 0.1\,\iMpc$,
which corresponds to $V_{\mathrm{eff}} \approx 6\,(\Gpc)^3$.

Using the resulting covariance matrix we compute the cumulative signal-to-noise per $x_2$-$x_3$ bin for the LOWZ
catalog over the range $0.03$ to $0.3\,\iMpc$, and compare this to the total signal-to-noise in the same
interval, which is shown in the lower panel of Fig.~\ref{fig:Q_SN}. The plot reveals that the elongated triangle
configurations carry the highest signal-to-noise, although squeezed shapes lag somewhat behind due to there
being fewer distinct configurations. The lowest signal-to-noise is found in triangle shapes that are close to
equilateral, but as there are many of those, they still comprise a significant fraction of the total
information content of the bispectrum.

\vspace*{-0.5em}
\subsection{Binning correction}
\label{sec:binning-correction}

The measurements of the power spectrum and bispectrum involve the average over finite bin widths. As long as our
observables do not vary strongly over the course of a given bin, we can compare the measurements with model
predictions evaluated at the center of the bin. This is typically sufficient for the power spectrum (except on
very large scales), but can amount to a major source of systematic error for the bispectrum~\cite{BerCroSco1206}.

In order to correct for this, we have to bin-average our theoretical models in precisely the same way as the
measurements are performed, which means
\begin{align}
  \label{eq:data.binavgB}
  B_{\mathrm{bin}}(k_1,k_2,k_3) &= \sum_{\B{q}_1 \in k_1} \sum_{\B{q}_2 \in
  k_2} \sum_{\B{q}_3 \in k_3} \frac{\delta_K(\B{q}_{123})}{N^T_{123}}\,B(q_1,q_2,q_3) \nonumber \\
  &\simeq \prod_{i=1}^3\int_{k_i} \mathrm{d}^3q_i\,
    \frac{\delta_D(\B{q}_{123})}{V_{B,123}}\,B(q_1,q_2,q_3)\,,
\end{align}
where the sums in the first line are taken over all Fourier modes $\B{q}_i$ whose magnitude satisfies
$k_i - \Delta k_B/2 \leq q_i \leq k_i + \Delta k_B/2$, $\delta_K$ denotes the Kronecker delta, and
$N^T_{123} = \prod_{i=1}^3 \sum_{\B{q}_i \in k_i} \delta_K(\B{q}_{123})$ are the number of fundamental triangles
per bin. In the second step we have approximated the discrete sums by integrals over spherical shells centered
on $k_i$ and made the identification $V_{B,123} = k_f^6\,N_{123}^T$. Computing the integrals in
Eq.~(\ref{eq:data.binavgB}) for the one-loop matter bispectrum of the LOWZ catalog and comparing to the
evaluation at the center of the bins, we can demonstrate the impact of ignoring the binning effect. This is
shown in the left panel of Fig.~\ref{fig:triangle_binning_correction}, which plots the relative difference
between the corrected and uncorrected model predictions, averaged over $k_1$ from $0.02\,\iMpc$ to
$0.2\,\iMpc$. We see that the impact is most severe for (nearly) collinear triangle configurations, where the
relative difference can be as large as $\sim 20\,\%$ in the squeezed limit, which is comparable to or even
larger than our measurement uncertainties ($\sim 10\,\%$ on average). For other triangle configurations the difference quickly drops to the order of
$\sim 1\,\%$ and becomes minimal for equilateral shapes.

Beyond tree-level the exact integration becomes computationally very expensive and so we face two options:
either we discard all collinear configurations, which should be disfavored as they carry the highest
signal-to-noise (see Fig.~\ref{fig:Q_SN}), or we find an acceptable approximation. One such approximation was
explored in \cite{SefCroDes1008,OddSefPor2003} and defines a triplet of effective $(k_1,k_2,k_3)$-modes, which
are constructed from averages over all fundamental triangles that can be realized on the discrete Fourier grid,
and at which the theory predictions will be computed instead of the bin centers. This has the advantage that no
additional model evaluations are necessary and it has been shown that it improves the accuracy to a level of a
few percent. In this work we pursue an alternative, which approximates Eq.~(\ref{eq:data.binavgB}) by
interpolating over a set of triplets $\B{Q} = (Q_1,Q_2,Q_3)$ that are chosen such that the three-dimensional
integration region for each triangle configuration is covered entirely by a group of tetrahedra
${\cal T}_{123} = {\cal T}(k_1,k_2,k_3)$. The integration over $\B{q}_1$, $\B{q}_2$ and $\B{q}_3$ subject to the
Delta function constraint can therefore be replaced by a sum over these tetrahedra, each integrated over its
entire volume. Using linear (Delaunay) interpolation to obtain the value of the bispectrum inside a given
tetrahedron ${\cal T}$ based on its four edge points $\B{Q}_{{\cal T}}^{(1)},\ldots,\B{Q}_{{\cal T}}^{(4)}$, we
can write
\begin{widetext}
  \begin{align}
    B_{\mathrm{bin}}(k_1,k_2,k_3) &= \frac{48\pi^2}{V_{B,123}} \sum_{{\cal T} \in {\cal T}_{123}} V_{{\cal T}} \int_0^1
    \mathrm{d}\lambda_3 \int_0^{1-\lambda_3} \mathrm{d}\lambda_2 \int_0^{1-\lambda_2-\lambda_3}
    \mathrm{d}\lambda_1\,\left[\prod_{i=1}^3\sum_{j=1}^4 \lambda_j\,Q_{{\cal
    T},i}^{(j)}\right]\,\left[\sum_{j=1}^4 \lambda_j\,B(Q_{{\cal T},1}^{(j)},Q_{{\cal T},2}^{(j)},Q_{{\cal
    T},3}^{(j)})\right] \nonumber \\ 
    &= \frac{48\pi^2}{V_{B,123}} \sum_{{\cal T} \in {\cal T}_{123}} \sum_{j=1}^4\,V_{{\cal T}}\,\alpha_j(\B{Q}_{{\cal T}}^{(1)},
      \ldots,\B{Q}_{{\cal T}}^{(4)})\, B(Q_{{\cal T},1}^{(j)},Q_{{\cal T},2}^{(j)},Q_{{\cal T},3}^{(j)})\,, \label{eq:data.Bbin_delaunay1}
  \end{align}
\end{widetext}
where $\lambda_j$ denote the barycentric coordinates (note that $\lambda_4 = 1 - \sum_{j=1}^3\lambda_j$),
$V_{{\cal T}}$ is the volume of the tetrahedron, and $\alpha_j$ are interpolation polynomials depending only on
the four edge points, which can be computed analytically. We can further simplify
Eq.~(\ref{eq:data.Bbin_delaunay1}) by rearranging the summations as a single sum over all unique edge points,
and summarizing all bispectrum configurations into the vectors $\B{B}_{\mathrm{bin}}(k_1,k_2,k_3)$ and
$\B{B}(Q_1,Q_2,Q_3)$, which are connected by the matrix multiplication
\begin{equation}
  \label{eq:data.Bbin_delaunay2}
  \B{B}_{\mathrm{bin}}(k_1,k_2,k_3) = \B{M}_{\mathrm{bin}} \cdot \B{B}(Q_1,Q_2,Q_3)\,.
\end{equation}
The matrix $\B{M}_{\mathrm{bin}}$ is determined by the values of the interpolation polynomials and as it does
not depend on cosmology, it only has to be computed once. The binning correction therefore amounts to the
bispectrum evaluation at all unique edge points and multiplication with $\B{M}_{\mathrm{bin}}$, which can be
implemented as a fast sparse matrix product. Along with this paper we provide a simple \texttt{Python}
package\footnote{\url{https://github.com/aegge/BispTools}} that produces the list of all unique tetrahedra edge points and computes the binning
matrix given the bin width and $k_{\mathrm{min}}$, $k_{\mathrm{max}}$ values of the measurements.

Applying this procedure to the one-loop matter bispectrum and comparing to the exact integration, we obtain the
middle panel of Fig.~\ref{fig:triangle_binning_correction}, which shows that the binning effect on the collinear
configurations has been significantly reduced. The relative difference for these triangle shapes is now of the
same order as for the equilateral ones and generally sub-percent. The largest deviation occurs for the most
squeezed configuration in the $k_1$ interval considered, but is no larger than $\sim 1.5\,\%$, which is already
well below the measurement uncertainties that we use in this work. Since we are not going to vary cosmological
parameters in our model fits below, we can afford to bin-average at least the tree-level model predictions
exactly. As can be seen from the right-hand panel in Fig.~\ref{fig:triangle_binning_correction} this improves
the agreement further, with the majority of configurations displaying relative differences smaller than
$0.1\,\%$.

\begin{table*}
  \centering
  \caption{Upper and lower limits of uniform prior distributions for the complete set of model parameters. For
    the higher-derivative and next-to-leading noise parameters we use an arbitrary normalization scale, which is
    fixed to $k_{\mathrm{HD}} = 0.4\,\iMpc$.}
  \begin{ruledtabular}
    \begin{tabular}{ccccccccccc}
      \multirow{2}{*}{Catalog} & \multirow{2}{*}{$b_1$} & \multirow{2}{*}{$b_2$} &
      \multirow{2}{*}{$\gamma_2$, $\gamma_{21}$} & \multirow{2}{*}{$\gamma_3$} & \multirow{2}{*}{$b_3$} &
               $\gamma_2^{\times}$, $\gamma_{21}^{\times}$,  & $\bar{n}\,N_{P,0}$, & \multirow{2}{*}{$\bar{n}\,M_{B,0}$}  &
       \multirow{2}{*}{$\beta_{B,a/\ldots/e}$} & $\bar{n}\,N_{P,2}$,
                                                                          $\bar{n}^2 N_{B,2}$, \Tstrut \\
                               & & & & & & $\gamma_{211}$, $\gamma_{22}$, $\gamma_{31}$ & $\bar{n}^2 N_{B,0}$ & && 
       $\bar{n}\,M_{B,2a/b}$ \Bstrut\\  
      \hline
      MGS & [0.5, 3] & [-7, 7] & [-4, 4] & [-30, 30] & [-80, 80] & [-50, 50] & [-1, 1] & [-2, 2] &
            [-100, 100] $k_{\mathrm{HD}}^{-2}$ & [-50, 50] $k_{\mathrm{HD}}^{-2}$ \Tstrut \\[0.1em]
      LOWZ & [1, 4] & \dittotikz & \dittotikz & \dittotikz & \dittotikz & \dittotikz & \dittotikz &
      [-5, 5] & \dittotikz & \dittotikz \\[0.1em]
      CMASS & [1, 4] & \dittotikz & \dittotikz & \dittotikz & \dittotikz & \dittotikz & \dittotikz &
      [-4 ,4] & \dittotikz & \dittotikz \\[0.1em]
      HALO1 & [0.5, 3] & \dittotikz & \dittotikz & \dittotikz & \dittotikz & \dittotikz & \dittotikz &
      [-2.1, 2.1] & \dittotikz & \dittotikz \\[0.1em]
      HALO2 & [1.5, 4.5] & \dittotikz & \dittotikz & \dittotikz & \dittotikz & \dittotikz & \dittotikz &
      [-8.5, 8.5] & \dittotikz & \dittotikz \\[0.1em]
      HALO3 & [1.7, 3.7] & \dittotikz & \dittotikz & \dittotikz & \dittotikz & \dittotikz & \dittotikz &
      [-7.2, 7.2] & \dittotikz & \dittotikz \\[0.1em]
      HALO4 & [2.5, 6.5] & [0,10] & \dittotikz & \dittotikz & \dittotikz & \dittotikz & \dittotikz \Bstrut &
      [-12.7, 12.7] & \dittotikz & \dittotikz \Bstrut \\ 
    \end{tabular}
  \end{ruledtabular}
  \label{tab:priors}
\end{table*}

\vspace*{-1em}
\subsection{Likelihood function and prior probabilities}
\label{sec:fitting}

The posterior distributions of the model parameters are determined following a standard Bayesian inference
method, which requires us to define an appropriate likelihood function and prior probabilities for all
parameters. We assume that the data from a single realization of our catalogs is drawn from a multivariate
Gaussian,
\begin{equation}
  -2 \log{{\cal L}} = \sum_{i,j=1}^{N_{\mathrm{bin}}} \left(X_i - \mu_i\right)\,C_{X,ij}^{-1}\,\left(X_j - \mu_j\right)\,,
\end{equation}
where $X_i$ is the vector containing the $N_{\mathrm{bin}}$ measurements of the power spectrum and bispectrum,
either individually or combined, and $\mu_i$ denotes the corresponding model predictions, which have been
bin-averaged for the bispectrum. $C_{X,ij}$ is the covariance matrix obtained as described in
Sec.~\ref{sec:measurements}. Since the various realizations are statistically independent, we combine them
into a total likelihood by computing the product of the $N_R$ individual ones, such that
\begin{equation}
  \label{eq:data.loglik_total}
  \log{{\cal L}}_{\mathrm{tot}} = \frac{1}{N_R} \sum_{n=1}^{N_R} \log{{\cal L}_{(n)}}\,.
\end{equation}
The factor $1/N_R$ is needed to ensure that the sampling volume, as characterized by the measurement
uncertainties, is left unchanged by this combination.

When quoting $\chi^2$ values as a measure of the goodness-of-fit, we have to account for the fact that our
covariance matrices correspond to a fixed sampling volume and not the total combined volume of all
realizations. The fluctuations in the data are therefore smaller than expected, which we can correct for by
rescaling the $\chi^2$ by the factor $N_R/\eta$ (see Eq.~\ref{eq:eta}), and hence we compute
\begin{equation}
  \label{eq:data.chi2}
  \chi^2 = -\frac{2N_R}{\eta} \log{{\cal L}}_{\mathrm{tot}}\,.
\end{equation}
Some of the model fits presented in Sec.~\ref{sec:results} will make use of the matter bispectrum extracted
from the underlying N-body simulations in exchange for the tree-level or one-loop model. These measurements
contain fluctuations themselves and consequently account for a part of the scatter in the galaxy or halo
bispectrum measurements. For the analogue case of the power spectrum it has been shown in \cite{EggScoCro2006}
that this leads to a predictable reduction in the $\chi^2$ values and following the same reasoning one can show
that for the bispectrum this reduction is given by
\begin{equation}
  \label{eq:data.chi2corr}
  \Delta \chi^2 = - b_1^3 \sum_{i,j=1}^{N_\mathrm{bin}} \left(2 C_{B_g \times B_m,ij} -
    b_1^3\,C_{B_m,ij}\right)\,C_{B_g,ij}^{-1}\,,
\end{equation}
where $C_{B_g,ij}$ and $C_{B_m,ij}$ are the galaxy and matter bispectrum covariance matrices, respectively, and
$C_{B_g \times B_m,ij}$ their cross-covariance. As for $C_{B_g,ij}$ we take the matter auto and
cross-covariances to be diagonal with their elements given by the maximum between the measured variances and the
Gaussian predictions.

\begin{figure}
  \centering
  \includegraphics{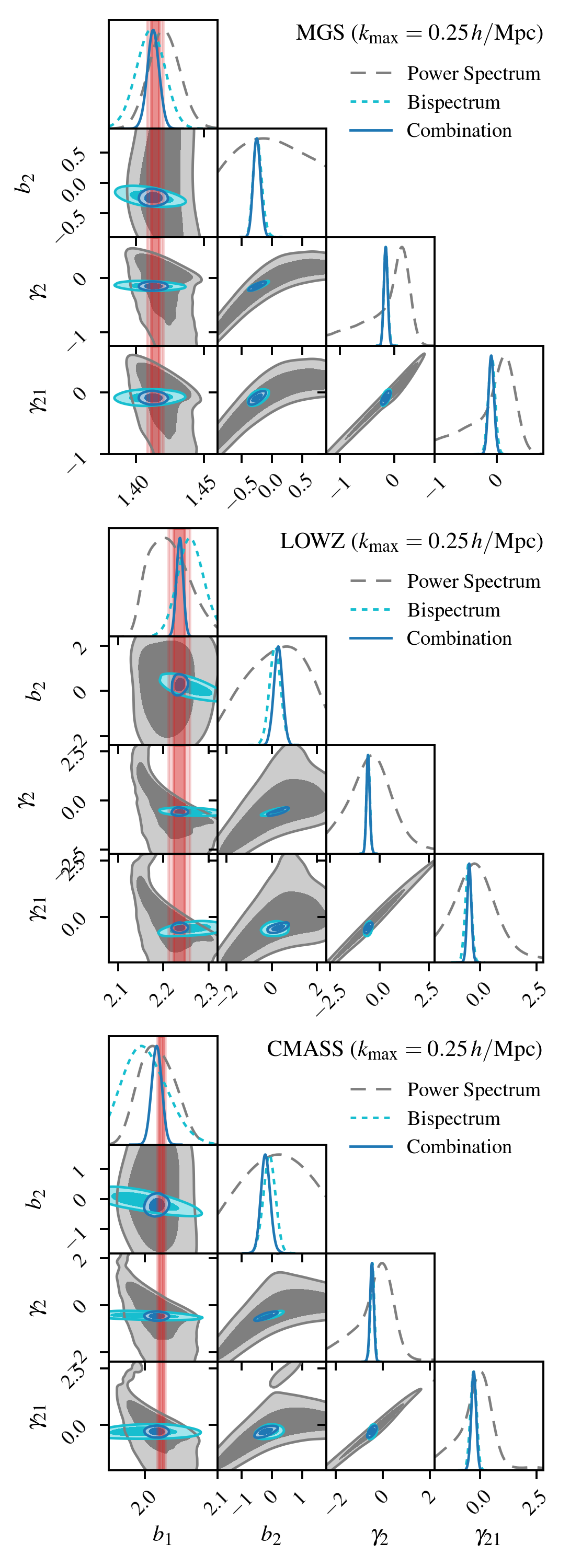}
  \caption{Marginalized posteriors for the common power spectrum and bispectrum parameter space, obtained from
    individual and joint fits to the three galaxy samples with $k_{\mathrm{max}} = 0.25\,\iMpc$. The red band
    indicates the fiducial value of $b_1$.}
  \label{fig:Pgg+Bggg_contours}
\end{figure}

We adopt flat prior probability distributions for all model parameters, with bounds that are symmetric around
zero (with the exception of $b_1$) and wide enough to minimize any prior dependence on the
posteriors\footnote{Note that when the maximum scale included in the model fits, $k_{\mathrm{max}}$, is small it is not possible
  to remove all prior dependence due to degeneracies and low sensitivities to certain parameters.}. The exact
values, most of which are identical across our various samples, can be found in Table~\ref{tab:priors}. The
lower bounds of the leading stochastic parameters --- $N_{P,0}$ for the power spectrum and $N_{B,0}$, as well as
$M_{B,0}$ for the bispectrum --- are motivated by the Poisson limit, ensuring that the overall stochastic
contribution is strictly positive (keeping in mind that the Poisson part has been subtracted from the
measurements). For the parameter $M_{B,0}$, which is multiplied by the matter power spectrum in our model, this
means that its lower bound should scale as $\sim b_1^2$ and the numbers quoted in Table~\ref{tab:priors} derive
from the fiducial $b_1$ values. The upper limits of these parameters are in principle not constrained, but based
on the study in \cite{EggScoCro2006} we do not expect large super-Poisson deviations and so we keep these priors
symmetric. In contrast to \cite{EggScoCro2006} we no longer use a Gaussian prior on the tidal bias parameter,
since the bispectrum data breaks the otherwise strong degeneracy between $\gamma_2$ and
$\gamma_{21}$.

Our fits are conducted by minimizing the negative total log-likelihood from Eq.~(\ref{eq:data.loglik_total})
using a Markov chain Monte Carlo (MCMC) approach. For each case we run several independent Markov chains with
different initial seeds and make sure they are converged according to the Gelman-Rubin criterion with $R < 0.01$
(see \cite{GelRub9201}) and that we reach a total number of 120,000 accepted steps. Afterwards these chains are
post-processed with \texttt{getdist} \cite{Lew2019} in order to extract the parameter posteriors and related
statistics.

\vspace*{-0.5em}
\section{Joint analysis of power spectrum and bispectrum at fixed cosmology}
\label{sec:results}

We are now going to present the results from jointly fitting the power spectrum and bispectrum model described
in Sec.~\ref{sec:model} to the measurements from our galaxy and halo catalogs. We keep all cosmological
parameters fixed in this section, which allows us to probe the galaxy bias model itself. Additionally, we
replace the contributions from the matter power spectrum and bispectrum (i.e., the terms multiplied by
$b_1^2$ and $b_1^3$, respectively) by their simulation measurements and so remove the leading model
uncertainties from stress-tensor corrections that originate from the nonlinear evolution of the matter field. We
are further going to use these results to derive rigorous constraints on the full bias parameter space and
compare them to PBS and coevolution relations.

\subsection{Consistency between power spectrum and bispectrum: a visual demonstration}
\label{sec:consistency}

Before delving into a more detailed test of the bias modeling and its regime of validity, we aim to give a
visual impression of the consistency between the power spectrum and bispectrum constraints and the importance of
the bias loop corrections in the bispectrum, which are included here for the first time.

To this end we consider for now a \emph{fiducial} model setup, in which we ignore all of the higher-derivative
and scale-dependent stochastic parameters (fixing their values to zero), but keep $N_{P,2}$ for the power
spectrum, as \cite{EggScoCro2006} demonstrated that its contribution is relevant for most of our samples. The
remaining parameters appearing in the general bias expansion up to fourth order are allowed to vary, which
results in a total of six model parameters for the power spectrum, 13 for the bispectrum, and 15 for their
combination (see Table~\ref{tab:num_parameters}).

The parameter constraints resulting from fits including modes up to $k_{\mathrm{max}} = 0.25\,\iMpc$ are shown
for the three galaxy samples in Fig.~\ref{fig:Pgg+Bggg_contours}. In order to focus on the consistency between
the power spectrum and bispectrum, we have limited these plots to the four parameters the two statistics have in
common, and marginalized over all remaining ones. Even for this choice of $k_{\mathrm{max}}$ the power spectrum
alone (gray contours) places only relatively weak constraints on the bias parameters, particularly on the three
higher-order parameters, which is primarily caused by the strong degeneracy between $\gamma_2$ and
$\gamma_{21}$. This degeneracy is due to the galaxy power spectrum being dominated by the one-point propagator
contribution, which is only sensitive to the combination $\gamma_{21} - 6\gamma_ 2/7$. As already discussed in
Sec.~\ref{sec:measurements}, the bispectrum's configuration dependence breaks these degeneracies, leading to
constraints on $b_2$, $\gamma_2$ and $\gamma_{21}$ (see light blue contours) that are tighter by more than an
order of magnitude. In addition, we see that for all three galaxy samples the posteriors from the individual
power spectrum and bispectrum fits are fully consistent. While combining both (blue contours) does not yield
further improvements for the higher-order parameters as they are dominated by the bispectrum, uncertainties on
the linear bias parameter are reduced by a factor four to seven. These constraints are in excellent agreement
with the large-scale measurements of $b_1$ from the galaxy-matter cross power spectrum (using the combined
volume of all simulations), which are shown as the red error bands in Fig.~\ref{fig:Pgg+Bggg_contours}.

\begin{figure*}
  \centering
  \includegraphics[width=0.97\textwidth]{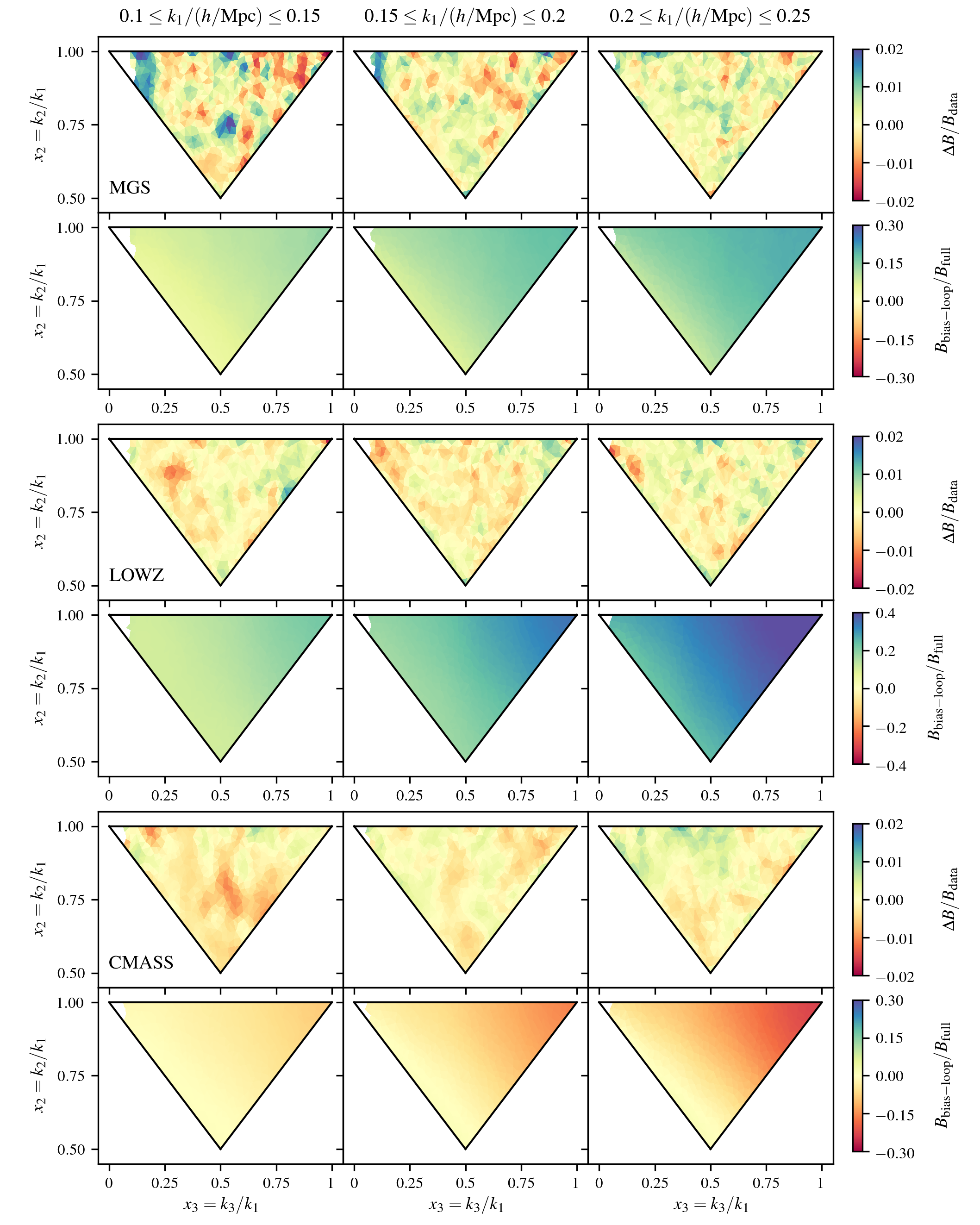}
  \caption{Relative difference between the bispectrum measurements and corresponding models using the best-fit
    parameters obtained from the joint power spectrum and bispectrum fit at $k_{\mathrm{max}} = 0.25\,\iMpc$
    (top rows of the three sub-panels); ratio of the bias loop contributions compared to the full model
    prediction using the same best-fit parameter sets (bottom rows). In both cases the results have been
    averaged over three consecutive $k_1$ bins of width $\Delta k_1 = 0.05\,\iMpc$ (see
    Eq.~\ref{eq:data.Qavg}).}
  \label{fig:bestfit_vs_data}
\end{figure*}

Apart from the parameter constraints themselves, it is illuminating to consider the residuals between the
measurements and the best-fit model predictions. Taking the latter from the joint fits at
$k_{\mathrm{max}} = 0.25\,\iMpc$ and averaging the relative difference
$(B_{\mathrm{model}}-B_{\mathrm{data}})/B_{\mathrm{data}}$ over three different $k_1$ bins, gives the upper rows
in the three sub-panels of Fig.~\ref{fig:bestfit_vs_data}. The first panel displays the results for MGS, where
the relative differences in the first bin from $0.1$ to $0.15\,\iMpc$ can grow as large as $\pm 2\,\%$ for
certain triangle configurations, but are generally at the level of $\sim 1\,\%$ or below for the two subsequent
bins. The fact that the agreement becomes better with an increasing scale of $k_1$ is simply because the fit was
performed at a $k_{\mathrm{max}}$ value larger than the scales involved in the first two bins, which have larger
measurement errors and thus less weight in the likelihood function. As we go up in redshift to the LOWZ and
CMASS samples, the relative differences become even smaller and in the latter case are well below the $1\,\%$
limit for the majority of triangle configurations in all three bins. We stress that this good match between
theory and measurement can be regarded as further evidence for the consistency of our power spectrum and
bispectrum models, since the best-fit parameters derive from their {\em joint fit} instead of the bispectrum alone.

\begin{table}
  \centering\setlength{\extrarowheight}{2pt}
  \caption{Number of fitting parameters for the power spectrum, bispectrum and their combination in different
    model configurations. The scale-dependent noise parameter for the power spectrum is included in all
    cases. The distinction between tree-level and one-loop only concerns the bispectrum.}
  \begin{ruledtabular}
    \begin{tabular}{cc*{4}{>{\centering\arraybackslash}p{0.8cm}}}
      \multirow{2}{*}{Model} & \multirow{2}{*}{Power Spec.} & \multicolumn{2}{c}{Bispectrum}  &
      \multicolumn{2}{c}{Combination}  \Tstrut \\ 
      & & tree & loop & tree & loop \Bstrut \\ \hline
      Fiducial          & \multirow{2}{*}{6} & 5  & 13 & 8  & 15 \Tstrut \\[-0.2em]
      Scale-dep. stoch. &                    & 8  & 16 & 11 & 18 \Tstrut\Bstrut\\[-0.2em]
      Higher-deriv.     & 7                  & 10 & 18 & 13 & 20 \Bstrut \\
    \end{tabular}
  \end{ruledtabular}
  \label{tab:num_parameters}
\end{table}

How important are the bias loop corrections to achieve this level of agreement? This is demonstrated by the
lower rows for each galaxy sample, where we plot the ratio of all loop contributions (except for the matter
part) over the full model prediction, using the same best-fit parameters as above. Already in the first
$k_1$-bin the bias loops contribute with $15$ - $20\,\%$ for MGS and LOWZ (somewhat less for CMASS) for the
nearly equilateral configurations, whereas their impact is less significant for collinear configurations. This
is not unexpected because for fixed $k_1$ the squeezed and folded configurations contain either one or two
shorter (and thus more linear) modes. The same trend continues and amplifies towards larger $k_1$ values,
reaching $15\,\%$ and beyond for most triangle shapes in case of MGS and even up to $40\,\%$ for LOWZ in the
last $k_1$-bin. The CMASS sample is the only one where the loop diagrams predominantly contribute negatively and
where the effect is more strongly skewed towards equilateral configurations, with collinear shapes remaining
mostly unchanged. Overall these plots show that the bias loop corrections quickly gain importance beyond
$k_{\mathrm{max}} \sim 0.15\,\iMpc$ and taking into account that our measurement uncertainties are of the order
$\sim 10\,\%$ they can no longer be ignored from that point onward. However, their reduced impact on collinear
triangles~\cite{EggScoSmi1906} suggests that for these particular configurations the validity of a given model can be extended to
larger values of $k_{\mathrm{max}}$ than for equilateral ones, analogously to the behavior of the matter bispectrum~\cite{ScoFri9907}. This motivates the introduction of a
configuration-dependent cutoff scale, but we leave a more detailed exploration of the benefits in connection to
this for a future study.

\subsection{Detailed test of the one-loop galaxy bias model}
\label{sec:detailed-model-test}

Having seen that loop corrections to the galaxy bispectrum become relevant starting from
$k_{\mathrm{max}} \sim 0.15\,\iMpc$, we now want to assess the validity of the tree-level or one-loop model in a
more quantitative manner.

\subsubsection{Performance metrics}
\label{sec:performance-metrics}

As already done in \cite{EggScoCro2006} we are going to estimate the range of validity by a combination of two
performance metrics, the figure of bias (FoB) and the goodness-of-fit. The former is defined as the deviation
between the posterior means $\bar{\theta}_{\alpha}$ of a set of parameters and their fiducial values
$\theta_{\mathrm{fid},\alpha}$, weighted by the inverse parameter covariance matrix:
\begin{equation}
  \label{eq:resultsA.fob}
  \mathrm{FoB} \equiv \left[\sum_{\alpha,\beta} \left(\bar{\theta}_{\alpha} -
      \theta_{\mathrm{fid},\alpha}\right)\,S_{\mathrm{tot},\alpha\beta}^{-1}\,\left(\bar{\theta}_{\beta} -
      \theta_{\mathrm{fid},\beta}\right)\right]^{1/2}\,.
\end{equation}
The total parameter covariance matrix,
$S_{\mathrm{tot},\alpha\beta} = S_{\alpha\beta} + S_{\mathrm{fid},\alpha\beta}$, is given by the sum of the
posterior covariance and measurement uncertainties in the fiducial values (in case they are not known with
complete certainty). In this section we only have a fiducial measurement of the linear bias parameter to compare
against\footnote{Mean and standard deviation of the fiducial measurement for all of our samples can be found in
  Table~1 of \cite{EggScoCro2006}.}, in which case the FoB simplifies to
$\left|\bar{b}_1 - b_{1,\mathrm{fid}}\right|/\sqrt{\sigma_{b_1}^2 + \sigma_{\mathrm{fid},b_1}^2}$. As a measure
of the goodness-of-fit we take the $\chi^2$ computed following the description in Sec.~\ref{sec:fitting} and
compare its value to the confidence limits of a $\chi^2$-distribution with
\begin{equation}
  \label{eq:resultsA.dof}
  \mathrm{dof} = N_R \times N_{\mathrm{bin}} - N_p
\end{equation}
degrees of freedom, where $N_p$ are the number of free fitting parameters. Either metric on its own is
insufficient to faithfully judge the validity of the model, since an acceptable $\chi^2$ might hide a biased
recovery of parameters, whereas the FoB can be subject to posterior projection effects (especially when only
based on a small subset of the fitting parameters) or can accidentally be low. For that reason we define the
validity range $k_{\dagger}$ as the cutoff scale at which the combination of the two exceeds a critical value in
comparison to their respective $95\,\%$ limits,
\begin{equation}
  \label{eq:resultsA.kdagger}
  \frac{\mathrm{FoB}(k_{\dagger})}{\mathrm{FoB}_{95\%}} + \frac{\chi^2(k_{\dagger}) -
    \mathrm{dof}(k_{\dagger})}{\chi^2_{95\%}(k_{\dagger}) - \mathrm{dof}(k_{\dagger})} = \sigma_{\mathrm{crit}}
\end{equation}
where $\mathrm{FoB}_{95\%} = 2$ when based on a single parameter and we set
$\sigma_{\mathrm{crit}} = 1$\footnote{Note that this definition differs slightly from the one introduced in
  \cite{EggScoCro2006}. Even though they do not have a significant impact on the results, the changes were made
  in order to treat the FoB and goodness-of-fit metrics on equal footing.}.

Finally, using a third metric --- the figure of merit (FoM) --- we contrast the validity scales of various
modeling assumptions in terms of their constraining power. This allows us to determine whether there is a
benefit of adding complexity to the model, while unlocking the information from more nonlinear scales. We define
the FoM as the inverse of the posterior volume enclosed by the $68\,\%$ confidence limit, normalized by the
fiducial parameter values, so that
\begin{equation}
  \label{eq:resultsA.fom}
  \mathrm{FoM} \equiv \frac{1}{\sqrt{\det
      \left[S_{\alpha\beta}/\left(\theta_{\mathrm{fid},\alpha}\,\theta_{\mathrm{fid},\beta}\right)\right]}}\,.
\end{equation}
As for the FoB, in this section we measure the FoM solely in terms of the linear bias parameter, and therefore the
FoM is given by $b_{1,\mathrm{fid}}/\sigma_{b_1}$.

\begin{figure*}
  \centering
  \includegraphics{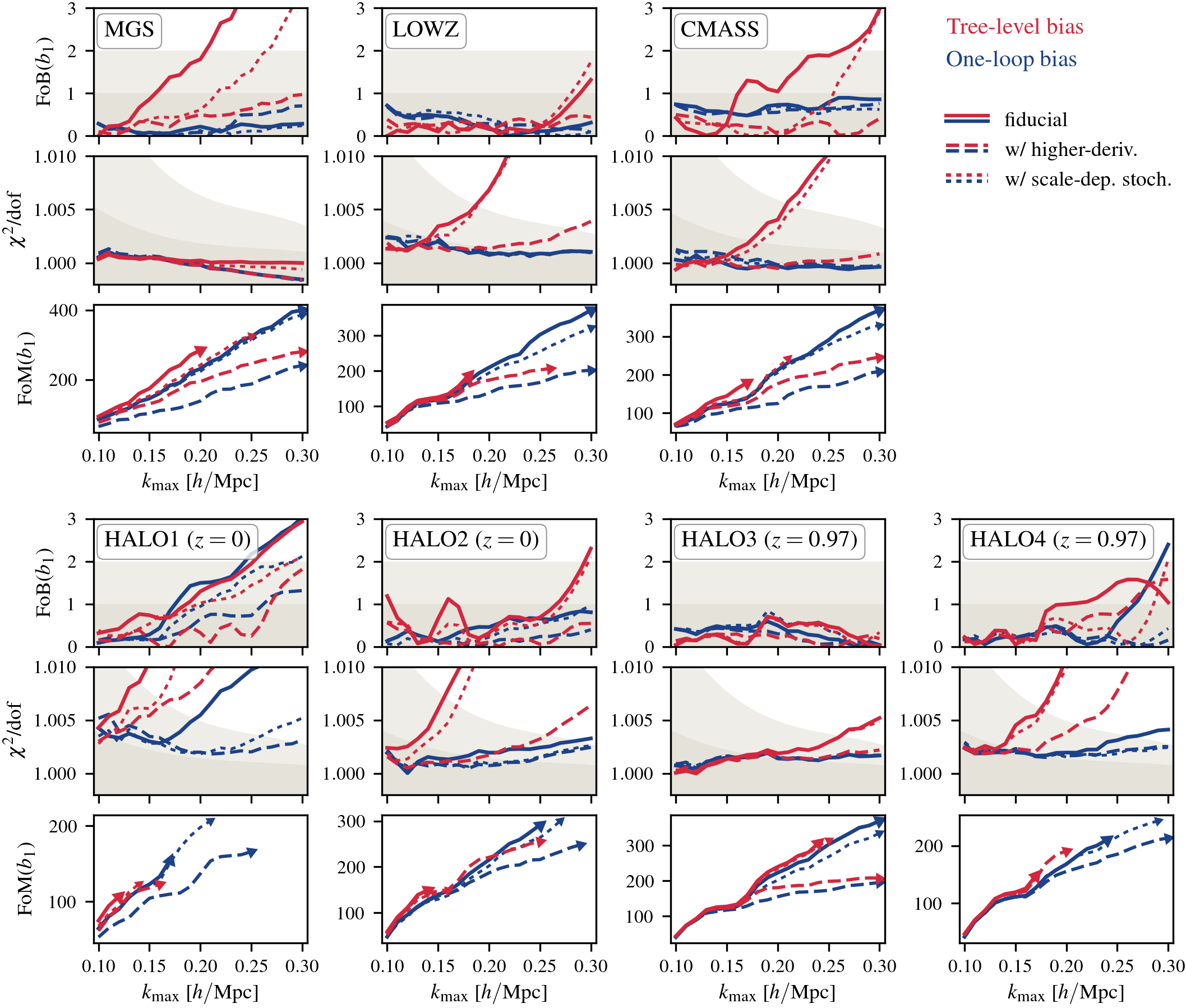}
  \caption{Figure of bias (FoB), goodness-of-fit and figure of merit (FoM) for joint fits of the galaxy or halo
    power spectrum and bispectrum as a function of the maximum $k$-mode allowed to participate in the
    fit. Differently colored lines indicate whether bias loop corrections in the bispectrum model have been
    included (blue) or not (red). Solid lines correspond to a bispectrum model that includes neither
    higher-derivative terms, nor scale-dependent stochasticity, while dashed lines account for the former and
    dotted lines for the latter (see Table~\ref{tab:num_parameters} for the number of fitting parameters in
    these cases and note that the power spectrum model always includes the scale-dependent stochastic term). The
    FoM is truncated at the estimated validity scale of the respective model, indicated by an arrowhead
    symbol. Grey shaded areas depict the $68\,\%$ and $95\,\%$ confidence limits.}
  \label{fig:fob_PB_trueM}
\end{figure*}

In Fig.~\ref{fig:fob_PB_trueM} we plot all three of these metrics derived from joint fits of the power spectrum
and bispectrum with various $k_{\mathrm{max}}$ values for both, the galaxy and halo samples. The power spectrum
model is fixed and includes all relevant loop corrections as well as the scale-dependent stochastic parameter
$N_{P,2}$, but we distinguish between different modeling configurations for the bispectrum: tree-level bias
terms only (red lines), and the full one-loop model presented in Sec.~\ref{sec:MPformalism} (blue lines). In
both cases we further consider the same fiducial setup from Sec.~\ref{sec:consistency} without
higher-derivative terms and scale-dependent stochasticity, but now also allow separately for either of these
effects, depicted by the dashed and dotted lines. Note that when including the higher-derivative terms in the
bispectrum model, we also include the corresponding term in the power spectrum, although it does not enter with
a free parameter since we have eliminated the stress-tensor corrections (see
Sec.~\ref{sec:high-deriv-bias}). The varying number of fitting parameters in each of these cases is given by
the last two columns of Table~\ref{tab:num_parameters}. For easier visual comparison between the models we have
evaluated $k_{\dagger}$ according to Eq.~(\ref{eq:resultsA.kdagger}) and stopped plotting the FoM at that scale,
which is indicated by an arrowhead symbol.

\vspace*{-1em}
\subsubsection{Fiducial case}
\label{sec:fiducial-case}

Starting with the fiducial case we observe that ignoring the bias loop corrections generally diminishes the
agreement with the measurements --- as is evident from the $\chi^2/\mathrm{dof}$ panels --- and for some of the
samples leads to a clearly biased estimation of $b_1$. The validity scales are therefore significantly reduced
compared to the one-loop bias model and suggest a break-down of the theory description soon after
$k_{\mathrm{max}} = 0.17\,\iMpc$. This is consistent with our previous discussion on the importance of the loop
corrections based on the best-fit results alone and besides shows that they cannot be adequately absorbed by the
tree-level terms. The one-loop model, on the other hand, remains valid over the entire tested range of scales up
to $k_{\mathrm{max}} = 0.3\,\iMpc$, with the exception of some of the halo samples, in particular HALO1 (where
it fails before $0.2\,\iMpc$). Most importantly, there is a clear benefit from including the bias loops and
extending the model further into the nonlinear regime, since this leads to a greatly improved FoM (up to a
factor 1.5 to 2) despite the increased number of nuisance parameters.

\subsubsection{Extensions to higher-derivatives and scale-dependent stochasticity}
\label{sec:model-extensions}

Inclusion of the five higher-derivative contributions yields an enhanced model performance for the tree-level
bias case, extending its validity for most samples to a scale equivalent with that of the fiducial one-loop
model (at least within the range of $k_{\mathrm{max}}$ that we consider). Although its goodness-of-fit becomes
worse when including more nonlinear scales, this seems to imply that the higher-derivative parameters
efficiently absorb the neglected bias contributions. However, marginalization over these additional parameters
greatly reduces the FoM, which stays well below what we obtain for the fiducial one-loop model for all samples,
even though the latter has a larger number of free parameters in total. The same effect can be observed when
higher-derivatives and one-loop bias terms are taken into account simultaneously, suggesting that for our
samples there is no advantage to be gained from doing so.

The extension to scale-dependent stochasticity has little impact on the validity ranges for the tree-level case,
but in combination with one-loop bias produces very similar $\chi^2$ values as the extension to
higher-derivatives, and thus slightly increases the scales before the model breaks down for several samples
(HALO1, HALO2 and HALO4). Moreover, the three extra parameters do not strongly penalize the FoM, such that for
these samples we are able to achieve a higher overall constraining power. For that reason we are going to
continue to work with the scale-dependent noise model for these cases in the following sections.

\subsection{Constraints on galaxy bias parameters}
\label{sec:galaxy-bias-constraints}

\begin{figure}
  \centering
  \includegraphics{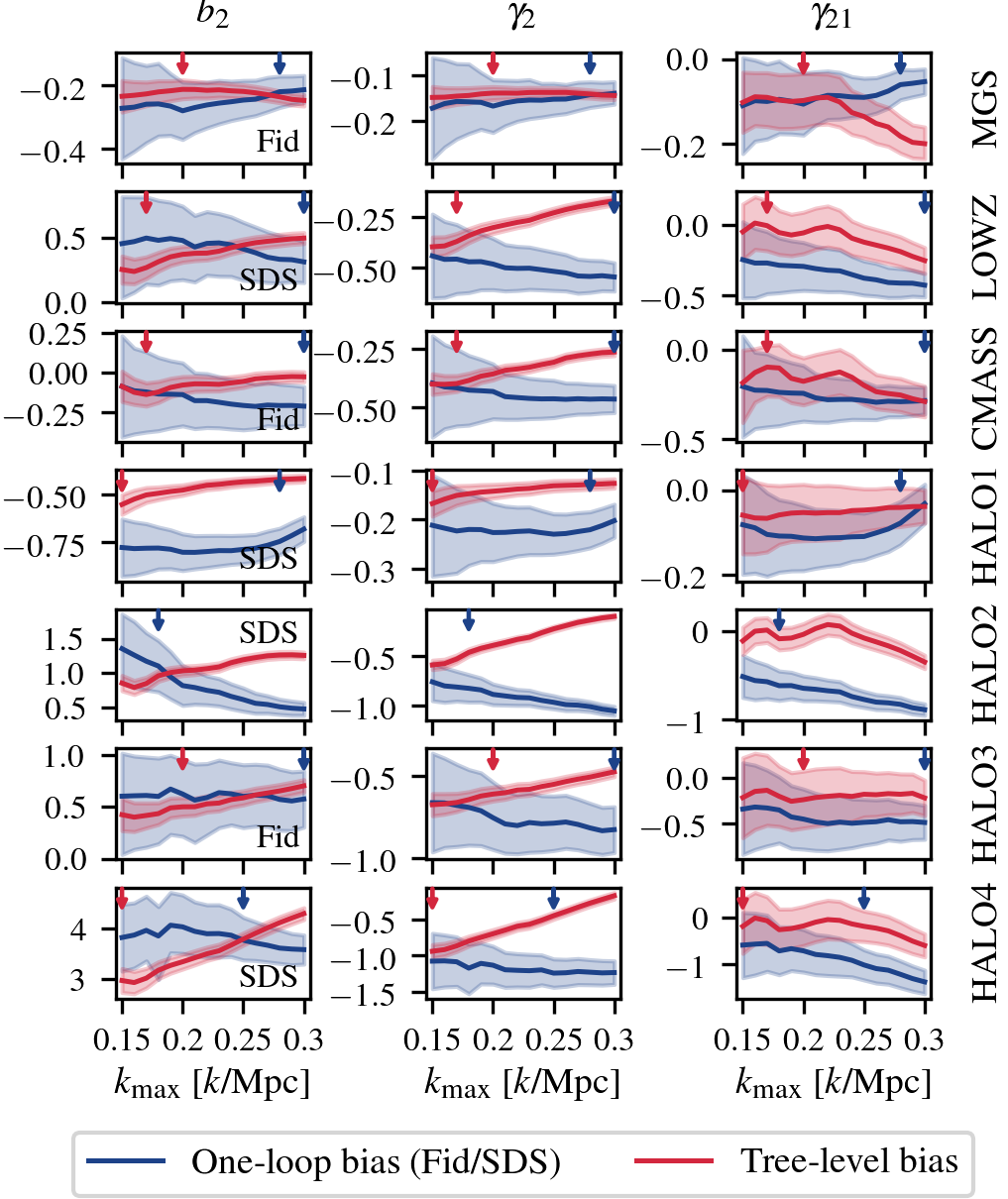}
  \caption{Constraints on $b_2$, $\gamma_2$ and $\gamma_{21}$ from joint fits of the power spectrum and
    bispectrum as a function of the cutoff scale $k_{\mathrm{max}}$. The blue error bands represent the results
    from the one-loop bias model in the fiducial (Fid) configuration or including scale-dependent stochasticity
    (SDS) as indicated in the panels for each sample. Red error bands show the fiducial tree-level bias model
    for comparison. Arrowheads of matching colors mark the scale at which we identify a trend in the constraints
    of one of the parameters (see text). For tree-level results on HALO2 this happens before $k_{\mathrm{max}}=0.15\,\iMpc$.}
  \label{fig:b2g2g21vskmax}
\end{figure}

\subsubsection{Dependence on cutoff $k_{\mathrm{max}}$}
\label{sec:cutoff-dependence}

In this section we consider in greater detail the constraints on the various bias parameters obtained from the
best-performing models identified above. We begin with the three higher-order bias parameters that appear both
in the power spectrum and bispectrum, and are thus the parameters (apart from $b_1$) which are most strictly
constrained. These are $b_2$, $\gamma_2$ and $\gamma_{21}$, whose mean posterior values from a joint fit along
with their 1-$\sigma$ uncertainties, indicated by the shaded error band, are shown for each sample in
Fig.~\ref{fig:b2g2g21vskmax}.

The results are plotted as a function of the maximum $k$-mode allowed in the fit, and the blue lines correspond
to either the fiducial one-loop model (Fid) or with the inclusion of scale-dependent stochasticity (SDS),
depending on whichever configuration performed better according to Fig.~\ref{fig:fob_PB_trueM} (note, however,
that for LOWZ we instead show the constraints from the one-loop SDS model for reasons becoming clear in a
moment). While the uncertainties decrease with increasing $k_{\mathrm{max}}$, we clearly see that there are
little to no shifts in the parameter mean values for the majority of cutoff scales and samples. The most obvious
one occurs for HALO2, where the constraints, particularly on $b_2$, are driven towards smaller values starting
from a scale shortly after $k_{\mathrm{max}} = 0.15\,\iMpc$. Encountering shifts in the recovered parameters as
one ventures deeper into the nonlinear regime signifies that the model is attempting to compensate potentially
missing contributions and can therefore be regarded as another indicator for a breakdown of its validity. The
fact that we do not observe such strong shifts in the one-loop model is in very good agreement with our
conclusions drawn in Sec.~\ref{sec:detailed-model-test} based on the FoB and goodness-of-fit. For comparison,
Fig.~\ref{fig:b2g2g21vskmax} also shows the analogous results for the fiducial tree-level model, in which case
the parameter shifts are simultaneously much stronger and arise at smaller $k_{\mathrm{max}}$ for each of the
samples. Again, this is consistent with our analysis in Sec.~\ref{sec:detailed-model-test}. We stress that
the qualitative behavior of all the remaining parameters not shown in the plot is very similar.

In the following we would like to contrast our bias measurements with the coevolution and PBS relations and for
a stringent comparison it is important to choose a $k_{\mathrm{max}}$ value at which no significant shift in the
constraints has occurred yet. To this end we check whether the mean posterior value of a given parameter and at
a given cutoff scale is consistent with the $68\,\%$ confidence intervals of all previous $k_{\mathrm{max}}$,
starting from $0.1\,\iMpc$. We follow this procedure for every parameter that is being varied in the model and
from all of these determine the maximum scale at which the criterion is still satisfied, yielding a unique scale
for each combination of model and sample. This scale is indicated by the arrowhead at the top of each panel in
Fig.~\ref{fig:b2g2g21vskmax} using matching colors to distinguish between the one-loop or tree-level model, and
we see that this quantitative assessment confirms our purely visual discussion from above. Finally, we note that
in case of LOWZ this analysis revealed stronger parameter shifts for the fiducial one-loop model compared to the
extension including scale-dependent stochasticity. Since it allows us to use a larger $k_{\mathrm{max}}$ value for
the constraints presented in the next section, we adopt the latter for this particular sample.

\subsubsection{Comparison to coevolution and PBS relations}
\label{sec:comp-co-evol}

\begin{figure*}
  \centering
  \includegraphics{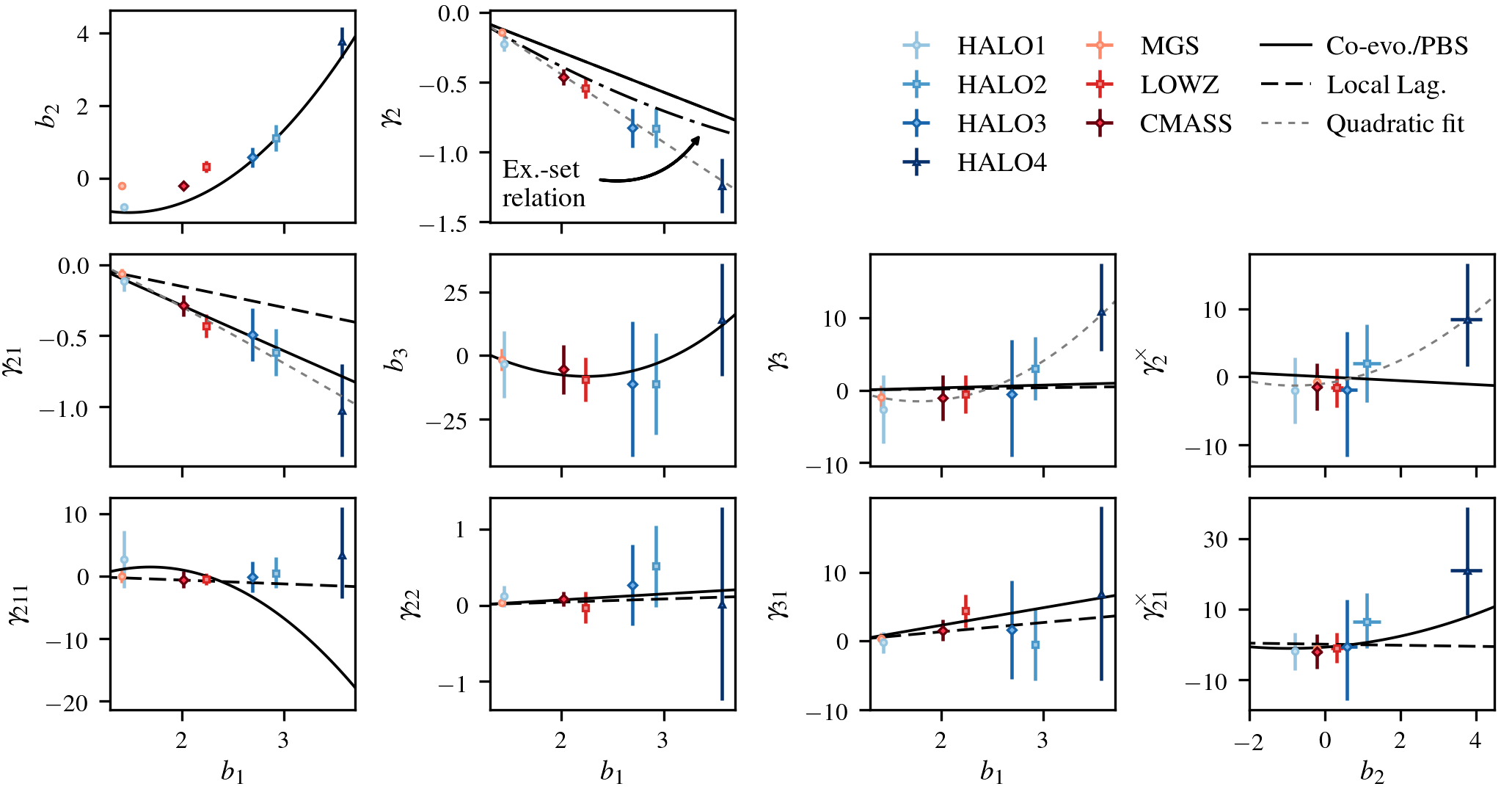}
  \caption{Constraints on galaxy bias parameters, plotted against the linear bias parameter, or nonlinear bias
    parameter $b_2$ in case of $\gamma_2^{\times}$ and $\gamma_{21}^{\times}$ (last column). The constraints
    derive from a joint power spectrum and bispectrum fit using the one-loop model with or without
    scale-dependent stochasticity (indicated for each sample in Fig.~\ref{fig:b2g2g21vskmax}) at a conservative
    choice of $k_{\mathrm{max}}$ values (see text). Dashed lines correspond to the local Lagrangian
    approximation, while solid lines show either the more general coevolution relations or PBS predictions (in
    case of $b_2$ and $b_3$). The dot-dashed line for $\gamma_2$ is the excursion-set inspired relation used in
    \cite{EggScoCro2006}, and the light dotted lines are quadratic fits that were used to plot coevolution
    relations depending on more than one parameter as a function of $b_1$ only.}
  \label{fig:bias_constraints}
\end{figure*}

As we have seen in Sec.~\ref{sec:consistency}, the combination of power spectrum and bispectrum allows us to
place considerably stronger constraints on $b_2$, $\gamma_2$ and $\gamma_{21}$ than the power spectrum alone. In
addition, through the one-loop corrections to the bispectrum we are sensitive to the remaining third-order bias
parameters and even the fourth-order parameters associated with NLE operators. It is interesting to examine
whether these constraints enable us to shed more light on the coevolution and halo-calibrated PBS relations
presented in Sec.~\ref{sec:coevolution}, which are often employed in the analyses of real survey data.

We use the one-loop model with or without the addition of scale-dependent stochasticity (as indicated in
Fig.~\ref{fig:b2g2g21vskmax}) and choose to display the constraints at different $k_{\mathrm{max}}$ values for
each sample. In order to be conservative we take the minimum between the model validity scale determined in
Sec.~\ref{sec:detailed-model-test} and the scale before parameter shifts appear as found in
Sec.~\ref{sec:cutoff-dependence}, which leads to the results presented in Fig.~\ref{fig:bias_constraints},
and the values reported in Table~\ref{tab:constraints_b2g2g21} and \ref{tab:constraints_b3rd4th}. The
parameter constraints are shown as functions of $b_1$, except for $\gamma_2^{\times}$ and
$\gamma_{21}^{\times}$, which are plotted against $b_2$ instead.

\begin{table}
  \centering 
  \caption{Constraints on $b_2$, $\gamma_2$ and $\gamma_{21}$ from a joint analysis of the power spectrum and
    bispectrum. The fiducial values for $b_1$ can be found in \cite{EggScoCro2006}.}
  \begin{ruledtabular}
    \begin{tabular}{cccc}
      Sample & $b_2$ & $\gamma_2$ & $\gamma_{21}$  \Tstrut\Bstrut \\  \hline
      MGS & $-0.22 \pm 0.05$ & $-0.14 \pm 0.03$ & $-0.060 \pm 0.04$ \Tstrut \\
      LOWZ & $0.3 \pm 0.2$ & $-0.54 \pm 0.07$ & $-0.43 \pm 0.08$ \\
      CMASS & $-0.2 \pm 0.1$ & $-0.46 \pm 0.06$ & $-0.29 \pm 0.08$ \\
      HALO1 & $-0.80 \pm 0.09$ & $-0.23 \pm 0.05$ & $-0.11 \pm 0.07$ \\
      HALO2 & $1.1 \pm 0.4$ & $-0.83 \pm 0.14$ & $-0.62 \pm 0.17$ \\
      HALO3 & $0.6 \pm 0.3$ & $-0.83 \pm 0.14$ & $-0.49 \pm 0.19$ \\
      HALO4 & $3.8 \pm 0.4$ & $-1.2 \pm 0.2$ & $-1.0 \pm 0.3$ \Bstrut
    \end{tabular}
  \end{ruledtabular}
  \label{tab:constraints_b2g2g21}
\end{table}

Beginning with the two second-order parameters, $b_2$ and $\gamma_2$, we see that the four halo sample
constraints on the former are in very good agreement with the PBS prediction from \cite{LazWagBal1602} (shown by
the black solid line). This prediction was calibrated against measurements from their own halo catalogs using
separate universe simulations, which means that arriving at the same results albeit with an entirely different
approach and different halo catalogs, is further convincing proof of the robustness of our model and
methodology. Curiously, we find a significantly larger (in terms of our measurement uncertainties) $b_2$ for all
three galaxy samples than the PBS prediction at the equivalent $b_1$ values. For the tidal bias parameter we
find very obvious deviations from the local Lagrangian approximation (in this case identical with the general
coevolution relation shown by the solid line), demonstrating this trend more clearly than the previous power
spectrum and bispectrum studies in \cite{SheChaSco1304,SaiBalVla1405}. Moreover, this conclusion is fully consistent with the
two analyses \cite{LazSch1712,AbiBal1807}, which measured bias parameters through cross-correlations between the
halo density and second- and third-order fields representing the various operators appearing in the bias
expansion (see Eq.~\ref{eq:model.expansion}). However, we note that in the high-mass (large $b_1$) limit our
$\gamma_2$ measurements are in tension with the excursion-set relation from \cite{SheChaSco1304}, which is not
the case for those reported in \cite{LazSch1712} (cf. Fig.~1 of \cite{EggScoCro2006}), but seems to be in good
agreement with \cite{AbiBal1807}.

Moving on to the first nonlocal parameter, $\gamma_{21}$, we obtain an equally evident deviation from the local
Lagrangian approximation (dashed line). This was also reported in \cite{LazSch1712,AbiBal1807} (for the
equivalent parameters $b_{\mathrm{td}}$ and $b_{\Gamma_3}$, respectively), but is in contrast with
\cite{SaiBalVla1405}, who found their results ($b_{\mathrm{3nl}}$ in their notation) to be consistent with a
local Lagrangian bias expansion. Since they only considered the tree-level bispectrum, $\gamma_{21}$ only enters
through the one-loop power spectrum, where it is partly degenerate with a potential higher-derivative
contribution. This was not included in their model, and so \cite{LazSch1712} argued that their study likely
produced biased measurements. We did not take into account higher-derivative contributions to derive the
constraints shown in Fig.~\ref{fig:bias_constraints} either, but account for scale-dependent stochasticity
through $N_{P,2}$, which the power spectrum is actually more sensitive to, as we showed in
\cite{EggScoCro2006}, and which also correlates strongly with $\gamma_{21}$ (see
Sec.~\ref{sec:param-correlations}). In addition, we include the bispectrum loop corrections, which grant
further constraining power on $\gamma_{21}$, and prevent a slight over-estimation that occurs for the tree-level
model even when $N_{P,2}$ is being varied (as we always do). According to Fig.~\ref{fig:b2g2g21vskmax} this is
the case for most samples when comparing the tree-level and one-loop constraints at the respective scales before
we identify a cutoff dependence.

After this discussion on the robustness of our $\gamma_{21}$ measurements, it is interesting to see that they
are in excellent agreement with the solid line, which depicts $\gamma_{21}$ in the general coevolution
assumption, i.e. $\gamma_{21,{\cal L}} = 0$ and subsequent conserved evolution. This relation depends on both,
$b_1$ and $\gamma_2$ (see Eq.~\ref{eq:model.g21evo}), and in order to plot it as a function of $b_1$ only, we
have fitted a simple quadratic form to the $\gamma_2(b_1)$ data from all samples, which is shown by the light
dotted line in the top right panel. This agreement is remarkable as it suggests that the leading NLE operator is
absent from the initial bias expansion and its effect only comes to bear through gravitational
evolution. Although a similar trend was already present in \cite{LazSch1712,AbiBal1807}, our reduced errors on
$\gamma_{21}$ have allowed for a more precise test of this assertion. However, one should keep in mind that such
tests are to be regarded in the context of the adopted statistical uncertainties, i.e. effective volume, which
we assumed to be $V_{\mathrm{eff}} = 6\,(\Gpc)^3$ --- significantly larger than the redshift slices analyzed in
past galaxy surveys or what is expected in upcoming ones.

The constraints on all of the remaining bias parameters come solely from the loop corrections of the bispectrum
and are therefore much less stringent, and in many cases still consistent with zero. As above we compare the
results for each parameter with the respective local Lagrangian or coevolution relations and the PBS prediction
in case of $b_3$, and apply the same strategy to plot these relations as single functions of $b_1$ and $b_2$,
which requires quadratic fits to the data from $\gamma_{21}$, $\gamma_3$ and $\gamma_2^{\times}$ (see
Sec.~\ref{sec:coevolution} for the dependencies on these parameters). In general, we find sensible results
that follow these relations fairly closely, though we lack the constraining power to distinguish clearly between
the local Lagrangian approximation and general coevolution, like we could for $\gamma_{21}$. The only exception
seems to be the fourth-order parameter $\gamma_{211}$, which shows a strong deviation from the latter for the
highly-biased halo samples. However, this should be considered with caution as the coevolution relation is
dominated by the quadratic fit to $\gamma_3$, which in turn is influenced strongly by the last data point for
the HALO4 sample. It will be interesting to combine our joint power spectrum and bispectrum fits here with the
large-scale trispectrum, where the four third-order bias parameters appear at leading order and should thus
show a boost in sensitivity. 

Finally, let us return to the question why there is an evident difference between the galaxy and halo samples in
the trend for $b_2$, whereas they are identical for $\gamma_2$ and $\gamma_{21}$. Since the galaxy samples were
generated using an HOD approach, it is illuminating to consider the relation between the galaxy and halo bias
parameters in the large-scale limit. Given a mean halo mass function, $\bar{n}_h(M_h)$, and mean occupation
function of central and satellite galaxies, $\left<N_g(M_h)\right>$, we can write a galaxy bias parameter
$b_{{\cal O},g}$ associated to some operator ${\cal O}$ of the bias expansion as \cite[e.g.,][]{SefSco0503}
\begin{equation}
  \label{eq:resultsA.bHOD}
  b_{{\cal O},g} = \frac{1}{\bar{n}_g} \int \D{\ln{M_h}} \, \bar{n}_h(M_h)\,\left<N_g(M_h)\right>\,b_{{\cal
      O},h}(M_h)\,,
\end{equation}
where $\bar{n}_g = \int \D{\ln{M_h}}\,\bar{n}_h(M_h)\,\left<N_g(M_h)\right>$ is the mean galaxy number
density. If the halo bias for this particular operator obeys a linear relationship with a set of other
bias parameters (with numerical coefficients $\alpha_i$),
\begin{equation}
  \begin{split}
    b_{{\cal O},h}(M_h) &= \alpha_1\,b_{{\cal O}_1,h}(M_h) + \alpha_2\,b_{{\cal O}_2,h}(M_h) + \ldots \\
    &= F\left[b_{{\cal O}_1,h}(M_h),\,b_{{\cal O}_2,h}(M_h),\,\ldots\right]\,,
  \end{split}
\end{equation}
then it follows immediately from Eq.~(\ref{eq:resultsA.bHOD}) that the HOD galaxy bias parameter must follow the
same relation, i.e. $b_{{\cal O},g} = F\left[b_{{\cal O}_1,g},\,b_{{\cal O}_2,g},\,\ldots\right]$. From
Fig.~\ref{fig:bias_constraints} we see that the halo sample constraints on $\gamma_2$ and $\gamma_{21}$ are well
described by linear functions of $b_1$, but for $b_2$ this is clearly invalid. Assuming that we can take these
measurements as representative for the general trend at arbitrary halo masses, our results for the galaxy
samples appear to be sensible. In particular, the fast growth of $b_2$ with increasing $b_1$ should yield a
stronger weighting of the high-mass halos in Eq.~(\ref{eq:resultsA.bHOD}) and thus larger values for $b_{2,g}$.

\subsubsection{Parameter correlations}
\label{sec:param-correlations}

As mentioned before, the $\gamma_2$ and $\gamma_{21}$ terms contribute only as the combination
$\gamma_{21} - 6\gamma_2/7$ to the evolved one-point propagator, leading to a strong degeneracy between these
two parameters when fitting the power spectrum alone. Let us now consider whether there are equally strong
correlations in the joint power spectrum and bispectrum parameter space.

\begin{figure}
  \centering
  \includegraphics[width=\columnwidth]{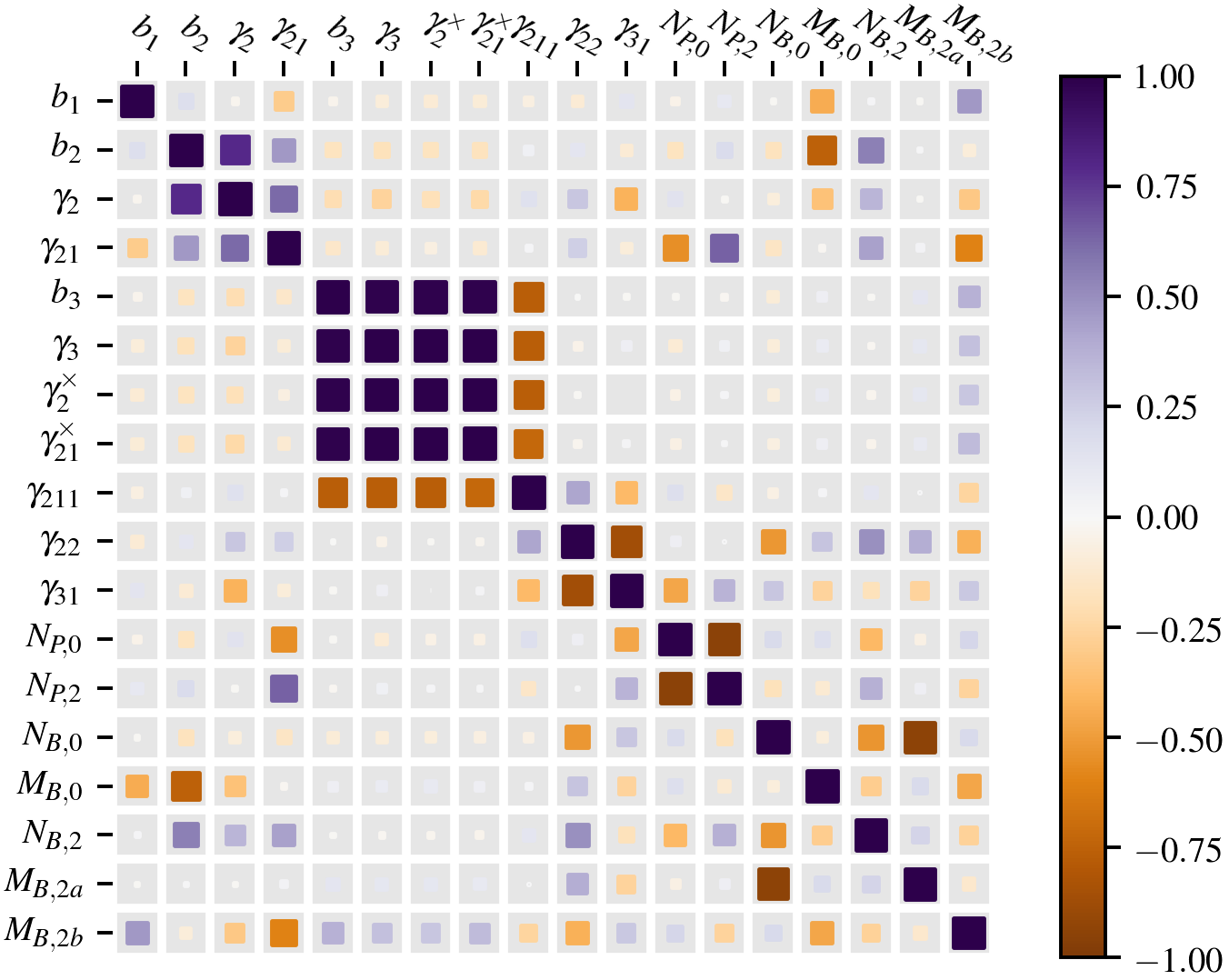}
  \caption{Full parameter correlation matrix for a joint power spectrum and bispectrum fit to the LOWZ data
    sample at $k_{\mathrm{max}} = 0.3\,\iMpc$, including all scale-dependent stochastic parameters. Note that
    the size of the squares scales with the absolute value of the correlation coefficients.}
  \label{fig:par_corr}
\end{figure}

In Fig.~\ref{fig:par_corr} we show the full parameter correlation matrix $S_{\alpha\beta}$ originating from
fitting the scale-dependent stochastic model for the power spectrum and bispectrum to the LOWZ data with cutoff
scale $k_{\mathrm{max}} = 0.3\,\iMpc$, which is qualitatively representative for the remaining samples and for
different cutoff scales. The two most striking features in this plot are the correlations between the three
higher-order bias parameters appearing both in the power spectrum and bispectrum --- $b_2$, $\gamma_2$ and
$\gamma_{21}$ --- and among the three remaining third-order parameters, in addition to $\gamma_{21}^{\times}$
and $\gamma_{211}$ at fourth order. As explained before, the correlation between $\gamma_2$ and $\gamma_{21}$ is
expected based on their identical contribution to the one-point propagator, but the addition of the bispectrum
partially breaks this degeneracy and so decreases the correlation coefficient to $\sim 0.7$ from $\sim 1$, which
we would obtain for the power spectrum alone. The correlation between $b_2$ and $\gamma_2$ is equally strong,
but its origin is less obvious. It might arise as a consequence of the non-zero spherical average of the
second-order Galileon, i.e. $\overline{{\cal G}}_2 = -2/3\,\delta^2$ \cite{EggScoSmi1906}, which would imply a
dependence on the combination $b_2 - 4\gamma_2/3$. We indeed find this to be in good agreement with the
degeneracy directions obtained for the MGS, HALO1 and HALO4 samples, while all other samples display trends that
are $25$ - $35\,\%$ steeper.

\begin{figure}
  \centering
  \includegraphics[width=\columnwidth]{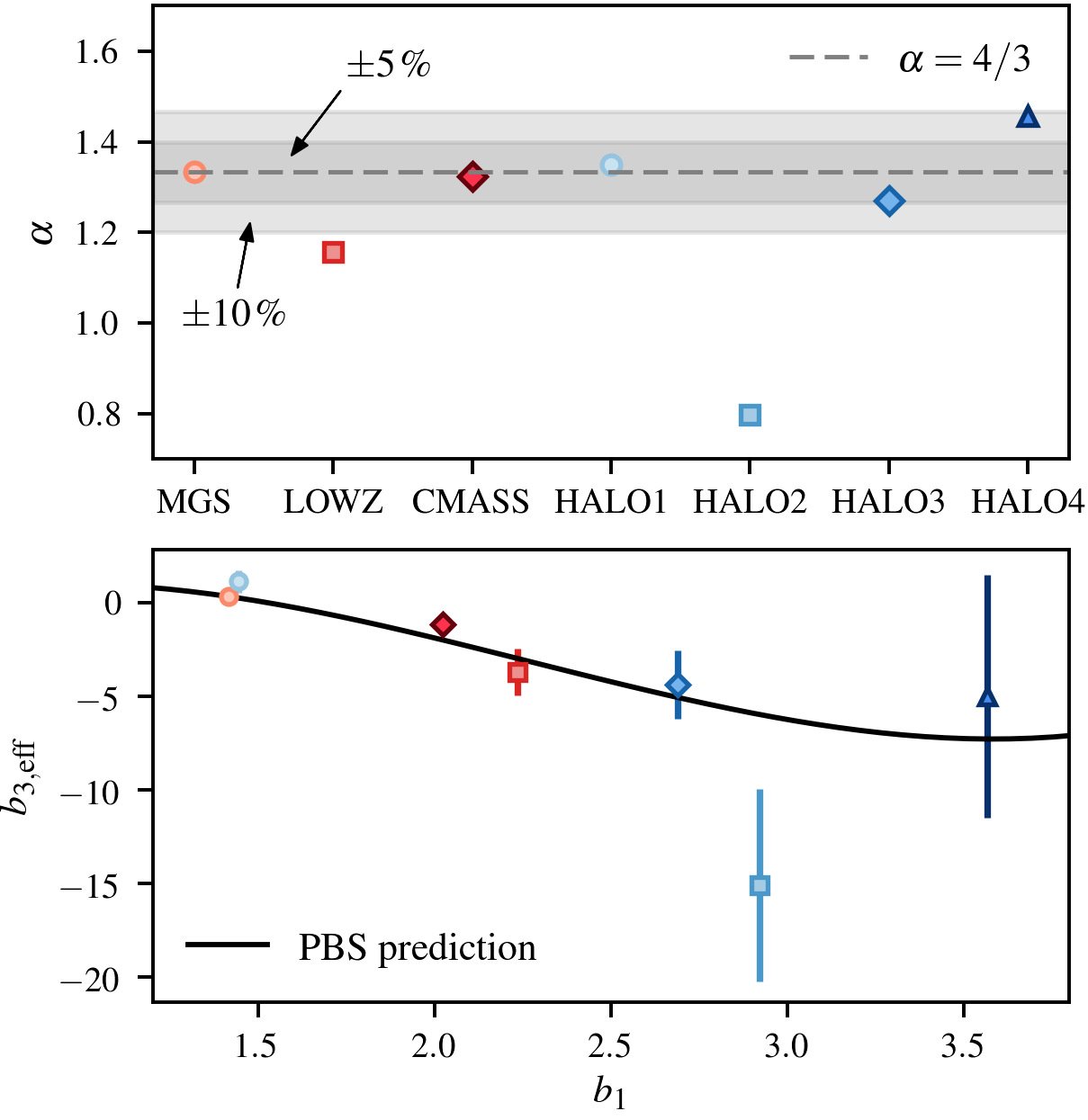}
  \caption{Upper panel: estimated slope $\alpha$ in the relation $b_{3,\mathrm{eff}} = b_3 - 4\gamma_2^{\times}
    + \alpha\,\gamma_3$, compared to the value implied by the spherically averaged bias expansion (dashed
    line); the gray bands indicate $5$ and $10\,\%$ deviations from this value. Lower panel: constraints on
    $b_{3,\mathrm{eff}}$ with $\alpha = 4/3$, plotted as a function of the fiducial linear bias of each sample,
    and compared to the PBS prediction (solid line).}
  \label{fig:g3slope_b3eff_constraints}
\end{figure}

Moving on to the second block of correlated parameters, we first note that strong correlations between
$\gamma_2^{\times}$ and $\gamma_{21}^{\times}$, as well as between $\gamma_3$ and $\gamma_{211}$ are to be
expected because they contribute the same scale-dependent terms to the evolved two-point propagators (analogous
to the $\gamma_2$ and $\gamma_{21}$ terms in the one-point propagator). In particular, one can show that
\begin{equation}
  \Gamma_{g,\gamma_{21}^{\times}}^{(2)} = -\frac{7}{6}\Gamma_{g,\gamma_{2}^{\times}}^{(2)}\,, \quad \text{and}
  \quad \Gamma_{g,\gamma_{211}}^{(2)} = \frac{7}{9}\Gamma_{g,\gamma_{3}}^{(2)}\,,
\end{equation}
which indicates that the bispectrum is mostly sensitive to the combinations
$\gamma_{21}^{\times} - 6\gamma_2^{\times}/7$ and $\gamma_{211} + 9\gamma_3/7$, provided that the third-order
parameters are not strongly constrained themselves. Note that the opposite signs in these combinations also
explain the correlation or anti-correlation of the two parameter pairs in Fig.~\ref{fig:par_corr}. The
additional correlations with the remaining parameters from the same block are induced by a strong degeneracy
between $b_3$, $\gamma_3$ and $\gamma_2^{\times}$. This degeneracy can be described by a single principle
component, which means that there must be two parameter combinations that are much tighter constrained than the
three parameters individually. Such combinations can be determined from the parameter correlation matrix, but in
order to gain some insight into their potential origin and whether they are universal across the various
samples, let us again consider the spherical average of the bias expansion. Using that
$\overline{{\cal G}_3} = 2/9\,\delta^3$ \cite{EggScoSmi1906} we see that in this case the three parameters
effectively appear in the combination
\begin{equation}
  \label{eq:resultsA.b3g3g2x_degeneracy}
  b_{3,\mathrm{eff}} = b_3 - 4\gamma_2^{\times} + \frac{4}{3}\gamma_3\,.
\end{equation}
By defining the new parameter $\tilde{b}_3 \equiv b_3 - 4\gamma_2^{\times}$ we can measure the coefficient in
front of $\gamma_3$ in this equation from the parameter covariance matrix as follows
\begin{equation}
  \alpha = - \frac{S_{\tilde{b}_3,\gamma_3}}{S_{\gamma_3,\gamma_3}}\,.
\end{equation}
The results are shown in the upper panel of Fig.~\ref{fig:g3slope_b3eff_constraints} for each of the samples,
and demonstrate that all except HALO2 indeed follow the degeneracy implied by
Eq.~(\ref{eq:resultsA.b3g3g2x_degeneracy}) within $\sim 10\,\%$. Our constraints on $b_{3,\mathrm{eff}}$ (using
the same $k_{\mathrm{max}}$ values as in Sec.~\ref{sec:comp-co-evol}), which are plotted in the lower panel
of Fig.~\ref{fig:g3slope_b3eff_constraints} against $b_1$, are thus much narrower than for the individual
parameters (cf. Fig.~\ref{fig:bias_constraints}), and are found to be inconsistent with zero at the $68\,\%$
confidence level in all cases except HALO4. Comparing with the PBS relation after accounting for $\gamma_2$ and
$\gamma_{21}$ in Eq.~(\ref{eq:model.b3PBS}) as discussed in Sec.~\ref{sec:comp-co-evol}, we obtain good
agreement --- only for HALO2 we measure a somewhat lower value. We find that the second well constrained
parameter combination depends more sensitively on the sample, but the results so far suggest that a reasonable
simplification of the parameter space could be to fix $b_3$ using the PBS relation, and fixing either $\gamma_3$
or $\gamma_2^{\times}$ to zero or to the respective coevolution relation, while leaving the third parameter
free.

\begin{figure*}
  \centering
  \includegraphics[width=0.8\textwidth]{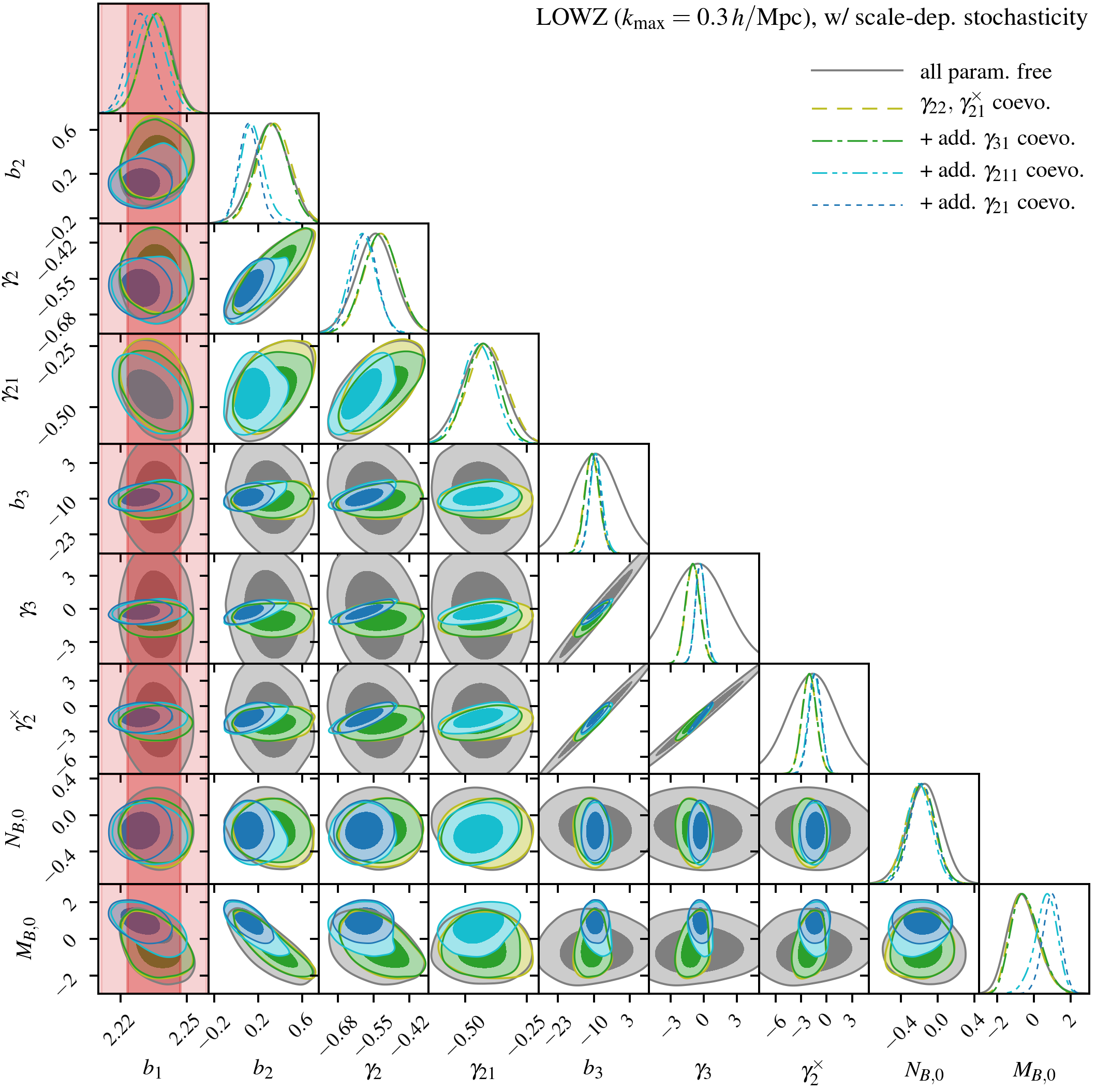}
  \caption{$68\,\%$ and $95\,\%$ posterior contours for model parameters fitted to the LOWZ catalog at
    $k_{\mathrm{max}} = 0.3\,\iMpc$ (including scale-dependent stochasticity); parameters not shown have been
    marginalized over. Gray contours (solid lines) derive from a fit where all model parameters are being
    varied, while the various colored contours apply coevolution relations to an increasing number of bias
    parameters. Starting with $\gamma_{22}$ and $\gamma_{21}^{\times}$ (light green, dashed) an additional
    parameter is fixed up to a total of five (dark blue, dotted). The red error band in the first column
    indicates the $1\sigma$ and $2\sigma$ uncertainties on the fiducial measurement of $b_1$.}
  \label{fig:coevo_contours}
\end{figure*}

The correlation structure among the stochastic parameters and their cross-correlations with parameters from the
general bias expansion appears more complex. One prominent feature is the anti-correlation between $N_{P,0}$ and
$N_{P,2}$, which is consistent with the expected limiting behavior of the stochasticity power spectrum,
$\lim_{k \to 0}{C_{gg}(k)} = 0$, and as shown in \cite{EggScoCro2006} it is well described by the linear
relationship
\begin{equation}
  \label{eq:resultsA.N2N0}
  N_{P,2}(N_{P,0}) \approx - 0.42\,\frac{N_{P,0}}{k_{\mathrm{HD}}^2}
\end{equation}
with $k_{\mathrm{HD}} = 0.4\,\iMpc$. We also note the degeneracy between $N_{P,2}$ and $\gamma_{21}$, which
shows that ignoring the scale-dependent stochasticity in the power spectrum can lead to a biased estimation of
$\gamma_{21}$ and in particular invalidate the agreement of the latter with the general coevolution assumption
as discussed in Sec.~\ref{sec:comp-co-evol}. In the high-$k$ limit the stochasticity bispectrum should
equally vanish, which would imply the existence of similar features for the corresponding noise
parameters. Fig.~\ref{fig:par_corr} indeed reveals anti-correlations between $N_{B,0}$ and $N_{B,2}$, as well as
$M_{B,0}$ and $M_{B,2b}$, although they are less significant than for the power spectrum and do not appear
across all samples. Further studies, possibly by utilizing also the cross bispectra with the matter field, are
required to confirm whether this is truly the case. The strong anti-correlation between $N_{B,0}$ and
$M_{B,2a}$, on other hand, is most likely caused by the $1/k^2$ scaling of the power spectrum for modes where
the scale-dependent stochastic contributions become relevant, such that
$N_{B,0} + M_{B,2a} \left[k_1^2\,P_{mm}(k_1) + \text{cyc.}\right] \approx N_{B,0} + \alpha\,M_{B,2a}$ for some
(positive) constant $\alpha$. Since we do not find strong constraints on $M_{B,2a}$ individually for most
samples, we could make the simplification, $M_{B,2a} = 0$, and thus effectively consider both parameters
together.

\subsection{Reducing the parameter space}
\label{sec:coevo}

Having found good agreement with the coevolution and PBS relations, as well as strong correlations among
subsets of the full parameter space, we now consider whether these results offer possibilities to reduce the
total number of free model parameters without diminishing our estimated validity ranges.

To begin with, we make the assumption that various fourth-order bias parameters are consistent with coevolution
[see Eqs.~(\ref{eq:model.g21xevo}) - (\ref{eq:model.g31evo})], which implies that they are absent from the
initial bias expansion, and hence, $\gamma_{21,{\cal L}}^{\times} = 0$ etc. The corresponding late-time,
Eulerian, bias parameters still depend on the values of other bias parameters, such as $b_1$ and
beyond\footnote{A dependence on $b_1$ alone is only the case in the local Lagrangian approximation, which is
  clearly incorrect following our discussion in Sec.~\ref{sec:comp-co-evol}.}, and for that reason they are
not completely held fixed, although they no longer appear as independent fitting parameters. In
Fig.~\ref{fig:coevo_contours} we show a subset of the posteriors (all parameters not shown are marginalized
over) obtained from fits to the LOWZ catalog at $k_{\mathrm{max}} = 0.3\,\iMpc$ and using the scale-dependent
stochastic bispectrum model. The gray (and largest) contours represent the $68\,\%$ and $95\,\%$ confidence
levels when all model parameters are being varied, whereas all of the subsequent colored contours employ
coevolution relations for an increasing number of parameters, as indicated.

From this plot we see that fixing $\gamma_{22}$, $\gamma_{21}^{\times}$ and $\gamma_{31}$ (green, dot-dashed)
only has a significant impact on the posteriors for $b_3$, $\gamma_3$ and $\gamma_2^{\times}$, but they remain
fully consistent with the original results. The narrowing of the constraints for these three parameters is not
surprising, because from Sec.~\ref{sec:param-correlations} we know that $\gamma_{2}^{\times}$ and
$\gamma_{21}^{\times}$ are strongly correlated, meaning that once the latter is not considered as an independent
parameter anymore, the former is automatically better constrained. This improvement additionally propagates to
$b_3$ and $\gamma_3$ due to their degeneracy with $\gamma_2^{\times}$. Since $\gamma_{211}$ and
$\gamma_{21}^{\times}$ are also correlated, an interesting case is the simultaneous assumption of coevolution
for both of these parameters, which might be inconsistent with their degeneracy direction. As shown by the light
blue (double dot-dashed) contours, this does not appear to be the case: even though slight parameter shifts
occur, most notably for $b_2$ and the noise parameter $M_{B,0}$, the posteriors do not become inconsistent at a
level greater than $1\sigma$. Remarkably, apart from a minor decrease in the uncertainties, having fixed all
fourth-order parameters has not affected the results for $\gamma_{21}$, so that its agreement with coevolution
as found in Sec.~\ref{sec:comp-co-evol} should still be valid. This is explicitly verified by the blue
(dashed) contours, where all NLE operators are assumed to absent from the initial bias expansion.

\begin{figure*}
  \centering
  \includegraphics{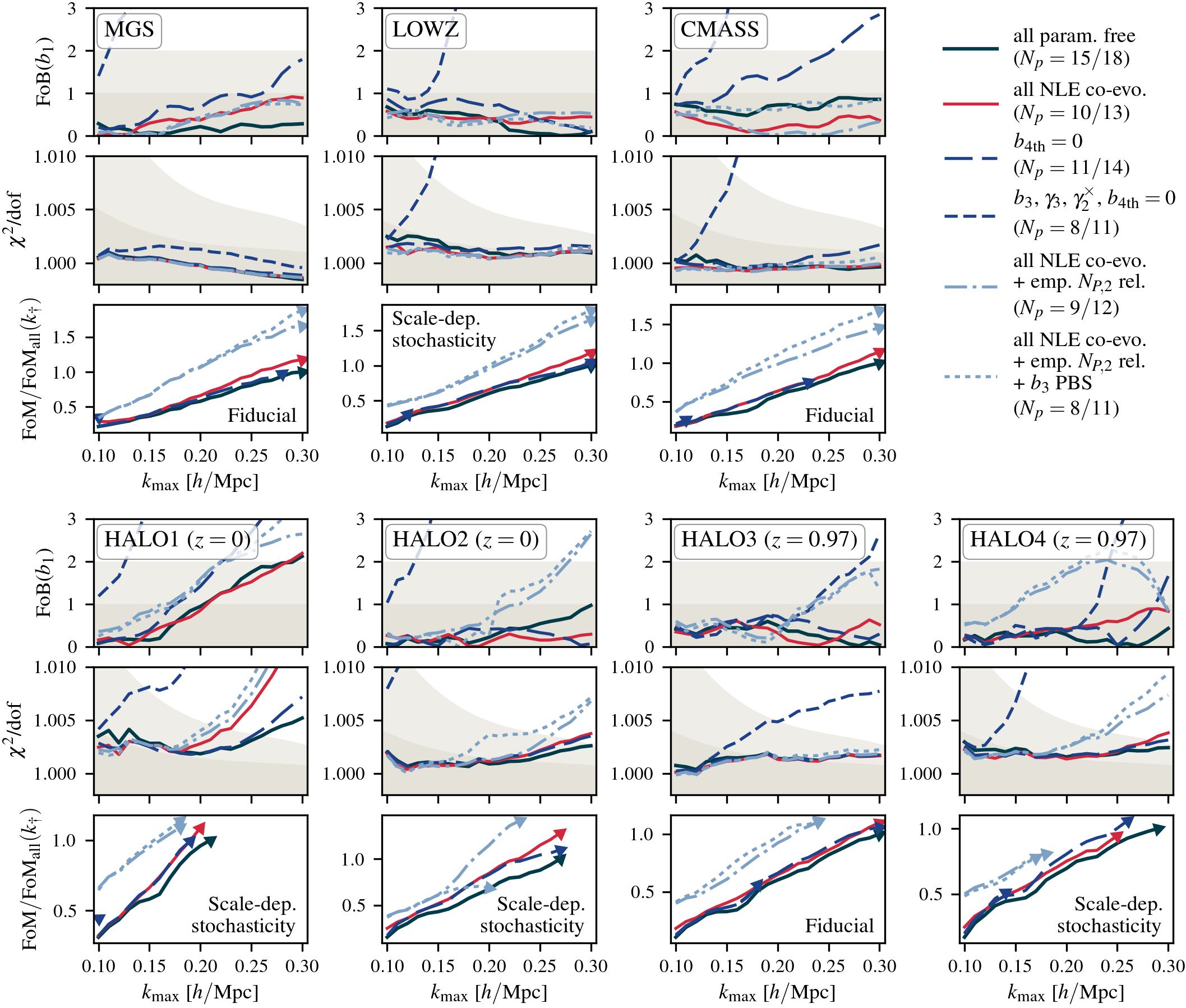}
  \caption{FoB, goodness-of-fit and FoM (normalized by the FoM of the full parameter space model at its validity
    scale) as a function of $k_{\mathrm{max}}$. Different line styles correspond to different assumptions on
    subsets of the participating bias operators: coevolution for all five NLE operators (red, solid), removal
    of the four fourth-order operators (blue, long-dashed), removal of fourth-order operators and $b_3$,
    $\gamma_3$ and $\gamma_2^{\times}$ (blue, short-dashed), NLE coevolution with $b_3$ fixed using the PBS
    relation and with or without the empirical relation for $N_{P,2}$ (light blue, dotted and dot-dashed). Each
    set of panels indicates whether scale-dependent stochastic terms have been included or not and the legend
    lists the number of free fitting parameters.}
  \label{fig:fob_PB_trueM_coevo}
\end{figure*}

For all other samples and at cutoff scales smaller than those where we detect significant shifts (see
Sec.~\ref{sec:cutoff-dependence}) we get a qualitatively very similar picture. To provide a full overview of
these results, in Fig.~\ref{fig:fob_PB_trueM_coevo} we plot the three performance metrics from
Sec.~\ref{sec:performance-metrics} as a function of $k_{\mathrm{max}}$ and compare the case where all five
bias parameters associated to NLE operators are fixed by means of coevolution relations (red lines) with the
previous results (black lines), using either the fiducial or scale-dependent stochastic bispectrum model. We
find that this five-dimensional reduction of the parameter space has no major negative impact on the FoB or the
goodness-of-fit, and thus on the deduced model validity ranges. Only for the HALO4 sample the model fails at a
slightly smaller $k_{\mathrm{max}}$ value, which can be traced back to the fact that fixing $\gamma_{211}$ and
$\gamma_{21}^{\times}$ simultaneously becomes increasingly inconsistent on small scales, as is also indicated by
Fig.~\ref{fig:bias_constraints}. On the other hand, the FoM, which in Fig.~\ref{fig:fob_PB_trueM_coevo} is
normalized with respect to the FoM at the validity scale of the full parameter space model, shows only little
improvements up to $15\,\%$ to $20\,\%$. These improvements mainly derive from being able to fix $\gamma_{21}$
because of its correlation with $b_1$ (cf. Fig.~\ref{fig:par_corr}). 

We contrast the performance of this coevolution model with simply ignoring the four fourth-order operators,
i.e., setting the corresponding bias parameters to zero, and the more drastic case, where we ignore all bias
operators in the bispectrum that do not contribute to the power spectrum (but we keep all remaining terms,
i.e. $b_2$, $\gamma_2$ and $\gamma_{21}$, including their loop corrections). These two cases are respectively
indicated by the blue long- and short-dashed lines, and we see that the former also does not lead to any
diminished model validity ranges. However, for most samples we find that the constraints on $\gamma_{21}$ are
consistently driven towards smaller values, while those for $\gamma_2$ stay constant, which means that they
become in tension with the coevolution relation. Fixing $\gamma_{21}$ in that way is therefore no longer a
generally applicable assumption, which is why we loose its positive benefit on the FoM. On the contrary, the
second considered case is strongly disfavored, as is evident both from the FoB as well as the goodness-of-fit,
and gives rise to validity scales that are even smaller than when all bias loop corrections to the bispectrum
are ignored (cf. Fig.~\ref{fig:fob_PB_trueM}). This is consistent with our analysis in Sec.
\ref{sec:param-correlations}, where we found that two parameter combinations involving $b_3$, $\gamma_3$ and
$\gamma_2^{\times}$ are well constrained, and strongly disfavored to be zero.

Finally, we test whether additional parameters can be fixed in the NLE coevolution model, in particular whether
we can exploit the degeneracy between $N_{P,2}$ and $N_{P,0}$, and the PBS relation for the effective parameter
$b_{3,\mathrm{eff}}$. The former case is shown by the light-blue, dot-dashed lines in
Fig.~\ref{fig:fob_PB_trueM_coevo}, where we made use of Eq.~(\ref{eq:resultsA.N2N0}) in order to remove
$N_{P,2}$ as an independent parameter. This reduction does not diminish the applicable range of the model with
the only critical exception being the HALO4 sample, for which we already reported in \cite{EggScoCro2006} a
departure from the relation in Eq.~(\ref{eq:resultsA.N2N0}). However, being able to fix $N_{P,2}$ brings
substantial improvements in the FoM and thus on the constraints on $b_1$ for all samples considered. Although
$N_{P,2}$ and $b_1$ do not appear correlated in the full parameter space according to Fig.~\ref{fig:par_corr},
we find that this correlation is induced once the coevolution assumptions are applied to the NLE
operators. Even in this considerably more constrained model, the PBS relation for the parameter combination
$b_{3,\textrm{eff}}$ is still an excellent description, as demonstrated by the light-blue, dotted lines. It
gives rise to a slight further increase in FoM, most notably for the galaxy samples, without compromising in the
validity range. The model performance for the HALO2 sample is poorer, due to the disagreement between the
measured value of $b_{3,\mathrm{eff}}$ and the PBS relation that we found earlier (see
Fig.~\ref{fig:g3slope_b3eff_constraints}).

\section{Joint analysis with varying amplitude of fluctuations}
\label{sec:joint-analysis-with-As}

A crucial advantage of combining the power spectrum and bispectrum is the ability to break degeneracies between
cosmological and bias parameters. One of the most prominent degeneracies in the power spectrum concerns the
amplitude of fluctuations, $A_s$, and the linear bias parameter, since they appear as the combination
$b_1\,\sqrt{A_s}$ in the leading contribution to the power spectrum. Although this degeneracy is somewhat reduced
by information from nonlinear scales, as well as for analyses in redshift space, the bispectrum's different
dependence on $A_s$ and $b_1$ allows to separate their effects much more clearly.

It is therefore interesting to investigate whether our previous results on the model performances still
hold once we include $A_s$ in the fitting procedure, and whether the one-loop bispectrum can yield
improved constraints on $A_s$ despite its increased parameter space. We focus here on $A_s$ while leaving other
cosmological parameters fixed, since $A_s$, like the bias parameters, appears as a coefficient in front of
contributions that can be tabulated, making its variation in the MCMC computationally very cheap.

\subsection{Methodology}
\label{sec:As_methodology}

Varying the amplitude of fluctuations means that we can no longer follow our previous approach of using the
measured matter bispectrum since the leading-order and nonlinear contributions scale differently with $A_s$. In
order to obtain an optimal description of the matter bispectrum for testing the bias modeling in this case
nonetheless, we compute its one-loop expression from Eq.~(\ref{eq:model.Bg1loop}) including stress-tensor 
corrections as described in Sec.~\ref{sec:high-deriv-bias}. We then determine the four stress-tensor
parameters $\beta_{B,a/\ldots/d}$ by fitting this model to the measured matter bispectrum with varying cutoff
scales, $k_{\mathrm{max}}$, using a covariance matrix that is constructed in the same way as for the galaxy and
halo samples (see Sec.~\ref{sec:measurements}). At a cutoff scale at which the reduced $\chi^2$ of these fits
exceeds the $95\,\%$ confidence limit we find their best-fit values and use them from here on in all subsequent
analyses involving the one-loop galaxy or halo bispectrum, keeping them fixed unless stated otherwise. The
values at the five different redshifts of our samples, as well as the cutoff scale of the corresponding fit are
given in Table~\ref{tab:stress-tensor_best-fit}. For the power spectrum we instead model the matter component
using the response function formalism \cite{BerTarNis1401,NisBerTar1611}, as implemented in the
\texttt{RESPRESSO} package \cite{NisBerTar1712}, which was found to yield the best results in the comparison of
different matter models presented in \cite{EggScoCro2006}.

\begin{table}
  \centering 
  \caption{Best-fit values of the stress-tensor parameters at the various redshifts of our simulations (see
    also Sec.~\ref{sec:galaxy-halo-catalogs}), resulting from fits of the matter bispectrum up to a scale
    $k_{\mathrm{max}}$ where the reduced $\chi^2$ is smaller than the corresponding $95\,\%$ confidence
    limit. All parameters $\beta_{B,a/\ldots/d}$ are given in units of $k_{\mathrm{HD}}^{-2}$ with
    $k_{\mathrm{HD}} = 0.4\,h/\mathrm{Mpc}$.}
  \begin{ruledtabular}
    \begin{tabular}{cccccc}
      Simulation & \multirow{2}{*}{$\beta_{B,a}$} & \multirow{2}{*}{$\beta_{B,b}$} &
        \multirow{2}{*}{$\beta_{B,c}$} & \multirow{2}{*}{$\beta_{B,d}$}  & \multirow{2}{*}{$k_{\mathrm{max}}$
         [$h/\mathrm{Mpc}$]} \Tstrut \\ redshift & & & & & \Bstrut \\  \hline
      0.0 & 0.144 & 0.361 & -0.287 & -0.135 & 0.16 \Tstrut \\
      0.132 & -2.58 & 1.75 & -0.588 & -0.506 & 0.23 \\
      0.342 & -1.31 & 0.958 & -0.530 & -0.114 & 0.19 \\
      0.57 & -0.256 & 0.403 & -0.235 & -0.0597 & 0.20 \\
      0.97 & -0.0421 & 0.232 & 0.0287 & -0.163 & 0.23 \Bstrut
    \end{tabular}
  \end{ruledtabular}
  \label{tab:stress-tensor_best-fit}
\end{table}

The strong degeneracies between the amplitude of fluctuations and the bias parameters can complicate the MCMC
procedure, which is why we choose to sample combinations of the two. In particular, we use the prescription
\begin{equation}
  \label{eq:resultsB.bn_rescaled}
  b_{n\mathrm{th}} \quad \rightarrow \quad b_{n\mathrm{th}}\,A_s^{n/2}
\end{equation}
for a generic $n$th order bias parameter, such that our new linear bias parameter becomes $b_1\,\sqrt{A_s}$, the
new tidal bias parameter becomes $\gamma_2\,A_s$ etc., which removes the main degeneracies. We keep the same
prior distributions for these combinations as for the original parameters given in Table~\ref{tab:priors} and
vary $A_s$ within a uniform prior with bounds $[0.086, 1.95]$. 

\subsection{Results}
\label{sec:As_results}

\subsubsection{Comparison of validity and merit}
\label{sec:comp-valid-merit}

\begin{figure*}
  \centering
  \includegraphics{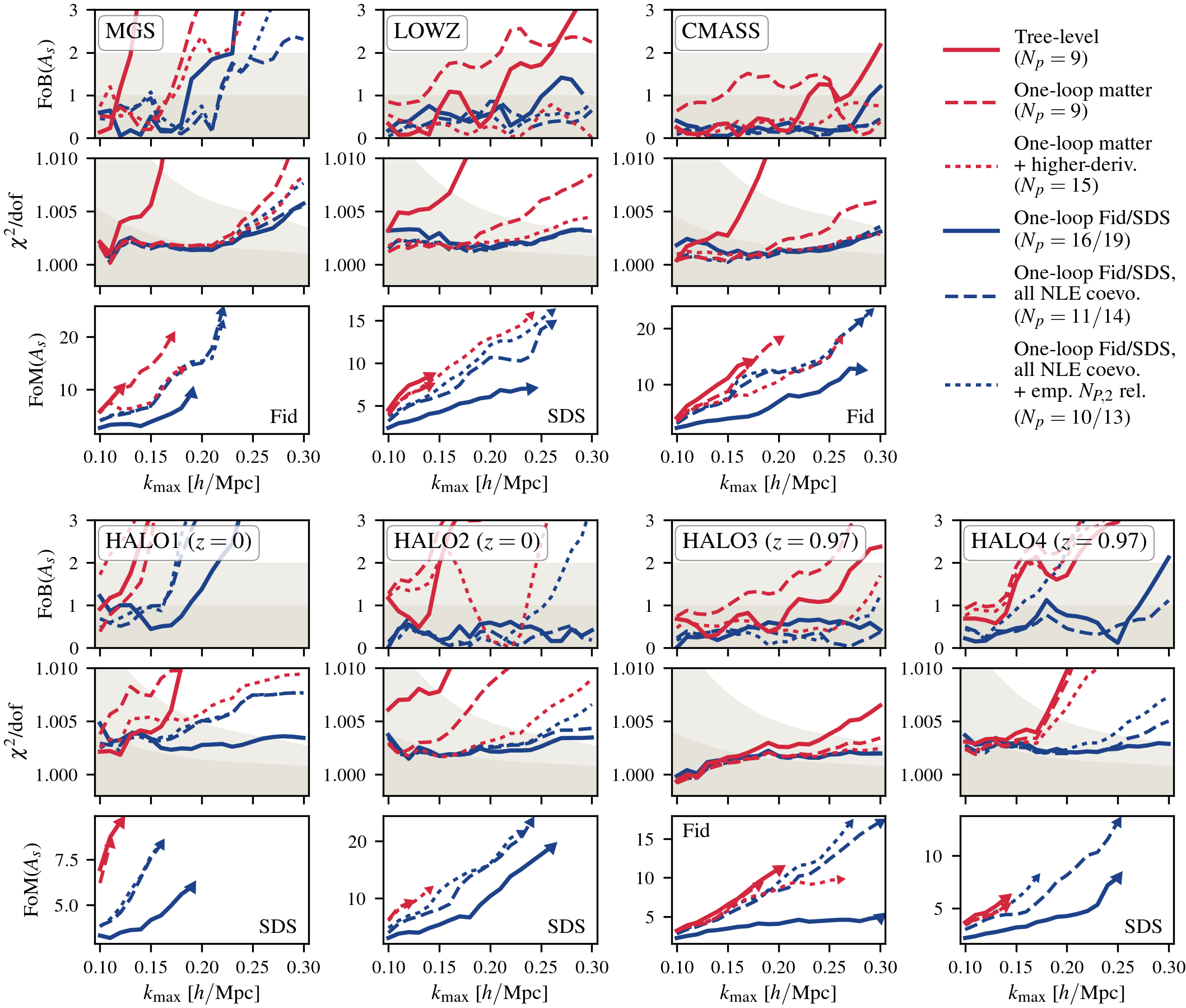}
  \caption{Same as Fig.~\ref{fig:fob_PB_trueM}, although the FoB and FoM are computed with respect to $A_s$
    instead of $b_1$. The power spectrum model includes loop corrections and is the same for all shown
    results. The bispectrum models corresponding to lines in red color only contain tree-level bias terms, but
    unlike the solid line, dashed and dotted ones make use of the one-loop matter bispectrum (see
    Sec.~\ref{sec:As_methodology}). All blue lines account for the full one-loop expressions with or without
    scale-dependent stochasticity (SDS/Fid, respectively) and different line styles distinguish various
    assumptions on a subset of the parameters. The total number of free parameters $N_p$ is given in the plot
    legend.}
  \label{fig:fob_PB_As}
\end{figure*}

To begin with, let us compare the achievable constraining power on $A_s$ for various previously discussed
modeling options of the bispectrum, while keeping the power spectrum model fixed. We evaluate the
goodness-of-fit, FoB and FoM for these cases as before, but with a focus on the amplitude of fluctuations by
defining the latter two metrics with respect to $A_s$ instead of $b_1$, and plot the results as a function of
cutoff scale in Fig.~\ref{fig:fob_PB_As}. The solid red and blue lines indicate the tree-level and one-loop
galaxy bispectrum, where the latter includes scale-dependent stochastic corrections for the same samples as
above, and all participating bias parameters are allowed to vary. We see that the former typically becomes
insufficient at scales beyond $k_{\mathrm{max}} \sim 0.15\,\iMpc$, while the consideration of nonlinear
corrections extends the validity to $0.2\,\iMpc$, and for most samples even to $0.25\,\iMpc$ and beyond
(validity ranges are shown by the arrowhead symbols in the FoM panels as before). This is largely consistent
with our previous findings for fixed $A_s$, though we note that the scales at which the one-loop model breaks
down are slightly reduced, possibly due to inaccuracies in the matter bispectrum\footnote{Since we now model the
  matter bispectrum, the correction of the $\chi^2$ values according to Eq.~(\ref{eq:data.chi2corr}) is no
  longer necessary. The fact that we generally find good agreement between the $\chi^2$ values in
  Fig.~\ref{fig:fob_PB_As} and those reported earlier (with the exception of MGS for large $k_{\mathrm{max}}$)
  can be regarded as an \emph{a posteriori} validation of our applied correction term.}. Comparing the maximum
FoM for these two cases reveals that for the majority of samples the benefit of extending the one-loop model
deeper into the nonlinear regime is not enough to overcome the penalizing effect from having to marginalize over
an increased number of bias parameters.

However, this situation changes once we apply coevolution relations to all of the NLE bias operators (blue
dashed), which we validated in Sec.~\ref{sec:coevo} for fixed $A_s$. Again we find that this assumption has
virtually no impact on the estimated validity ranges across all samples, but we observe a larger decrease of the
uncertainties on $A_s$ than previously for $b_1$. That in turn leads to a clearly improved maximal FoM for the
one-loop model, achieving constraints on $A_s$ that are tighter by factors of about 1.5 (CMASS) up to 2.4 (MGS)
than what can be obtained from the tree-level model. Only for the HALO1 sample does the one-loop model fail at a
$k_{\mathrm{max}}$ that is too low to guarantee an enhanced constraining power. Using additionally the empirical
relation for the scale-dependent stochastic parameter of the power spectrum, $N_{P,2}$, does not give rise to
the same significant boost in FoM for $A_s$ as for $b_1$ (c.f. Sec.~\ref{sec:coevo}), as shown by the blue
dotted line. Apart from the HALO4 sample, the performance is generally very similar to the previous case,
although some slight improvements can be observed, including for LOWZ and CMASS.

\begin{figure*}
  \centering
  \includegraphics{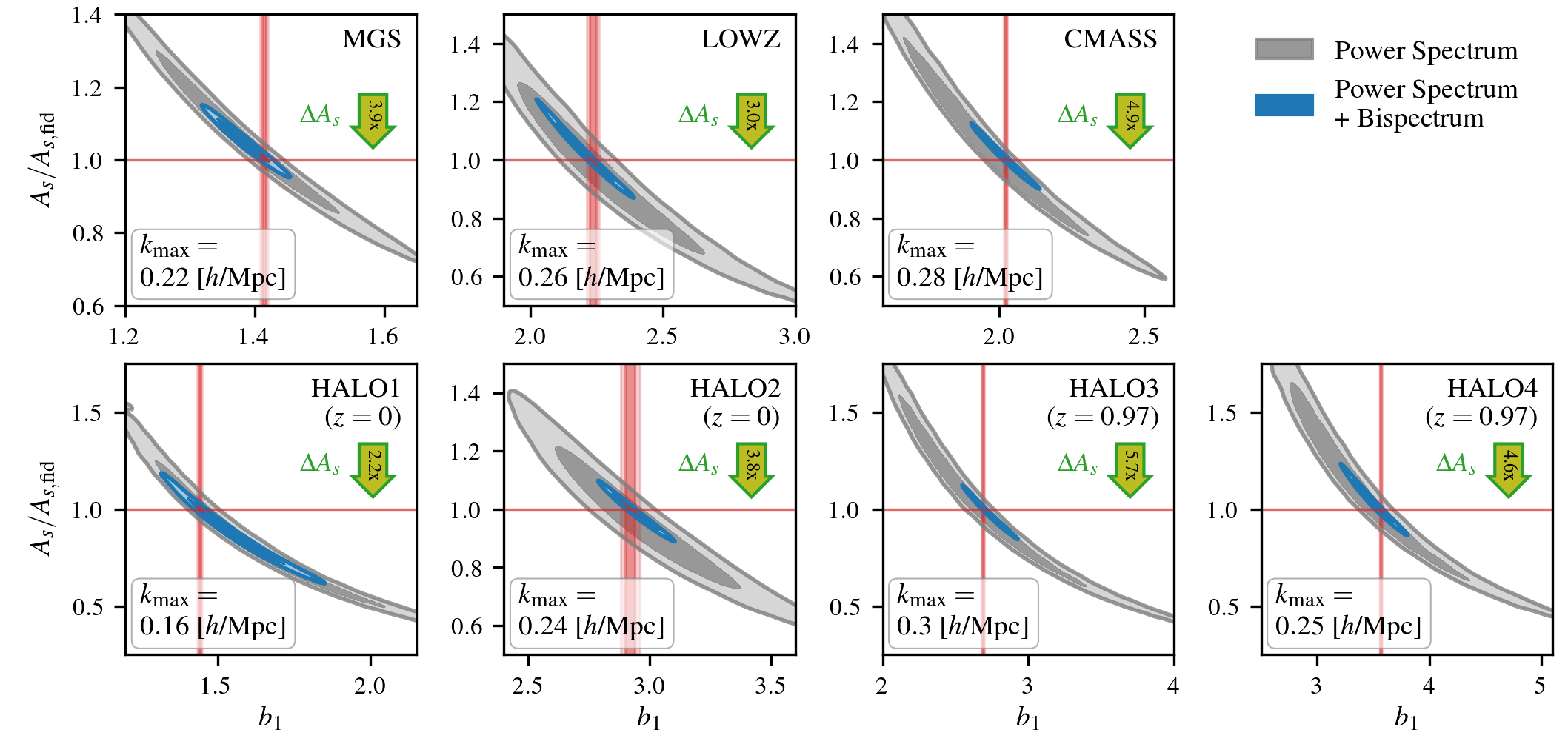}
  \caption{$68\,\%$ and $95\,\%$ confidence limits on $A_s$ and $b_1$ (rescaled to the original linear bias
    parameter by inverting Eq.~\ref{eq:resultsB.bn_rescaled}) from the power spectrum alone (gray contours) and
    from the combination with the bispectrum (blue contours) using the full one-loop model with NLE terms
    assumed to follow the coevolution relations. In both cases the constraints have been evaluated at the
    maximal scale where the latter is still valid according to Fig.~\ref{fig:fob_PB_As}. Red lines (error bands)
    indicate fiducial parameter values (and their uncertainties) and the improvement in the constraint on $A_s$
    from adding the bispectrum is given by the green arrow.}
  \label{fig:PB_respresso_As_vs_b1}
\end{figure*}

We also explore whether we can augment the pure tree-level model represented by the red solid line. To that end,
we consider two possibilities where we substitute the tree-level matter bispectrum by its one-loop expression
with 1) fixed stress-tensor contributions, identical to how we treat the full one-loop model (red
dashed), and 2) all higher-derivative parameters allowed to vary (red dotted)\footnote{Note that because the
  higher-derivative parameters now also partially encompass deviations from a non-vanishing stress-tensor, we
  can no longer exploit the relationship between $\beta_P$ and $\beta_{B,a/b}$ (see discussion in
  Sec.~\ref{sec:high-deriv-bias}). For that reason all six parameters are being varied in this case.}. While
the first option does improve the goodness-of-fit, the FoB becomes worse for many samples, indicating that the
fit adapts $A_s$ to scale the matter loop correction up or down in an attempt to account for the lacking
 bias loop corrections. As a result, the overall validity range and thus the maximal FoM is not
significantly enhanced compared to the plain tree-level model --- only MGS stands out in this regard, where the
uncertainties on $A_s$ shrink by a factor $\sim 2$. The addition of the higher-derivative terms, considered in
the second option, yields further improvements in the goodness-of-fit, and at least for all three galaxy samples
does not strongly bias the recovery of $A_s$. Interestingly, in contrast to our results in
Sec.~\ref{sec:detailed-model-test}, here we do not find an equally severe decrease in constraining power from
marginalizing over the additional parameters, which renders the maximal FoM for the LOWZ and CMASS samples
comparable to the one-loop model with fixed NLE bias operators. However, for all other samples this is not the
case, so this result might be fortuitous and should be interpreted carefully. Moreover, the fact that the
validity ranges are consistently lower than for the one-loop model, despite having a larger parameter space,
suggests that the higher-derivative terms mainly act as a convenient way of absorbing the bias loop
corrections. This conclusion is also supported by various higher-derivative parameters attaining a significant
dependence on $k_{\mathrm{max}}$.

\subsubsection{Improvements over the power spectrum}
\label{sec:impr-over-power}

After finding the validity ranges for the joint power spectrum and bispectrum analysis, we can turn to the
question of how much extra information on $A_s$ we gain compared to the power spectrum alone. For that purpose
we show in Fig.~\ref{fig:PB_respresso_As_vs_b1} the $68\,\%$ and $95\,\%$ confidence limits on $A_s$ and $b_1$ (rescaled back to the original linear
bias parameter) obtained for these two cases, where we have used the one-loop bispectrum model with fixed NLE
bias operators and similarly assumed coevolution for $\gamma_{21}$ (the only contributing NLE operator) in the
individual power spectrum analysis. The constraints are respectively plotted at the maximum valid
$k_{\mathrm{max}}$ value of the joint fit as determined from the data shown in Fig.~\ref{fig:fob_PB_As}. For
each of the samples the power spectrum contours (gray) tightly follow the expected degeneracy direction,
$b_1\,\sqrt{A_s} = \mathrm{const.}$, which prevents any better constraint on $A_s$. Inclusion of an increasing
number of bispectrum triangle configurations (larger $k_{\mathrm{max}}$), however, breaks this degeneracy more
and more efficiently, which leads to substantial reductions in the uncertainties. As indicated for $A_s$ by the
green arrows in the plot, these range between factors of 2.2 for HALO1 and 5.7 for HALO3. Moreover, we have
checked that enlarging the cutoff scale for the power spectrum fit, e.g. to $k_{\mathrm{max}} = 0.3\,\iMpc$, has
little to no impact on this outcome, and $\Delta A_s$ generally shrinks by at least a factor of 3 (except for
HALO1).  

Nevertheless, as mentioned at the beginning of this section, these results are not fully representative in the
sense that this analysis was performed in real opposed to redshift space, where the combination of power
spectrum multipoles already helps in disentangling $b_1$ from $A_s$ (assuming $\Lambda$CDM, otherwise from a
combination of $A_s$ and the growth rate of structure). In addition, correlations between different triangle
configurations that we have ignored in the covariance matrix of the bispectrum, could further diminish the
improvement factors quoted in Fig.~\ref{fig:PB_respresso_As_vs_b1}. On the other hand, similar correlations in
the power spectrum part of the covariance matrix would act in the opposite direction and the cross-covariance
between both statistics, which was also not included in our analysis, has been shown to have a strong
\emph{positive} effect on constraining the amplitude of fluctuations \cite{SefCroPue0607,ByuEggReg1710}. Our
results are therefore at the very least indicative of the large potential of the bispectrum, and in particular
we have shown here for the first time that this potential is not heavily diminished by the increased parameter
space that is required for describing the bispectrum on mildly nonlinear scales.

\section{Conclusions}
\label{sec:conclusions}

We have analyzed real-space measurements of the power spectrum and bispectrum from a large set of galaxy and
halo mock catalogs, and performed fits of their joint likelihood using models build from perturbation theory. In
particular, we have accounted here for the first time for all next-to-leading (one-loop) corrections due to
galaxy bias in the bispectrum, which puts its description on equal footing with models routinely used in the
analysis of the power spectrum \cite{GilNorVer1407,BeuSeoSai1704,GriSanSal1705,IvaSimZal1909,AmiGleKok1909,TroSanAsg2001}, and allowed us to
include more nonlinear triangle shapes than in previous studies
\cite{FelFriFry0102,SaiBalVla1405,GilPerVer1606,GilPerVer1702,SleEisBeu1706,PeaSam1808,SugSaiBeu2102}. Since modeling the nonlinear
corrections requires the introduction of additional bias parameters, and thus a larger number of nuisance
parameters that need to be marginalized over, our analysis pursued two main goals: 1) identifying whether the
information from nonlinear scales in the bispectrum offsets the penalty of the extra parameters, and 2) testing
the applicability and universality of coevolution and peak-background split (PBS) relations in order to reduce
the bias parameter space.

\subsection{At fixed cosmology}
\label{sec:conclusions_fixed_cosmo}

Holding initially all cosmological parameters fixed at their fiducial values, we have found that the one-loop
galaxy bias corrections to the bispectrum become relevant at scales $k \gtrsim 0.17\,\iMpc$. This is largely
consistent across our various samples, for each of which we adopted statistical uncertainties corresponding to
an effective survey volume of $6\,(\Gpc)^3$, and also agrees qualitatively with the scale at which analogous
corrections start to significantly affect the power spectrum. The posteriors of both statistics remain fully
compatible and based on the goodness-of-fit, as well as the accurate recovery of the linear bias parameter
$b_1$, we have quantitatively ascertained that joint fits of the power spectrum and bispectrum can be applied
over the entire tested range of scales, i.e. $k_{\mathrm{max}} = 0.3\,\iMpc$, for most of the samples. Notably,
we gain a factor of 1.5 to 2 when contrasting the constraining power on $b_1$ at the maximum scale at which our
metrics regard the model as valid with that of a model excluding loop corrections. This indicates a clear
benefit from the information on nonlinear scales, despite the larger nuisance parameter space.

From the same analysis we have carefully extracted measurements of all bias parameters up to fourth order, and
for the parameters shared between the power spectrum and bispectrum --- the nonlinear and tidal biases $b_2$ and
$\gamma_2$, as well as the nonlocal parameter $\gamma_{21}$ --- these range among the most stringent that have
been reported so far. Comparing these measurements with the commonly applied local Lagrangian approximations for
$\gamma_2$ and $\gamma_{21}$, we have found clear deviations that are in agreement with the results of other,
independent analyses \cite{LazSch1712,AbiBal1807}. This suggests the presence of Lagrangian tidal bias and,
especially, the importance of including its effect in joint analyses of the power spectrum and bispectrum (see
also \cite{OddSefPor2003}). At the high-mass end of our halo samples (large $b_1$), we have also detected a
departure from the excursion-set prediction of \cite{SheChaSco1304} for $\gamma_2$, which was working well when considering
only data from the power spectrum \cite{EggScoCro2006}.  On the other hand, for all of our samples $\gamma_{21}$
is in excellent agreement with the more general coevolution relation, which takes the non-zero Lagrangian tidal
bias into account. The PBS relation of \cite{LazWagBal1602} for $b_2$ (calibrated for dark matter halos) matches
very well with the results obtained from the halo samples, but disagrees strongly with those from all three
galaxy catalogs, which we have attributed to the nonlinearity of the $b_2(b_1)$ relation.

Constraints on higher-order bias parameters are generally much weaker and individually often consistent with
zero, since they only enter as nonlinear corrections to the bispectrum. However, an analysis of the parameter
correlation matrix revealed that there are two very well constrained combinations of third-order parameters ---
one of them in good agreement with the PBS relation for $b_3$ --- which implies that they cannot simply be
dropped from the model.

Taking these results as guidance in order to find suitable reductions of the parameter space, we have shown that
the application of coevolution relations to all galaxy bias contributions that are generated by nonlocal
gravitational evolution (NLE bias operators corresponding to $\gamma_{21}$ and all four fourth-order parameters)
presents an exceptionally good approximation for our data. The estimated validity ranges are not diminished,
while the uncertainties on $b_1$ shrink by $15$ - $20\,\%$. Further improvements can be achieved by fixing the
scale-dependent stochastic component in the power spectrum using the empirical relation provided in
\cite{EggScoCro2006}. Hence, with these approximations, modeling the bispectrum at one-loop order only takes three
more parameters in excess of those already required by the one-loop power spectrum.

\subsection{Varying $A_s$}
\label{sec:conclusions_varying-As}

In a second part of our analysis we have allowed the amplitude of fluctuations, $A_s$, to vary alongside the
bias parameters. Judging the model performance in that case by goodness-of-fit and an accurate estimation of
$A_s$, we have found slightly reduced validity ranges for both the tree-level and one-loop bias descriptions
compared to the results at fixed cosmology, although the latter remains reliable up to at least
$k_{\mathrm{max}} = 0.25\,\iMpc$ for the majority of samples. We have further shown that the penalty on $A_s$
from marginalizing over bias parameters is more severe than on $b_1$, leading for some samples to comparable or
even worse constraints on $A_s$ from the one-loop model than what we obtain from the tree-level bias
description.

However, we have also demonstrated that the coevolution relations for NLE operators continue to hold and their
application turns the situation clearly in favor of the one-loop model. In particular, it improves the
statistical uncertainties on $A_s$ by factors of 1.5 to 2.4 compared to an analysis constrained to larger scales
using the tree-level model. In comparison with the power spectrum alone the joint analysis shrinks the
constraints on $A_s$ by factors of $\sim 4$ - 6 for most samples, and evidently highlights its value for
breaking the prominent $b_1$ - $A_s$ degeneracy. Even though this degeneracy is already partly broken by the
power spectrum monopole and quadrupole in a redshift-space analysis, likely leading to less pronounced
improvements than those quoted above, the bispectrum will still be instrumental in disentangling the effects of
$A_s$ and the growth rate of structures, which we will consider in future work.

\subsection{On scale-dependent stochastic and higher-derivative contributions}
\label{sec:conclusions_k2n+hd}

We have also analyzed the effect of including scale-dependent stochastic (three extra parameters) and
higher-derivative (five extra parameters) contributions, both to the tree-level and one-loop galaxy bias model
of the bispectrum. While the performance of the tree-level model is not significantly enhanced when augmented by
scale-dependent stochasticity, we have seen that it leads to slightly improved results and less
$k_{\mathrm{max}}$ dependent fitting parameters for the one-loop model when applied to the more heavily biased
samples. This is in accordance with the same finding for the power spectrum presented in \cite{EggScoCro2006}.

At fixed cosmology higher-derivative terms can extend the validity of the tree-level bias description far into
the nonlinear regime for multiple samples, which implies that they are able to efficiently absorb the missing
bias loop corrections. However, this comes at the expense of a significantly increased uncertainty on $b_1$,
such that in this case there is little benefit in going beyond the large-scale analysis. Interestingly, the same
effect is not as severe with varying $A_s$, giving minimum measurement errors on $A_s$ that are larger than
those obtained by the one-loop model by only a few per-cent for the galaxy samples. On the contrary, for the
halo samples we did not find the same improvements from the inclusion of higher-derivative contributions, so we
caution that their ability to absorb and marginalize over loop corrections may well depend on the particular
sample under consideration.

\vspace*{2em}
\acknowledgements

The authors thank Cristiano Porciani and Emiliano Sefusatti for useful discussions. AE acknowledges support from
the European Research Council (grant number ERC-StG-716532-PUNCA), RES from the STFC (grant number ST/P000525/1,
ST/T000473/1), MC and AP from the Spanish Ministry of Science MINECO under grant PGC2018-102021, and AGS from
the Excellence Cluster ORIGINS, which is funded by the Deutsche Forschungsgemeinschaft (DFG, German Research
Foundation) under Germany’s Excellence Strategy - EXC-2094 - 390783311. This research made use of matplotlib, a
Python library for publication quality graphics \cite{Hunter:2007}.

\appendix

\vspace*{2em}
\section{FOURTH-ORDER BIAS OPERATORS}
\label{sec:fourth-order-bias}

The one-loop bispectrum receives contributions from four fourth-order bias operators, whose expressions are
collected in the following. As explained in Sec.~\ref{sec:bias-expansion}, each of them can be written in terms
of Galileons and higher-order LPT potentials:
\begin{align}
  \mathrm{i)} \quad & \delta\,{\cal G}_2(\varphi_2,\varphi_1)\,, \label{eq:app_operatrors.dG21} \\
  \mathrm{ii)} \quad & {\cal G}_3(\varphi_2,\varphi_1,\varphi_1)\,, \\
  \mathrm{iii)} \quad & {\cal G}_2(\varphi_2,\varphi_2)\,, \\
  \mathrm{iv)} \quad & {\cal G}_2(\varphi_3,\varphi_1) = \frac{1}{18} {\cal G}_2(\varphi_3^{(a)},\varphi_1) +
                       \frac{5}{42} {\cal G}_2(\varphi_3^{(b)},\varphi_1) \nonumber \\ &\hspace*{5.6em}+
                                                                                         \frac{1}{14} \nabla_i 
                       \left(\B{\nabla} \times \B{A}_3\right)_j \nabla_{ij} \varphi_1\,,
\end{align}
where the last operator is a combination of the two scalar third-order potentials $\varphi_3^{(a)}$ and
$\varphi_3^{(b)}$, and the vector potential $\B{A}_3$ \cite{EggScoSmi1906}. They are defined as
\begin{align}
  \nabla^2\,\varphi_3^{(a)} &= -{\cal G}_3(\varphi_1)\,, \\
  \nabla^2\,\varphi_3^{(b)} &= -{\cal G}_2(\varphi_2,\varphi_1)\,, \\
  \nabla^2  \B{A}_3 &= - \hat{e}_i\, \epsilon_{ijk}\,\left(\nabla_{jl}\,\varphi_1\right)\,\left(\nabla_{kl}\,\varphi_2\right)\,,
\end{align}
where $\epsilon_{ijk}$ is the fully anti-symmetric Levi-Civita symbol and $\hat{e}_i$ denotes the unit vector in
$i$-direction.

In Fourier space we can express any of these operators as the integral
\begin{equation}
  \begin{split}
      {\cal O}^{(4)}(\B{k}) = \; &(2\pi)^3 \int_{\B{k}_1,\ldots,\B{k}_4} \delta_D(\B{k}-\B{k}_{1234})\, \\ &\times\,{\cal
        K}^{(4)}_{{\cal O}}(\B{k}_1,\ldots,\B{k}_4) \prod_{i=1}^4 \delta_L(\B{k}_i)\,,
  \end{split}
\end{equation}
and the kernel functions corresponding to the four operators are given by \cite{EggScoSmi1906}
\begin{widetext}
\begin{align}
  {\cal K}^{(4)}_{\delta{\cal G}_2(\varphi_2,\varphi_1)}(\B{k}_1,\B{k}_2,\B{k}_3,\B{k}_4) &=
                                                                                            \frac{1}{12}\Big[K(\B{k}_1,\B{k}_{23})\,K(\B{k}_2,\B{k}_3) + \text{sym.}(12)\Big]\,, \\
      {\cal K}^{(4)}_{{\cal G}_3(\varphi_2,\varphi_1,\varphi_1)}(\B{k}_1,\B{k}_2,\B{k}_3,\B{k}_4) &=
    \frac{1}{6}\Big[L(\B{k}_1,\B{k}_2,\B{k}_{34})\,K(\B{k}_3,\B{k}_4) + \text{sym.}(6)\Big]\,, \\
    {\cal K}^{(4)}_{{\cal G}_2(\varphi_2,\varphi_2)}(\B{k}_1,\B{k}_2,\B{k}_3,\B{k}_4) &=
    \frac{1}{3}\Big[K(\B{k}_{12},\B{k}_{34})\,K(\B{k}_1,\B{k}_2)\,K(\B{k}_3,\B{k}_4) + \text{sym.}(3)\Big]\,, \\
    {\cal K}^{(4)}_{{\cal G}_2(\varphi_3,\varphi_1)}(\B{k}_1,\B{k}_2,\B{k}_3,\B{k}_4) &=
    \frac{1}{12}\left[\frac{1}{18}K(\B{k}_1,\B{k}_{234})\left(\frac{15}{7}\,K(\B{k}_{23},\B{k}_4)\,K(\B{k}_2,\B{k}_3) -
        L(\B{k}_2,\B{k}_3,\B{k}_4)\right)\right. \nonumber \\
    &\hspace{1.2em}\left.+\frac{1}{14}\Big(M(\B{k}_1,\B{k}_{23},\B{k}_{4},\B{k}_{234}) -
      M(\B{k}_1,\B{k}_{234},\B{k}_{23},\B{k}_4) \Big) K(\B{k}_2,\B{k}_3) +
      \text{sym.}(12)\right]\,, \label{eq:app_operators.KG31}
\end{align}
\end{widetext}
where $\mathrm{sym.}(n)$ stands for the $n-1$ terms that need to be added in order to symmetrize the expressions
over the four participating wave vectors. In Eq.~(\ref{eq:app_operators.KG31}) we have further made use of the
kernel function
\begin{equation}
  M(\B{k}_1,\B{k}_2,\B{k}_3,\B{k}_4) \equiv \frac{(\B{k}_1 \cdot \B{k}_2)\,(\B{k}_2 \cdot \B{k}_3)\,(\B{k}_3
    \cdot \B{k}_4)\,(\B{k}_4 \cdot \B{k}_1)}{(k_1\,k_2\,k_3\,k_4)^2}\,.
\end{equation}
Finally, when computing the loop corrections for the two-point propagator (see Eq.~\ref{eq:model.MP2}), we need
to make sure to only include the ``finite'' part of the kernels and subtract any potential sensitivities to the
nonlinear regime in the large-scale limit (which are absorbed by the bias parameters). This only occurs for the
first operator (Eq.~\ref{eq:app_operatrors.dG21}) and one can show that
\begin{equation}
  \begin{split}
        {\cal K}_{\delta{\cal G}_2(\varphi_2,\varphi_1)}^{(4,\mathrm{lim})}(\B{k_1},\B{k}_2,\B{q}) &\equiv
      \lim_{\B{q} \to 0} {\cal K}_{\delta{\cal G}_2(\varphi_2,\varphi_1)}^{(4)}(\B{k}_1,\B{k}_2,\B{q},-\B{q}) \\
      &= \frac{1}{3} K(\B{k}_1,\B{q})\,K(\B{k}_2,\B{q}) \\ &\hspace*{1em}+ \frac{1}{6}
      K(\B{k}_1,\B{k}_2)\,K(\B{k}_1+\B{k}_2,\B{q})\,,
  \end{split}
\end{equation}
and so we define
\begin{equation}
  {\cal K}_{\delta{\cal G}_2(\varphi_2,\varphi_1)}^{(4,F)} \equiv   {\cal
    K}_{\delta{\cal G}_2(\varphi_2,\varphi_1)}^{(4)} -   {\cal K}_{\delta{\cal
      G}_2(\varphi_2,\varphi_1)}^{(4,\mathrm{lim})}\,,
\end{equation}
where we have suppressed the argument $(\B{k}_1,\B{k}_2,\B{q},-\B{q})$.

\begin{table*}
  \centering 
  \caption{Constraints on third- and fourth-order bias parameters from a joint analysis of the power spectrum and
    bispectrum with cutoff scales as described in Sec.~\ref{sec:galaxy-bias-constraints}.}
  \begin{ruledtabular}
    \begin{tabular}{cccccccc}
      Sample & $b_3$ & $\gamma_3$ & $\gamma_{2}^{\times}$ & $\gamma_{211}$ & $\gamma_{22}$ & $\gamma_{31}$ &
                                                                                                             $\gamma_{21}^{\times}$
                                                                                                             \Tstrut\Bstrut
      \\  \hline
      MGS & $-1.5 \pm 4$ & $-0.91 \pm 2$ & $-0.76 \pm 2$ & $0.0098 \pm 1$ & $0.039 \pm 0.05$ & $0.39 \pm 0.8$ &
                                                                                                                $-1.4 \pm 2$ \Tstrut \\
LOWZ & $-9.5 \pm 9$ & $-0.54 \pm 3$ & $-1.6 \pm 3$ & $-0.48 \pm 0.9$ & $-0.03 \pm 0.2$ & $4.4 \pm 2$ & $-1 \pm 4$ \\
CMASS & $-5.5 \pm 10$ & $-1 \pm 3$ & $-1.4 \pm 3$ & $-0.52 \pm 1$ & $0.086 \pm 0.1$ & $1.6 \pm 2$ & $-2.1 \pm 5$ \\
HALO1 & $-3.4 \pm 10$ & $-2.6 \pm 5$ & $-2 \pm 5$ & $2.7 \pm 5$ & $0.12 \pm 0.1$ & $-0.19 \pm 2$ & $-2 \pm 5$ \\
HALO2 & $-11 \pm 20$ & $3 \pm 4$ & $2 \pm 6$ & $0.43 \pm 2$ & $0.51 \pm 0.5$ & $-0.53 \pm 5$ & $6.4 \pm 8$ \\
HALO3 & $-11 \pm 30$ & $-0.57 \pm 8$ & $-1.9 \pm 9$ & $-0.1 \pm 3$ & $0.27 \pm 0.5$ & $1.7 \pm 7$ & $-0.79 \pm 10$ \\
HALO4 & $14 \pm 20$ & $11 \pm 6$ & $8.5 \pm 8$ & $3.4 \pm 7$ & $0.017 \pm 1$ & $6.9 \pm 10$ & $21 \pm 20$ \Bstrut
    \end{tabular}
  \end{ruledtabular}
  \label{tab:constraints_b3rd4th}
\end{table*}

\section{EVOLUTION OF CONSERVED TRACERS}
\label{sec:evolotion_tracers}

To complete the expressions given in Sec.~\ref{sec:MPformalism}, we here briefly review the general solution
of the coupled equations for matter and tracer densities, $\delta$ and $\delta_g$, as well as matter velocity
$\theta$, where we assume the conservation of tracers and the absence of velocity bias ($\theta_g =
\theta$). For further information, see \cite{Sco0101,Chan:2012,EggScoSmi1906}.

We start by introducing the triplet
\begin{equation}
  \Psi_a(\B{k},\tau) \equiv \Big[\delta(\B{k},\tau),\, \theta(\B{k},\tau)/f {\cal H},\,
  \delta_g(\B{k},\tau)\Big]\,, 
\end{equation}
where $\tau$ is the conformal time, ${\cal H}$ the conformal Hubble rate, and $f$ the growth rate of
structures. Using this triplet and changing time variable to the logarithm of the linear growth factor, $\eta
\equiv \ln{D(\tau)}$, the continuity and Euler equations can be written in the following compact form
\begin{equation}
  \begin{split}
    &\frac{\partial \Psi_a(\B{k},\eta)}{\partial \eta} + \Omega_{ab}\,\Psi_b(\B{k},\eta) \\ &\hspace*{0.2em}= (2\pi)^3
    \int_{\B{k}_1,\B{k}_2} 
    \delta_D(\B{k}-\B{k}_{12})\,\gamma_{abc}(\B{k}_1,\B{k}_2) \\
    &\hspace*{0.3em}\times\,\Psi_b(\B{k}_1,\eta)\,\Psi_c(\B{k}_2,\eta)\,. \label{eq:app_evo.fluid_equations}
  \end{split}
\end{equation}
The couplings between densities and velocities at a linear level are described by the matrix
\begin{equation}
  \Omega_{ab} \equiv \frac{1}{2}
  \left[
    \begin{array}{ccc}
      0 & -2 & 0 \\
      -3 & 1 & 0 \\
      0 & -2 & 0
    \end{array}\right]\,,
\end{equation}
whereas nonlinear interactions are encoded in $\gamma_{abc}(\B{k}_1,\B{k}_2)$, whose only non-zero components
are given by
\begin{align}
  \gamma_{121} = \gamma_{323} &= \alpha(\B{k}_1,\B{k}_2) \equiv \frac{\B{k}_{12} \cdot \B{k}_1}{k_1^2}\,, \\
  \gamma_{222} &= \beta(\B{k}_1,\B{k}_2) \equiv \frac{k_{12}^2\,\left(\B{k}_1 \cdot \B{k}_2\right)}{2 k_1^2\,k_2^2}\,,
\end{align}
and $\gamma_{112}(\B{k}_1,\B{k}_2) = \gamma_{121}(\B{k}_2,\B{k}_1)$. Letting $\phi_a(\B{k})$ denote the initial
conditions for the three fields at $\eta=0$, it can be shown via Laplace transformation that
Eq.~(\ref{eq:app_evo.fluid_equations}) has an integral solution,
\begin{equation}
  \label{eq:app_evo.integral_solution}
  \begin{split}
      \Psi_a(\B{k},\eta) = \, &g_{ab}(\eta)\,\phi_b(\B{k}) + (2\pi)^3 \int_0^\eta \text{d}\eta'\,g_{ab}(\eta-\eta')
      \\ 
      &\hspace*{-3.5em}\times \int_{\B{k}_1,\B{k}_2}
      \delta(\B{k}-\B{k}_{12})\,\gamma_{bcd}(\B{k}_1,\B{k}_2)\,\Psi_c(\B{k}_1,\eta')\,\Psi_d(\B{k}_2,\eta')\,,
  \end{split}
\end{equation}
with $g_{ab}(\eta)$ denoting the linear (time) propagator, which is given by a combination of growing and
decaying modes:
\begin{align}
  g_{ab}(\eta) = \;
  &\frac{\mathrm{e}^{\eta}}{5} \left[
    \begin{array}{ccc}
      3 & 2 & 0 \\
      3 & 2 & 0 \\
      3 & 2 & 0
    \end{array}
  \right] - \frac{\mathrm{e}^{-3\eta/2}}{5}\left[
    \begin{array}{ccc}
      -2 & 2 & 0 \\
      3 & -3 & 0 \\
      -2 & 2 & 0 \\
    \end{array}
  \right] \nonumber \\
  &+\left[
    \begin{array}{ccc}
      0 & 0 & 0 \\
      0 & 0 & 0 \\
      -1 & 0 & 1
    \end{array}
  \right]\,.
\end{align}
Plugging in the multipoint propagator expansion for $\Psi_a$ and $\phi_a$ into
Eq.~(\ref{eq:app_evo.integral_solution}) one can derive a set of recursion relations at arbitrary order of
perturbation theory, which lead to the expressions in Eq.~(\ref{eq:model.MPrecursion}) to
(\ref{eq:model.MPloop}).

\section{CONSTRAINTS ON HIGHER-ORDER BIAS PARAMETERS}
\label{sec:constraints_bias_higherorder}

In Table~\ref{tab:constraints_b3rd4th} we have collected the remaining constraints on the third- and
fourth-order bias parameters for all of our samples. These values derive from the same joint power spectrum and
bispectrum analysis that was described in Sec.~\ref{sec:galaxy-bias-constraints} and thus complete
Table~\ref{tab:constraints_b2g2g21}.

\bibliography{refs}

\begin{thebibliography}{113}%
\makeatletter
\providecommand \@ifxundefined [1]{%
 \@ifx{#1\undefined}
}%
\providecommand \@ifnum [1]{%
 \ifnum #1\expandafter \@firstoftwo
 \else \expandafter \@secondoftwo
 \fi
}%
\providecommand \@ifx [1]{%
 \ifx #1\expandafter \@firstoftwo
 \else \expandafter \@secondoftwo
 \fi
}%
\providecommand \natexlab [1]{#1}%
\providecommand \enquote  [1]{``#1''}%
\providecommand \bibnamefont  [1]{#1}%
\providecommand \bibfnamefont [1]{#1}%
\providecommand \citenamefont [1]{#1}%
\providecommand \href@noop [0]{\@secondoftwo}%
\providecommand \href [0]{\begingroup \@sanitize@url \@href}%
\providecommand \@href[1]{\@@startlink{#1}\@@href}%
\providecommand \@@href[1]{\endgroup#1\@@endlink}%
\providecommand \@sanitize@url [0]{\catcode `\\12\catcode `\$12\catcode
  `\&12\catcode `\#12\catcode `\^12\catcode `\_12\catcode `\%12\relax}%
\providecommand \@@startlink[1]{}%
\providecommand \@@endlink[0]{}%
\providecommand \url  [0]{\begingroup\@sanitize@url \@url }%
\providecommand \@url [1]{\endgroup\@href {#1}{\urlprefix }}%
\providecommand \urlprefix  [0]{URL }%
\providecommand \Eprint [0]{\href }%
\providecommand \doibase [0]{http://dx.doi.org/}%
\providecommand \selectlanguage [0]{\@gobble}%
\providecommand \bibinfo  [0]{\@secondoftwo}%
\providecommand \bibfield  [0]{\@secondoftwo}%
\providecommand \translation [1]{[#1]}%
\providecommand \BibitemOpen [0]{}%
\providecommand \bibitemStop [0]{}%
\providecommand \bibitemNoStop [0]{.\EOS\space}%
\providecommand \EOS [0]{\spacefactor3000\relax}%
\providecommand \BibitemShut  [1]{\csname bibitem#1\endcsname}%
\let\auto@bib@innerbib\@empty
\bibitem [{\citenamefont {{Levi}}\ \emph {et~al.}(2013)\citenamefont {{Levi}},
  \citenamefont {{Bebek}}, \citenamefont {{Beers}}, \citenamefont {{Blum}},
  \citenamefont {{Cahn}}, \citenamefont {{Eisenstein}}, \citenamefont
  {{Flaugher}}, \citenamefont {{Honscheid}}, \citenamefont {{Kron}},
  \citenamefont {{Lahav}}, \citenamefont {{McDonald}}, \citenamefont {{Roe}},
  \citenamefont {{Schlegel}},\ and\ \citenamefont {{representing the DESI
  collaboration}}}]{DESI}%
  \BibitemOpen
  \bibfield  {author} {\bibinfo {author} {\bibfnamefont {M.}~\bibnamefont
  {{Levi}}}, \bibinfo {author} {\bibfnamefont {C.}~\bibnamefont {{Bebek}}},
  \bibinfo {author} {\bibfnamefont {T.}~\bibnamefont {{Beers}}}, \bibinfo
  {author} {\bibfnamefont {R.}~\bibnamefont {{Blum}}}, \bibinfo {author}
  {\bibfnamefont {R.}~\bibnamefont {{Cahn}}}, \bibinfo {author} {\bibfnamefont
  {D.}~\bibnamefont {{Eisenstein}}}, \bibinfo {author} {\bibfnamefont
  {B.}~\bibnamefont {{Flaugher}}}, \bibinfo {author} {\bibfnamefont
  {K.}~\bibnamefont {{Honscheid}}}, \bibinfo {author} {\bibfnamefont
  {R.}~\bibnamefont {{Kron}}}, \bibinfo {author} {\bibfnamefont
  {O.}~\bibnamefont {{Lahav}}}, \bibinfo {author} {\bibfnamefont
  {P.}~\bibnamefont {{McDonald}}}, \bibinfo {author} {\bibfnamefont
  {N.}~\bibnamefont {{Roe}}}, \bibinfo {author} {\bibfnamefont
  {D.}~\bibnamefont {{Schlegel}}}, \ and\ \bibinfo {author} {\bibnamefont
  {{representing the DESI collaboration}}},\ }\href@noop {} {\bibfield
  {journal} {\bibinfo  {journal} {arXiv e-prints}\ ,\ \bibinfo {eid}
  {arXiv:1308.0847}} (\bibinfo {year} {2013})},\ \Eprint
  {http://arxiv.org/abs/1308.0847} {arXiv:1308.0847 [astro-ph.CO]} \BibitemShut
  {NoStop}%
\bibitem [{\citenamefont {{Laureijs}}\ \emph {et~al.}(2011)\citenamefont
  {{Laureijs}}, \citenamefont {{Amiaux}}, \citenamefont {{Arduini}},
  \citenamefont {{Augu{\`e}res}}, \citenamefont {{Brinchmann}}, \citenamefont
  {{Cole}}, \citenamefont {{Cropper}}, \citenamefont {{Dabin}}, \citenamefont
  {{Duvet}}, \citenamefont {{Ealet}}, \citenamefont {{Garilli}}, \citenamefont
  {{Gondoin}}, \citenamefont {{Guzzo}}, \citenamefont {{Hoar}}, \citenamefont
  {{Hoekstra}},\ and\ \citenamefont {{Others}}}]{Euclid}%
  \BibitemOpen
  \bibfield  {author} {\bibinfo {author} {\bibfnamefont {R.}~\bibnamefont
  {{Laureijs}}}, \bibinfo {author} {\bibfnamefont {J.}~\bibnamefont
  {{Amiaux}}}, \bibinfo {author} {\bibfnamefont {S.}~\bibnamefont {{Arduini}}},
  \bibinfo {author} {\bibfnamefont {J.~L.}\ \bibnamefont {{Augu{\`e}res}}},
  \bibinfo {author} {\bibfnamefont {J.}~\bibnamefont {{Brinchmann}}}, \bibinfo
  {author} {\bibfnamefont {R.}~\bibnamefont {{Cole}}}, \bibinfo {author}
  {\bibfnamefont {M.}~\bibnamefont {{Cropper}}}, \bibinfo {author}
  {\bibfnamefont {C.}~\bibnamefont {{Dabin}}}, \bibinfo {author} {\bibfnamefont
  {L.}~\bibnamefont {{Duvet}}}, \bibinfo {author} {\bibfnamefont
  {A.}~\bibnamefont {{Ealet}}}, \bibinfo {author} {\bibfnamefont
  {B.}~\bibnamefont {{Garilli}}}, \bibinfo {author} {\bibfnamefont
  {P.}~\bibnamefont {{Gondoin}}}, \bibinfo {author} {\bibfnamefont
  {L.}~\bibnamefont {{Guzzo}}}, \bibinfo {author} {\bibfnamefont
  {J.}~\bibnamefont {{Hoar}}}, \bibinfo {author} {\bibfnamefont
  {H.}~\bibnamefont {{Hoekstra}}}, \ and\ \bibinfo {author} {\bibnamefont
  {{Others}}},\ }\href@noop {} {\bibfield  {journal} {\bibinfo  {journal}
  {arXiv e-prints}\ ,\ \bibinfo {eid} {arXiv:1110.3193}} (\bibinfo {year}
  {2011})},\ \Eprint {http://arxiv.org/abs/1110.3193} {arXiv:1110.3193
  [astro-ph.CO]} \BibitemShut {NoStop}%
\bibitem [{\citenamefont {{Maartens}}\ \emph {et~al.}(2015)\citenamefont
  {{Maartens}}, \citenamefont {{Abdalla}}, \citenamefont {{Jarvis}},\ and\
  \citenamefont {{Santos}}}]{SKA}%
  \BibitemOpen
  \bibfield  {author} {\bibinfo {author} {\bibfnamefont {R.}~\bibnamefont
  {{Maartens}}}, \bibinfo {author} {\bibfnamefont {F.~B.}\ \bibnamefont
  {{Abdalla}}}, \bibinfo {author} {\bibfnamefont {M.}~\bibnamefont {{Jarvis}}},
  \ and\ \bibinfo {author} {\bibfnamefont {M.~G.}\ \bibnamefont {{Santos}}},\
  }\href@noop {} {\bibfield  {journal} {\bibinfo  {journal} {arXiv e-prints}\
  ,\ \bibinfo {eid} {arXiv:1501.04076}} (\bibinfo {year} {2015})},\ \Eprint
  {http://arxiv.org/abs/1501.04076} {arXiv:1501.04076 [astro-ph.CO]}
  \BibitemShut {NoStop}%
\bibitem [{\citenamefont {{Sefusatti}}\ \emph {et~al.}(2006)\citenamefont
  {{Sefusatti}}, \citenamefont {{Crocce}}, \citenamefont {{Pueblas}},\ and\
  \citenamefont {{Scoccimarro}}}]{SefCroPue0607}%
  \BibitemOpen
  \bibfield  {author} {\bibinfo {author} {\bibfnamefont {E.}~\bibnamefont
  {{Sefusatti}}}, \bibinfo {author} {\bibfnamefont {M.}~\bibnamefont
  {{Crocce}}}, \bibinfo {author} {\bibfnamefont {S.}~\bibnamefont {{Pueblas}}},
  \ and\ \bibinfo {author} {\bibfnamefont {R.}~\bibnamefont {{Scoccimarro}}},\
  }\href {\doibase 10.1103/PhysRevD.74.023522} {\bibfield  {journal} {\bibinfo
  {journal} {\prd}\ }\textbf {\bibinfo {volume} {74}},\ \bibinfo {pages}
  {023522} (\bibinfo {year} {2006})},\ \Eprint
  {http://arxiv.org/abs/arXiv:astro-ph/0604505} {arXiv:astro-ph/0604505}
  \BibitemShut {NoStop}%
\bibitem [{\citenamefont {{Song}}\ \emph {et~al.}(2015)\citenamefont {{Song}},
  \citenamefont {{Taruya}},\ and\ \citenamefont {{Oka}}}]{SonTarOka1508}%
  \BibitemOpen
  \bibfield  {author} {\bibinfo {author} {\bibfnamefont {Y.-S.}\ \bibnamefont
  {{Song}}}, \bibinfo {author} {\bibfnamefont {A.}~\bibnamefont {{Taruya}}}, \
  and\ \bibinfo {author} {\bibfnamefont {A.}~\bibnamefont {{Oka}}},\ }\href
  {\doibase 10.1088/1475-7516/2015/08/007} {\bibfield  {journal} {\bibinfo
  {journal} {\jcap}\ }\textbf {\bibinfo {volume} {2015}},\ \bibinfo {eid} {007}
  (\bibinfo {year} {2015})},\ \Eprint {http://arxiv.org/abs/1502.03099}
  {arXiv:1502.03099 [astro-ph.CO]} \BibitemShut {NoStop}%
\bibitem [{\citenamefont {{Gagrani}}\ and\ \citenamefont
  {{Samushia}}(2017)}]{GagSam1705}%
  \BibitemOpen
  \bibfield  {author} {\bibinfo {author} {\bibfnamefont {P.}~\bibnamefont
  {{Gagrani}}}\ and\ \bibinfo {author} {\bibfnamefont {L.}~\bibnamefont
  {{Samushia}}},\ }\href {\doibase 10.1093/mnras/stx135} {\bibfield  {journal}
  {\bibinfo  {journal} {\mnras}\ }\textbf {\bibinfo {volume} {467}},\ \bibinfo
  {pages} {928} (\bibinfo {year} {2017})},\ \Eprint
  {http://arxiv.org/abs/1610.03488} {arXiv:1610.03488 [astro-ph.CO]}
  \BibitemShut {NoStop}%
\bibitem [{\citenamefont {{Byun}}\ \emph {et~al.}(2017)\citenamefont {{Byun}},
  \citenamefont {{Eggemeier}}, \citenamefont {{Regan}}, \citenamefont
  {{Seery}},\ and\ \citenamefont {{Smith}}}]{ByuEggReg1710}%
  \BibitemOpen
  \bibfield  {author} {\bibinfo {author} {\bibfnamefont {J.}~\bibnamefont
  {{Byun}}}, \bibinfo {author} {\bibfnamefont {A.}~\bibnamefont {{Eggemeier}}},
  \bibinfo {author} {\bibfnamefont {D.}~\bibnamefont {{Regan}}}, \bibinfo
  {author} {\bibfnamefont {D.}~\bibnamefont {{Seery}}}, \ and\ \bibinfo
  {author} {\bibfnamefont {R.~E.}\ \bibnamefont {{Smith}}},\ }\href {\doibase
  10.1093/mnras/stx1681} {\bibfield  {journal} {\bibinfo  {journal} {\mnras}\
  }\textbf {\bibinfo {volume} {471}},\ \bibinfo {pages} {1581} (\bibinfo {year}
  {2017})},\ \Eprint {http://arxiv.org/abs/1705.04392} {arXiv:1705.04392
  [astro-ph.CO]} \BibitemShut {NoStop}%
\bibitem [{\citenamefont {{Yankelevich}}\ and\ \citenamefont
  {{Porciani}}(2019)}]{YanPor1902}%
  \BibitemOpen
  \bibfield  {author} {\bibinfo {author} {\bibfnamefont {V.}~\bibnamefont
  {{Yankelevich}}}\ and\ \bibinfo {author} {\bibfnamefont {C.}~\bibnamefont
  {{Porciani}}},\ }\href {\doibase 10.1093/mnras/sty3143} {\bibfield  {journal}
  {\bibinfo  {journal} {\mnras}\ }\textbf {\bibinfo {volume} {483}},\ \bibinfo
  {pages} {2078} (\bibinfo {year} {2019})},\ \Eprint
  {http://arxiv.org/abs/1807.07076} {arXiv:1807.07076 [astro-ph.CO]}
  \BibitemShut {NoStop}%
\bibitem [{\citenamefont {{Agarwal}}\ \emph {et~al.}(2020)\citenamefont
  {{Agarwal}}, \citenamefont {{Desjacques}}, \citenamefont {{Jeong}},\ and\
  \citenamefont {{Schmidt}}}]{AgaDesJeo2007}%
  \BibitemOpen
  \bibfield  {author} {\bibinfo {author} {\bibfnamefont {N.}~\bibnamefont
  {{Agarwal}}}, \bibinfo {author} {\bibfnamefont {V.}~\bibnamefont
  {{Desjacques}}}, \bibinfo {author} {\bibfnamefont {D.}~\bibnamefont
  {{Jeong}}}, \ and\ \bibinfo {author} {\bibfnamefont {F.}~\bibnamefont
  {{Schmidt}}},\ }\href@noop {} {\bibfield  {journal} {\bibinfo  {journal}
  {arXiv e-prints}\ ,\ \bibinfo {eid} {arXiv:2007.04340}} (\bibinfo {year}
  {2020})},\ \Eprint {http://arxiv.org/abs/2007.04340} {arXiv:2007.04340
  [astro-ph.CO]} \BibitemShut {NoStop}%
\bibitem [{\citenamefont {{Chudaykin}}\ and\ \citenamefont
  {{Ivanov}}(2019)}]{ChuIva1911}%
  \BibitemOpen
  \bibfield  {author} {\bibinfo {author} {\bibfnamefont {A.}~\bibnamefont
  {{Chudaykin}}}\ and\ \bibinfo {author} {\bibfnamefont {M.~M.}\ \bibnamefont
  {{Ivanov}}},\ }\href {\doibase 10.1088/1475-7516/2019/11/034} {\bibfield
  {journal} {\bibinfo  {journal} {\jcap}\ }\textbf {\bibinfo {volume} {2019}},\
  \bibinfo {eid} {034} (\bibinfo {year} {2019})},\ \Eprint
  {http://arxiv.org/abs/1907.06666} {arXiv:1907.06666 [astro-ph.CO]}
  \BibitemShut {NoStop}%
\bibitem [{\citenamefont {{Hahn}}\ \emph {et~al.}(2020)\citenamefont {{Hahn}},
  \citenamefont {{Villaescusa-Navarro}}, \citenamefont {{Castorina}},\ and\
  \citenamefont {{Scoccimarro}}}]{HahVilCas2003}%
  \BibitemOpen
  \bibfield  {author} {\bibinfo {author} {\bibfnamefont {C.}~\bibnamefont
  {{Hahn}}}, \bibinfo {author} {\bibfnamefont {F.}~\bibnamefont
  {{Villaescusa-Navarro}}}, \bibinfo {author} {\bibfnamefont {E.}~\bibnamefont
  {{Castorina}}}, \ and\ \bibinfo {author} {\bibfnamefont {R.}~\bibnamefont
  {{Scoccimarro}}},\ }\href {\doibase 10.1088/1475-7516/2020/03/040} {\bibfield
   {journal} {\bibinfo  {journal} {\jcap}\ }\textbf {\bibinfo {volume}
  {2020}},\ \bibinfo {eid} {040} (\bibinfo {year} {2020})},\ \Eprint
  {http://arxiv.org/abs/1909.11107} {arXiv:1909.11107 [astro-ph.CO]}
  \BibitemShut {NoStop}%
\bibitem [{\citenamefont {{Kamalinejad}}\ and\ \citenamefont
  {{Slepian}}(2020)}]{KamSle2011}%
  \BibitemOpen
  \bibfield  {author} {\bibinfo {author} {\bibfnamefont {F.}~\bibnamefont
  {{Kamalinejad}}}\ and\ \bibinfo {author} {\bibfnamefont {Z.}~\bibnamefont
  {{Slepian}}},\ }\href@noop {} {\bibfield  {journal} {\bibinfo  {journal}
  {arXiv e-prints}\ ,\ \bibinfo {eid} {arXiv:2011.00899}} (\bibinfo {year}
  {2020})},\ \Eprint {http://arxiv.org/abs/2011.00899} {arXiv:2011.00899
  [astro-ph.CO]} \BibitemShut {NoStop}%
\bibitem [{\citenamefont {{Hahn}}\ and\ \citenamefont
  {{Villaescusa-Navarro}}(2020)}]{HahVil2012}%
  \BibitemOpen
  \bibfield  {author} {\bibinfo {author} {\bibfnamefont {C.}~\bibnamefont
  {{Hahn}}}\ and\ \bibinfo {author} {\bibfnamefont {F.}~\bibnamefont
  {{Villaescusa-Navarro}}},\ }\href@noop {} {\bibfield  {journal} {\bibinfo
  {journal} {arXiv e-prints}\ ,\ \bibinfo {eid} {arXiv:2012.02200}} (\bibinfo
  {year} {2020})},\ \Eprint {http://arxiv.org/abs/2012.02200} {arXiv:2012.02200
  [astro-ph.CO]} \BibitemShut {NoStop}%
\bibitem [{\citenamefont {{Scoccimarro}}\ \emph {et~al.}(2012)\citenamefont
  {{Scoccimarro}}, \citenamefont {{Hui}}, \citenamefont {{Manera}},\ and\
  \citenamefont {{Chan}}}]{ScoHuiMan1204}%
  \BibitemOpen
  \bibfield  {author} {\bibinfo {author} {\bibfnamefont {R.}~\bibnamefont
  {{Scoccimarro}}}, \bibinfo {author} {\bibfnamefont {L.}~\bibnamefont
  {{Hui}}}, \bibinfo {author} {\bibfnamefont {M.}~\bibnamefont {{Manera}}}, \
  and\ \bibinfo {author} {\bibfnamefont {K.~C.}\ \bibnamefont {{Chan}}},\
  }\href {\doibase 10.1103/PhysRevD.85.083002} {\bibfield  {journal} {\bibinfo
  {journal} {\prd}\ }\textbf {\bibinfo {volume} {85}},\ \bibinfo {eid} {083002}
  (\bibinfo {year} {2012})},\ \Eprint {http://arxiv.org/abs/1108.5512}
  {arXiv:1108.5512 [astro-ph.CO]} \BibitemShut {NoStop}%
\bibitem [{\citenamefont {{Sefusatti}}\ \emph {et~al.}(2012)\citenamefont
  {{Sefusatti}}, \citenamefont {{Crocce}},\ and\ \citenamefont
  {{Desjacques}}}]{SefCroDes1210}%
  \BibitemOpen
  \bibfield  {author} {\bibinfo {author} {\bibfnamefont {E.}~\bibnamefont
  {{Sefusatti}}}, \bibinfo {author} {\bibfnamefont {M.}~\bibnamefont
  {{Crocce}}}, \ and\ \bibinfo {author} {\bibfnamefont {V.}~\bibnamefont
  {{Desjacques}}},\ }\href {\doibase 10.1111/j.1365-2966.2012.21271.x}
  {\bibfield  {journal} {\bibinfo  {journal} {\mnras}\ }\textbf {\bibinfo
  {volume} {425}},\ \bibinfo {pages} {2903} (\bibinfo {year} {2012})},\ \Eprint
  {http://arxiv.org/abs/1111.6966} {arXiv:1111.6966 [astro-ph.CO]} \BibitemShut
  {NoStop}%
\bibitem [{\citenamefont {{Tellarini}}\ \emph {et~al.}(2016)\citenamefont
  {{Tellarini}}, \citenamefont {{Ross}}, \citenamefont {{Tasinato}},\ and\
  \citenamefont {{Wands}}}]{TelRosTas1606}%
  \BibitemOpen
  \bibfield  {author} {\bibinfo {author} {\bibfnamefont {M.}~\bibnamefont
  {{Tellarini}}}, \bibinfo {author} {\bibfnamefont {A.~J.}\ \bibnamefont
  {{Ross}}}, \bibinfo {author} {\bibfnamefont {G.}~\bibnamefont {{Tasinato}}},
  \ and\ \bibinfo {author} {\bibfnamefont {D.}~\bibnamefont {{Wands}}},\ }\href
  {\doibase 10.1088/1475-7516/2016/06/014} {\bibfield  {journal} {\bibinfo
  {journal} {\jcap}\ }\textbf {\bibinfo {volume} {2016}},\ \bibinfo {eid} {014}
  (\bibinfo {year} {2016})},\ \Eprint {http://arxiv.org/abs/1603.06814}
  {arXiv:1603.06814 [astro-ph.CO]} \BibitemShut {NoStop}%
\bibitem [{\citenamefont {{Karagiannis}}\ \emph {et~al.}(2018)\citenamefont
  {{Karagiannis}}, \citenamefont {{Lazanu}}, \citenamefont {{Liguori}},
  \citenamefont {{Raccanelli}}, \citenamefont {{Bartolo}},\ and\ \citenamefont
  {{Verde}}}]{KarLazLig1806}%
  \BibitemOpen
  \bibfield  {author} {\bibinfo {author} {\bibfnamefont {D.}~\bibnamefont
  {{Karagiannis}}}, \bibinfo {author} {\bibfnamefont {A.}~\bibnamefont
  {{Lazanu}}}, \bibinfo {author} {\bibfnamefont {M.}~\bibnamefont {{Liguori}}},
  \bibinfo {author} {\bibfnamefont {A.}~\bibnamefont {{Raccanelli}}}, \bibinfo
  {author} {\bibfnamefont {N.}~\bibnamefont {{Bartolo}}}, \ and\ \bibinfo
  {author} {\bibfnamefont {L.}~\bibnamefont {{Verde}}},\ }\href {\doibase
  10.1093/mnras/sty1029} {\bibfield  {journal} {\bibinfo  {journal} {\mnras}\
  }\textbf {\bibinfo {volume} {478}},\ \bibinfo {pages} {1341} (\bibinfo {year}
  {2018})},\ \Eprint {http://arxiv.org/abs/1801.09280} {arXiv:1801.09280
  [astro-ph.CO]} \BibitemShut {NoStop}%
\bibitem [{\citenamefont {{Karagiannis}}\ \emph {et~al.}(2020)\citenamefont
  {{Karagiannis}}, \citenamefont {{Fonseca}}, \citenamefont {{Maartens}},\ and\
  \citenamefont {{Camera}}}]{KarFonRoy2010}%
  \BibitemOpen
  \bibfield  {author} {\bibinfo {author} {\bibfnamefont {D.}~\bibnamefont
  {{Karagiannis}}}, \bibinfo {author} {\bibfnamefont {J.}~\bibnamefont
  {{Fonseca}}}, \bibinfo {author} {\bibfnamefont {R.}~\bibnamefont
  {{Maartens}}}, \ and\ \bibinfo {author} {\bibfnamefont {S.}~\bibnamefont
  {{Camera}}},\ }\href@noop {} {\bibfield  {journal} {\bibinfo  {journal}
  {arXiv e-prints}\ ,\ \bibinfo {eid} {arXiv:2010.07034}} (\bibinfo {year}
  {2020})},\ \Eprint {http://arxiv.org/abs/2010.07034} {arXiv:2010.07034
  [astro-ph.CO]} \BibitemShut {NoStop}%
\bibitem [{\citenamefont {{Moradinezhad Dizgah}}\ \emph
  {et~al.}(2020)\citenamefont {{Moradinezhad Dizgah}}, \citenamefont
  {{Biagetti}}, \citenamefont {{Sefusatti}}, \citenamefont {{Desjacques}},\
  and\ \citenamefont {{Nore{\~n}a}}}]{MorBiaSef2010}%
  \BibitemOpen
  \bibfield  {author} {\bibinfo {author} {\bibfnamefont {A.}~\bibnamefont
  {{Moradinezhad Dizgah}}}, \bibinfo {author} {\bibfnamefont {M.}~\bibnamefont
  {{Biagetti}}}, \bibinfo {author} {\bibfnamefont {E.}~\bibnamefont
  {{Sefusatti}}}, \bibinfo {author} {\bibfnamefont {V.}~\bibnamefont
  {{Desjacques}}}, \ and\ \bibinfo {author} {\bibfnamefont {J.}~\bibnamefont
  {{Nore{\~n}a}}},\ }\href@noop {} {\bibfield  {journal} {\bibinfo  {journal}
  {arXiv e-prints}\ ,\ \bibinfo {eid} {arXiv:2010.14523}} (\bibinfo {year}
  {2020})},\ \Eprint {http://arxiv.org/abs/2010.14523} {arXiv:2010.14523
  [astro-ph.CO]} \BibitemShut {NoStop}%
\bibitem [{\citenamefont {{Gil-Mar{\'\i}n}}\ \emph
  {et~al.}(2015{\natexlab{a}})\citenamefont {{Gil-Mar{\'\i}n}}, \citenamefont
  {{Nore{\~n}a}}, \citenamefont {{Verde}}, \citenamefont {{Percival}},
  \citenamefont {{Wagner}}, \citenamefont {{Manera}},\ and\ \citenamefont
  {{Schneider}}}]{GilNorVer1507}%
  \BibitemOpen
  \bibfield  {author} {\bibinfo {author} {\bibfnamefont {H.}~\bibnamefont
  {{Gil-Mar{\'\i}n}}}, \bibinfo {author} {\bibfnamefont {J.}~\bibnamefont
  {{Nore{\~n}a}}}, \bibinfo {author} {\bibfnamefont {L.}~\bibnamefont
  {{Verde}}}, \bibinfo {author} {\bibfnamefont {W.~J.}\ \bibnamefont
  {{Percival}}}, \bibinfo {author} {\bibfnamefont {C.}~\bibnamefont
  {{Wagner}}}, \bibinfo {author} {\bibfnamefont {M.}~\bibnamefont {{Manera}}},
  \ and\ \bibinfo {author} {\bibfnamefont {D.~P.}\ \bibnamefont
  {{Schneider}}},\ }\href {\doibase 10.1093/mnras/stv961} {\bibfield  {journal}
  {\bibinfo  {journal} {\mnras}\ }\textbf {\bibinfo {volume} {451}},\ \bibinfo
  {pages} {539} (\bibinfo {year} {2015}{\natexlab{a}})},\ \Eprint
  {http://arxiv.org/abs/1407.5668} {arXiv:1407.5668 [astro-ph.CO]} \BibitemShut
  {NoStop}%
\bibitem [{\citenamefont {{Gil-Mar{\'\i}n}}\ \emph
  {et~al.}(2017{\natexlab{a}})\citenamefont {{Gil-Mar{\'\i}n}}, \citenamefont
  {{Percival}}, \citenamefont {{Verde}}, \citenamefont {{Brownstein}},
  \citenamefont {{Chuang}}, \citenamefont {{Kitaura}}, \citenamefont
  {{Rodr{\'\i}guez-Torres}},\ and\ \citenamefont {{Olmstead}}}]{GilPerVer1702}%
  \BibitemOpen
  \bibfield  {author} {\bibinfo {author} {\bibfnamefont {H.}~\bibnamefont
  {{Gil-Mar{\'\i}n}}}, \bibinfo {author} {\bibfnamefont {W.~J.}\ \bibnamefont
  {{Percival}}}, \bibinfo {author} {\bibfnamefont {L.}~\bibnamefont {{Verde}}},
  \bibinfo {author} {\bibfnamefont {J.~R.}\ \bibnamefont {{Brownstein}}},
  \bibinfo {author} {\bibfnamefont {C.-H.}\ \bibnamefont {{Chuang}}}, \bibinfo
  {author} {\bibfnamefont {F.-S.}\ \bibnamefont {{Kitaura}}}, \bibinfo {author}
  {\bibfnamefont {S.~A.}\ \bibnamefont {{Rodr{\'\i}guez-Torres}}}, \ and\
  \bibinfo {author} {\bibfnamefont {M.~D.}\ \bibnamefont {{Olmstead}}},\ }\href
  {\doibase 10.1093/mnras/stw2679} {\bibfield  {journal} {\bibinfo  {journal}
  {\mnras}\ }\textbf {\bibinfo {volume} {465}},\ \bibinfo {pages} {1757}
  (\bibinfo {year} {2017}{\natexlab{a}})},\ \Eprint
  {http://arxiv.org/abs/1606.00439} {arXiv:1606.00439 [astro-ph.CO]}
  \BibitemShut {NoStop}%
\bibitem [{\citenamefont {{Slepian}}\ \emph
  {et~al.}(2017{\natexlab{a}})\citenamefont {{Slepian}}, \citenamefont
  {{Eisenstein}}, \citenamefont {{Beutler}}, \citenamefont {{Chuang}},
  \citenamefont {{Cuesta}}, \citenamefont {{Ge}}, \citenamefont
  {{Gil-Mar{\'\i}n}}, \citenamefont {{Ho}}, \citenamefont {{Kitaura}},
  \citenamefont {{McBride}}, \citenamefont {{Nichol}}, \citenamefont
  {{Percival}}, \citenamefont {{Rodr{\'\i}guez-Torres}}, \citenamefont
  {{Ross}}, \citenamefont {{Scoccimarro}}, \citenamefont {{Seo}}, \citenamefont
  {{Tinker}}, \citenamefont {{Tojeiro}},\ and\ \citenamefont
  {{Vargas-Maga{\~n}a}}}]{SleEisBeu1706}%
  \BibitemOpen
  \bibfield  {author} {\bibinfo {author} {\bibfnamefont {Z.}~\bibnamefont
  {{Slepian}}}, \bibinfo {author} {\bibfnamefont {D.~J.}\ \bibnamefont
  {{Eisenstein}}}, \bibinfo {author} {\bibfnamefont {F.}~\bibnamefont
  {{Beutler}}}, \bibinfo {author} {\bibfnamefont {C.-H.}\ \bibnamefont
  {{Chuang}}}, \bibinfo {author} {\bibfnamefont {A.~J.}\ \bibnamefont
  {{Cuesta}}}, \bibinfo {author} {\bibfnamefont {J.}~\bibnamefont {{Ge}}},
  \bibinfo {author} {\bibfnamefont {H.}~\bibnamefont {{Gil-Mar{\'\i}n}}},
  \bibinfo {author} {\bibfnamefont {S.}~\bibnamefont {{Ho}}}, \bibinfo {author}
  {\bibfnamefont {F.-S.}\ \bibnamefont {{Kitaura}}}, \bibinfo {author}
  {\bibfnamefont {C.~K.}\ \bibnamefont {{McBride}}}, \bibinfo {author}
  {\bibfnamefont {R.~C.}\ \bibnamefont {{Nichol}}}, \bibinfo {author}
  {\bibfnamefont {W.~J.}\ \bibnamefont {{Percival}}}, \bibinfo {author}
  {\bibfnamefont {S.}~\bibnamefont {{Rodr{\'\i}guez-Torres}}}, \bibinfo
  {author} {\bibfnamefont {A.~J.}\ \bibnamefont {{Ross}}}, \bibinfo {author}
  {\bibfnamefont {R.}~\bibnamefont {{Scoccimarro}}}, \bibinfo {author}
  {\bibfnamefont {H.-J.}\ \bibnamefont {{Seo}}}, \bibinfo {author}
  {\bibfnamefont {J.}~\bibnamefont {{Tinker}}}, \bibinfo {author}
  {\bibfnamefont {R.}~\bibnamefont {{Tojeiro}}}, \ and\ \bibinfo {author}
  {\bibfnamefont {M.}~\bibnamefont {{Vargas-Maga{\~n}a}}},\ }\href {\doibase
  10.1093/mnras/stw3234} {\bibfield  {journal} {\bibinfo  {journal} {\mnras}\
  }\textbf {\bibinfo {volume} {468}},\ \bibinfo {pages} {1070} (\bibinfo {year}
  {2017}{\natexlab{a}})}\BibitemShut {NoStop}%
\bibitem [{\citenamefont {{Slepian}}\ \emph
  {et~al.}(2017{\natexlab{b}})\citenamefont {{Slepian}}, \citenamefont
  {{Eisenstein}}, \citenamefont {{Brownstein}}, \citenamefont {{Chuang}},
  \citenamefont {{Gil-Mar{\'\i}n}}, \citenamefont {{Ho}}, \citenamefont
  {{Kitaura}}, \citenamefont {{Percival}}, \citenamefont {{Ross}},
  \citenamefont {{Rossi}}, \citenamefont {{Seo}}, \citenamefont {{Slosar}},\
  and\ \citenamefont {{Vargas-Maga{\~n}a}}}]{SleEisBro1708}%
  \BibitemOpen
  \bibfield  {author} {\bibinfo {author} {\bibfnamefont {Z.}~\bibnamefont
  {{Slepian}}}, \bibinfo {author} {\bibfnamefont {D.~J.}\ \bibnamefont
  {{Eisenstein}}}, \bibinfo {author} {\bibfnamefont {J.~R.}\ \bibnamefont
  {{Brownstein}}}, \bibinfo {author} {\bibfnamefont {C.-H.}\ \bibnamefont
  {{Chuang}}}, \bibinfo {author} {\bibfnamefont {H.}~\bibnamefont
  {{Gil-Mar{\'\i}n}}}, \bibinfo {author} {\bibfnamefont {S.}~\bibnamefont
  {{Ho}}}, \bibinfo {author} {\bibfnamefont {F.-S.}\ \bibnamefont {{Kitaura}}},
  \bibinfo {author} {\bibfnamefont {W.~J.}\ \bibnamefont {{Percival}}},
  \bibinfo {author} {\bibfnamefont {A.~J.}\ \bibnamefont {{Ross}}}, \bibinfo
  {author} {\bibfnamefont {G.}~\bibnamefont {{Rossi}}}, \bibinfo {author}
  {\bibfnamefont {H.-J.}\ \bibnamefont {{Seo}}}, \bibinfo {author}
  {\bibfnamefont {A.}~\bibnamefont {{Slosar}}}, \ and\ \bibinfo {author}
  {\bibfnamefont {M.}~\bibnamefont {{Vargas-Maga{\~n}a}}},\ }\href {\doibase
  10.1093/mnras/stx488} {\bibfield  {journal} {\bibinfo  {journal} {\mnras}\
  }\textbf {\bibinfo {volume} {469}},\ \bibinfo {pages} {1738} (\bibinfo {year}
  {2017}{\natexlab{b}})},\ \Eprint {http://arxiv.org/abs/1607.06097}
  {arXiv:1607.06097 [astro-ph.CO]} \BibitemShut {NoStop}%
\bibitem [{\citenamefont {{Pearson}}\ and\ \citenamefont
  {{Samushia}}(2018)}]{PeaSam1808}%
  \BibitemOpen
  \bibfield  {author} {\bibinfo {author} {\bibfnamefont {D.~W.}\ \bibnamefont
  {{Pearson}}}\ and\ \bibinfo {author} {\bibfnamefont {L.}~\bibnamefont
  {{Samushia}}},\ }\href {\doibase 10.1093/mnras/sty1266} {\bibfield  {journal}
  {\bibinfo  {journal} {\mnras}\ }\textbf {\bibinfo {volume} {478}},\ \bibinfo
  {pages} {4500} (\bibinfo {year} {2018})},\ \Eprint
  {http://arxiv.org/abs/1712.04970} {arXiv:1712.04970 [astro-ph.CO]}
  \BibitemShut {NoStop}%
\bibitem [{\citenamefont {{Smith}}\ \emph {et~al.}(2008)\citenamefont
  {{Smith}}, \citenamefont {{Sheth}},\ and\ \citenamefont
  {{Scoccimarro}}}]{SmiSheSco0807}%
  \BibitemOpen
  \bibfield  {author} {\bibinfo {author} {\bibfnamefont {R.~E.}\ \bibnamefont
  {{Smith}}}, \bibinfo {author} {\bibfnamefont {R.~K.}\ \bibnamefont
  {{Sheth}}}, \ and\ \bibinfo {author} {\bibfnamefont {R.}~\bibnamefont
  {{Scoccimarro}}},\ }\href {\doibase 10.1103/PhysRevD.78.023523} {\bibfield
  {journal} {\bibinfo  {journal} {\prd}\ }\textbf {\bibinfo {volume} {78}},\
  \bibinfo {eid} {023523} (\bibinfo {year} {2008})},\ \Eprint
  {http://arxiv.org/abs/0712.0017} {arXiv:0712.0017 [astro-ph]} \BibitemShut
  {NoStop}%
\bibitem [{\citenamefont {{Rampf}}\ and\ \citenamefont
  {{Wong}}(2012)}]{RamWon1206}%
  \BibitemOpen
  \bibfield  {author} {\bibinfo {author} {\bibfnamefont {C.}~\bibnamefont
  {{Rampf}}}\ and\ \bibinfo {author} {\bibfnamefont {Y.~Y.~Y.}\ \bibnamefont
  {{Wong}}},\ }\href {\doibase 10.1088/1475-7516/2012/06/018} {\bibfield
  {journal} {\bibinfo  {journal} {\jcap}\ }\textbf {\bibinfo {volume} {2012}},\
  \bibinfo {eid} {018} (\bibinfo {year} {2012})},\ \Eprint
  {http://arxiv.org/abs/1203.4261} {arXiv:1203.4261 [astro-ph.CO]} \BibitemShut
  {NoStop}%
\bibitem [{\citenamefont {{Bernardeau}}\ \emph {et~al.}(2012)\citenamefont
  {{Bernardeau}}, \citenamefont {{Crocce}},\ and\ \citenamefont
  {{Scoccimarro}}}]{BerCroSco1206}%
  \BibitemOpen
  \bibfield  {author} {\bibinfo {author} {\bibfnamefont {F.}~\bibnamefont
  {{Bernardeau}}}, \bibinfo {author} {\bibfnamefont {M.}~\bibnamefont
  {{Crocce}}}, \ and\ \bibinfo {author} {\bibfnamefont {R.}~\bibnamefont
  {{Scoccimarro}}},\ }\href {\doibase 10.1103/PhysRevD.85.123519} {\bibfield
  {journal} {\bibinfo  {journal} {\prd}\ }\textbf {\bibinfo {volume} {85}},\
  \bibinfo {eid} {123519} (\bibinfo {year} {2012})},\ \Eprint
  {http://arxiv.org/abs/1112.3895} {arXiv:1112.3895 [astro-ph.CO]} \BibitemShut
  {NoStop}%
\bibitem [{\citenamefont {{Assassi}}\ \emph {et~al.}(2014)\citenamefont
  {{Assassi}}, \citenamefont {{Baumann}}, \citenamefont {{Green}},\ and\
  \citenamefont {{Zaldarriaga}}}]{Assassi:2014}%
  \BibitemOpen
  \bibfield  {author} {\bibinfo {author} {\bibfnamefont {V.}~\bibnamefont
  {{Assassi}}}, \bibinfo {author} {\bibfnamefont {D.}~\bibnamefont
  {{Baumann}}}, \bibinfo {author} {\bibfnamefont {D.}~\bibnamefont {{Green}}},
  \ and\ \bibinfo {author} {\bibfnamefont {M.}~\bibnamefont {{Zaldarriaga}}},\
  }\href {\doibase 10.1088/1475-7516/2014/08/056} {\bibfield  {journal}
  {\bibinfo  {journal} {\jcap}\ }\textbf {\bibinfo {volume} {8}},\ \bibinfo
  {eid} {056} (\bibinfo {year} {2014})},\ \Eprint
  {http://arxiv.org/abs/1402.5916} {arXiv:1402.5916} \BibitemShut {NoStop}%
\bibitem [{\citenamefont {{Baldauf}}\ \emph {et~al.}(2015)\citenamefont
  {{Baldauf}}, \citenamefont {{Mercolli}}, \citenamefont {{Mirbabayi}},\ and\
  \citenamefont {{Pajer}}}]{BalMerMir1505}%
  \BibitemOpen
  \bibfield  {author} {\bibinfo {author} {\bibfnamefont {T.}~\bibnamefont
  {{Baldauf}}}, \bibinfo {author} {\bibfnamefont {L.}~\bibnamefont
  {{Mercolli}}}, \bibinfo {author} {\bibfnamefont {M.}~\bibnamefont
  {{Mirbabayi}}}, \ and\ \bibinfo {author} {\bibfnamefont {E.}~\bibnamefont
  {{Pajer}}},\ }\href {\doibase 10.1088/1475-7516/2015/05/007} {\bibfield
  {journal} {\bibinfo  {journal} {\jcap}\ }\textbf {\bibinfo {volume} {2015}},\
  \bibinfo {eid} {007} (\bibinfo {year} {2015})},\ \Eprint
  {http://arxiv.org/abs/1406.4135} {arXiv:1406.4135 [astro-ph.CO]} \BibitemShut
  {NoStop}%
\bibitem [{\citenamefont {{Angulo}}\ \emph
  {et~al.}(2015{\natexlab{a}})\citenamefont {{Angulo}}, \citenamefont
  {{Foreman}}, \citenamefont {{Schmittfull}},\ and\ \citenamefont
  {{Senatore}}}]{AngForSch1510}%
  \BibitemOpen
  \bibfield  {author} {\bibinfo {author} {\bibfnamefont {R.~E.}\ \bibnamefont
  {{Angulo}}}, \bibinfo {author} {\bibfnamefont {S.}~\bibnamefont {{Foreman}}},
  \bibinfo {author} {\bibfnamefont {M.}~\bibnamefont {{Schmittfull}}}, \ and\
  \bibinfo {author} {\bibfnamefont {L.}~\bibnamefont {{Senatore}}},\ }\href
  {\doibase 10.1088/1475-7516/2015/10/039} {\bibfield  {journal} {\bibinfo
  {journal} {Journal of Cosmology and Astro-Particle Physics}\ }\textbf
  {\bibinfo {volume} {2015}},\ \bibinfo {eid} {039} (\bibinfo {year}
  {2015}{\natexlab{a}})},\ \Eprint {http://arxiv.org/abs/1406.4143}
  {arXiv:1406.4143 [astro-ph.CO]} \BibitemShut {NoStop}%
\bibitem [{\citenamefont {{Lazanu}}\ \emph {et~al.}(2016)\citenamefont
  {{Lazanu}}, \citenamefont {{Giannantonio}}, \citenamefont {{Schmittfull}},\
  and\ \citenamefont {{Shellard}}}]{LazGiaSch1604}%
  \BibitemOpen
  \bibfield  {author} {\bibinfo {author} {\bibfnamefont {A.}~\bibnamefont
  {{Lazanu}}}, \bibinfo {author} {\bibfnamefont {T.}~\bibnamefont
  {{Giannantonio}}}, \bibinfo {author} {\bibfnamefont {M.}~\bibnamefont
  {{Schmittfull}}}, \ and\ \bibinfo {author} {\bibfnamefont {E.~P.~S.}\
  \bibnamefont {{Shellard}}},\ }\href {\doibase 10.1103/PhysRevD.93.083517}
  {\bibfield  {journal} {\bibinfo  {journal} {\prd}\ }\textbf {\bibinfo
  {volume} {93}},\ \bibinfo {eid} {083517} (\bibinfo {year} {2016})},\ \Eprint
  {http://arxiv.org/abs/1510.04075} {arXiv:1510.04075 [astro-ph.CO]}
  \BibitemShut {NoStop}%
\bibitem [{\citenamefont {{Hashimoto}}\ \emph {et~al.}(2017)\citenamefont
  {{Hashimoto}}, \citenamefont {{Rasera}},\ and\ \citenamefont
  {{Taruya}}}]{HasRasTar1708}%
  \BibitemOpen
  \bibfield  {author} {\bibinfo {author} {\bibfnamefont {I.}~\bibnamefont
  {{Hashimoto}}}, \bibinfo {author} {\bibfnamefont {Y.}~\bibnamefont
  {{Rasera}}}, \ and\ \bibinfo {author} {\bibfnamefont {A.}~\bibnamefont
  {{Taruya}}},\ }\href {\doibase 10.1103/PhysRevD.96.043526} {\bibfield
  {journal} {\bibinfo  {journal} {\prd}\ }\textbf {\bibinfo {volume} {96}},\
  \bibinfo {eid} {043526} (\bibinfo {year} {2017})},\ \Eprint
  {http://arxiv.org/abs/1705.02574} {arXiv:1705.02574 [astro-ph.CO]}
  \BibitemShut {NoStop}%
\bibitem [{\citenamefont {{Desjacques}}\ \emph
  {et~al.}(2018{\natexlab{a}})\citenamefont {{Desjacques}}, \citenamefont
  {{Jeong}},\ and\ \citenamefont {{Schmidt}}}]{DesJeoSch1812}%
  \BibitemOpen
  \bibfield  {author} {\bibinfo {author} {\bibfnamefont {V.}~\bibnamefont
  {{Desjacques}}}, \bibinfo {author} {\bibfnamefont {D.}~\bibnamefont
  {{Jeong}}}, \ and\ \bibinfo {author} {\bibfnamefont {F.}~\bibnamefont
  {{Schmidt}}},\ }\href {\doibase 10.1088/1475-7516/2018/12/035} {\bibfield
  {journal} {\bibinfo  {journal} {\jcap}\ }\textbf {\bibinfo {volume} {2018}},\
  \bibinfo {eid} {035} (\bibinfo {year} {2018}{\natexlab{a}})},\ \Eprint
  {http://arxiv.org/abs/1806.04015} {arXiv:1806.04015 [astro-ph.CO]}
  \BibitemShut {NoStop}%
\bibitem [{\citenamefont {{Eggemeier}}\ \emph {et~al.}(2019)\citenamefont
  {{Eggemeier}}, \citenamefont {{Scoccimarro}},\ and\ \citenamefont
  {{Smith}}}]{EggScoSmi1906}%
  \BibitemOpen
  \bibfield  {author} {\bibinfo {author} {\bibfnamefont {A.}~\bibnamefont
  {{Eggemeier}}}, \bibinfo {author} {\bibfnamefont {R.}~\bibnamefont
  {{Scoccimarro}}}, \ and\ \bibinfo {author} {\bibfnamefont {R.~E.}\
  \bibnamefont {{Smith}}},\ }\href {\doibase 10.1103/PhysRevD.99.123514}
  {\bibfield  {journal} {\bibinfo  {journal} {\prd}\ }\textbf {\bibinfo
  {volume} {99}},\ \bibinfo {eid} {123514} (\bibinfo {year} {2019})},\ \Eprint
  {http://arxiv.org/abs/1812.03208} {arXiv:1812.03208 [astro-ph.CO]}
  \BibitemShut {NoStop}%
\bibitem [{\citenamefont {{Desjacques}}\ \emph
  {et~al.}(2018{\natexlab{b}})\citenamefont {{Desjacques}}, \citenamefont
  {{Jeong}},\ and\ \citenamefont {{Schmidt}}}]{Desjacques:2018}%
  \BibitemOpen
  \bibfield  {author} {\bibinfo {author} {\bibfnamefont {V.}~\bibnamefont
  {{Desjacques}}}, \bibinfo {author} {\bibfnamefont {D.}~\bibnamefont
  {{Jeong}}}, \ and\ \bibinfo {author} {\bibfnamefont {F.}~\bibnamefont
  {{Schmidt}}},\ }\href {\doibase 10.1016/j.physrep.2017.12.002} {\bibfield
  {journal} {\bibinfo  {journal} {\physrep}\ }\textbf {\bibinfo {volume}
  {733}},\ \bibinfo {pages} {1} (\bibinfo {year}
  {2018}{\natexlab{b}})}\BibitemShut {NoStop}%
\bibitem [{\citenamefont {{Kaiser}}(1984)}]{Kaiser:1984}%
  \BibitemOpen
  \bibfield  {author} {\bibinfo {author} {\bibfnamefont {N.}~\bibnamefont
  {{Kaiser}}},\ }\href {\doibase 10.1086/184341} {\bibfield  {journal}
  {\bibinfo  {journal} {\apjl}\ }\textbf {\bibinfo {volume} {284}},\ \bibinfo
  {pages} {L9} (\bibinfo {year} {1984})}\BibitemShut {NoStop}%
\bibitem [{\citenamefont {{Coles}}(1993)}]{Coles:1993}%
  \BibitemOpen
  \bibfield  {author} {\bibinfo {author} {\bibfnamefont {P.}~\bibnamefont
  {{Coles}}},\ }\href {\doibase 10.1093/mnras/262.4.1065} {\bibfield  {journal}
  {\bibinfo  {journal} {\mnras}\ }\textbf {\bibinfo {volume} {262}},\ \bibinfo
  {pages} {1065} (\bibinfo {year} {1993})}\BibitemShut {NoStop}%
\bibitem [{\citenamefont {{Fry}}\ and\ \citenamefont
  {{Gaztanaga}}(1993)}]{FryGaz9308}%
  \BibitemOpen
  \bibfield  {author} {\bibinfo {author} {\bibfnamefont {J.~N.}\ \bibnamefont
  {{Fry}}}\ and\ \bibinfo {author} {\bibfnamefont {E.}~\bibnamefont
  {{Gaztanaga}}},\ }\href {\doibase 10.1086/173015} {\bibfield  {journal}
  {\bibinfo  {journal} {\apj}\ }\textbf {\bibinfo {volume} {413}},\ \bibinfo
  {pages} {447} (\bibinfo {year} {1993})},\ \Eprint
  {http://arxiv.org/abs/arXiv:astro-ph/9302009} {arXiv:astro-ph/9302009}
  \BibitemShut {NoStop}%
\bibitem [{\citenamefont {{Catelan}}\ \emph {et~al.}(1998)\citenamefont
  {{Catelan}}, \citenamefont {{Lucchin}}, \citenamefont {{Matarrese}},\ and\
  \citenamefont {{Porciani}}}]{CatLucMat9807}%
  \BibitemOpen
  \bibfield  {author} {\bibinfo {author} {\bibfnamefont {P.}~\bibnamefont
  {{Catelan}}}, \bibinfo {author} {\bibfnamefont {F.}~\bibnamefont
  {{Lucchin}}}, \bibinfo {author} {\bibfnamefont {S.}~\bibnamefont
  {{Matarrese}}}, \ and\ \bibinfo {author} {\bibfnamefont {C.}~\bibnamefont
  {{Porciani}}},\ }\href {\doibase 10.1046/j.1365-8711.1998.01455.x} {\bibfield
   {journal} {\bibinfo  {journal} {\mnras}\ }\textbf {\bibinfo {volume}
  {297}},\ \bibinfo {pages} {692} (\bibinfo {year} {1998})},\ \Eprint
  {http://arxiv.org/abs/arXiv:astro-ph/9708067} {arXiv:astro-ph/9708067}
  \BibitemShut {NoStop}%
\bibitem [{\citenamefont {{Catelan}}\ \emph {et~al.}(2000)\citenamefont
  {{Catelan}}, \citenamefont {{Porciani}},\ and\ \citenamefont
  {{Kamionkowski}}}]{Catelan:2000}%
  \BibitemOpen
  \bibfield  {author} {\bibinfo {author} {\bibfnamefont {P.}~\bibnamefont
  {{Catelan}}}, \bibinfo {author} {\bibfnamefont {C.}~\bibnamefont
  {{Porciani}}}, \ and\ \bibinfo {author} {\bibfnamefont {M.}~\bibnamefont
  {{Kamionkowski}}},\ }\href {\doibase 10.1046/j.1365-8711.2000.04023.x}
  {\bibfield  {journal} {\bibinfo  {journal} {\mnras}\ }\textbf {\bibinfo
  {volume} {318}},\ \bibinfo {pages} {L39} (\bibinfo {year} {2000})},\ \Eprint
  {http://arxiv.org/abs/astro-ph/0005544} {astro-ph/0005544} \BibitemShut
  {NoStop}%
\bibitem [{\citenamefont {{McDonald}}\ and\ \citenamefont
  {{Roy}}(2009)}]{McDonald:2009}%
  \BibitemOpen
  \bibfield  {author} {\bibinfo {author} {\bibfnamefont {P.}~\bibnamefont
  {{McDonald}}}\ and\ \bibinfo {author} {\bibfnamefont {A.}~\bibnamefont
  {{Roy}}},\ }\href {\doibase 10.1088/1475-7516/2009/08/020} {\bibfield
  {journal} {\bibinfo  {journal} {\jcap}\ }\textbf {\bibinfo {volume} {8}},\
  \bibinfo {eid} {020} (\bibinfo {year} {2009})},\ \Eprint
  {http://arxiv.org/abs/0902.0991} {arXiv:0902.0991 [astro-ph.CO]} \BibitemShut
  {NoStop}%
\bibitem [{\citenamefont {{Matsubara}}(2011)}]{Matsubara:2011}%
  \BibitemOpen
  \bibfield  {author} {\bibinfo {author} {\bibfnamefont {T.}~\bibnamefont
  {{Matsubara}}},\ }\href {\doibase 10.1103/PhysRevD.83.083518} {\bibfield
  {journal} {\bibinfo  {journal} {\prd}\ }\textbf {\bibinfo {volume} {83}},\
  \bibinfo {eid} {083518} (\bibinfo {year} {2011})},\ \Eprint
  {http://arxiv.org/abs/1102.4619} {arXiv:1102.4619 [astro-ph.CO]} \BibitemShut
  {NoStop}%
\bibitem [{\citenamefont {{Guo}}\ and\ \citenamefont
  {{Jing}}(2009)}]{GuoJin0909}%
  \BibitemOpen
  \bibfield  {author} {\bibinfo {author} {\bibfnamefont {H.}~\bibnamefont
  {{Guo}}}\ and\ \bibinfo {author} {\bibfnamefont {Y.~P.}\ \bibnamefont
  {{Jing}}},\ }\href {\doibase 10.1088/0004-637X/702/1/425} {\bibfield
  {journal} {\bibinfo  {journal} {\apj}\ }\textbf {\bibinfo {volume} {702}},\
  \bibinfo {pages} {425} (\bibinfo {year} {2009})},\ \Eprint
  {http://arxiv.org/abs/0907.0282} {arXiv:0907.0282 [astro-ph.CO]} \BibitemShut
  {NoStop}%
\bibitem [{\citenamefont {{Pollack}}\ \emph {et~al.}(2012)\citenamefont
  {{Pollack}}, \citenamefont {{Smith}},\ and\ \citenamefont
  {{Porciani}}}]{Pollack:2012}%
  \BibitemOpen
  \bibfield  {author} {\bibinfo {author} {\bibfnamefont {J.~E.}\ \bibnamefont
  {{Pollack}}}, \bibinfo {author} {\bibfnamefont {R.~E.}\ \bibnamefont
  {{Smith}}}, \ and\ \bibinfo {author} {\bibfnamefont {C.}~\bibnamefont
  {{Porciani}}},\ }\href {\doibase 10.1111/j.1365-2966.2011.20279.x} {\bibfield
   {journal} {\bibinfo  {journal} {\mnras}\ }\textbf {\bibinfo {volume}
  {420}},\ \bibinfo {pages} {3469} (\bibinfo {year} {2012})},\ \Eprint
  {http://arxiv.org/abs/1109.3458} {arXiv:1109.3458} \BibitemShut {NoStop}%
\bibitem [{\citenamefont {{Chan}}\ \emph {et~al.}(2012)\citenamefont {{Chan}},
  \citenamefont {{Scoccimarro}},\ and\ \citenamefont {{Sheth}}}]{Chan:2012}%
  \BibitemOpen
  \bibfield  {author} {\bibinfo {author} {\bibfnamefont {K.~C.}\ \bibnamefont
  {{Chan}}}, \bibinfo {author} {\bibfnamefont {R.}~\bibnamefont
  {{Scoccimarro}}}, \ and\ \bibinfo {author} {\bibfnamefont {R.~K.}\
  \bibnamefont {{Sheth}}},\ }\href {\doibase 10.1103/PhysRevD.85.083509}
  {\bibfield  {journal} {\bibinfo  {journal} {\prd}\ }\textbf {\bibinfo
  {volume} {85}},\ \bibinfo {eid} {083509} (\bibinfo {year} {2012})},\ \Eprint
  {http://arxiv.org/abs/1201.3614} {arXiv:1201.3614 [astro-ph.CO]} \BibitemShut
  {NoStop}%
\bibitem [{\citenamefont {{Baldauf}}\ \emph {et~al.}(2012)\citenamefont
  {{Baldauf}}, \citenamefont {{Seljak}}, \citenamefont {{Desjacques}},\ and\
  \citenamefont {{McDonald}}}]{Baldauf:2012}%
  \BibitemOpen
  \bibfield  {author} {\bibinfo {author} {\bibfnamefont {T.}~\bibnamefont
  {{Baldauf}}}, \bibinfo {author} {\bibfnamefont {U.}~\bibnamefont {{Seljak}}},
  \bibinfo {author} {\bibfnamefont {V.}~\bibnamefont {{Desjacques}}}, \ and\
  \bibinfo {author} {\bibfnamefont {P.}~\bibnamefont {{McDonald}}},\ }\href
  {\doibase 10.1103/PhysRevD.86.083540} {\bibfield  {journal} {\bibinfo
  {journal} {\prd}\ }\textbf {\bibinfo {volume} {86}},\ \bibinfo {eid} {083540}
  (\bibinfo {year} {2012})},\ \Eprint {http://arxiv.org/abs/1201.4827}
  {arXiv:1201.4827 [astro-ph.CO]} \BibitemShut {NoStop}%
\bibitem [{\citenamefont {{Sheth}}\ \emph {et~al.}(2013)\citenamefont
  {{Sheth}}, \citenamefont {{Chan}},\ and\ \citenamefont
  {{Scoccimarro}}}]{SheChaSco1304}%
  \BibitemOpen
  \bibfield  {author} {\bibinfo {author} {\bibfnamefont {R.~K.}\ \bibnamefont
  {{Sheth}}}, \bibinfo {author} {\bibfnamefont {K.~C.}\ \bibnamefont {{Chan}}},
  \ and\ \bibinfo {author} {\bibfnamefont {R.}~\bibnamefont {{Scoccimarro}}},\
  }\href {\doibase 10.1103/PhysRevD.87.083002} {\bibfield  {journal} {\bibinfo
  {journal} {\prd}\ }\textbf {\bibinfo {volume} {87}},\ \bibinfo {eid} {083002}
  (\bibinfo {year} {2013})},\ \Eprint {http://arxiv.org/abs/1207.7117}
  {arXiv:1207.7117 [astro-ph.CO]} \BibitemShut {NoStop}%
\bibitem [{\citenamefont {{Pollack}}\ \emph {et~al.}(2014)\citenamefont
  {{Pollack}}, \citenamefont {{Smith}},\ and\ \citenamefont
  {{Porciani}}}]{PolSmiPor1405}%
  \BibitemOpen
  \bibfield  {author} {\bibinfo {author} {\bibfnamefont {J.~E.}\ \bibnamefont
  {{Pollack}}}, \bibinfo {author} {\bibfnamefont {R.~E.}\ \bibnamefont
  {{Smith}}}, \ and\ \bibinfo {author} {\bibfnamefont {C.}~\bibnamefont
  {{Porciani}}},\ }\href {\doibase 10.1093/mnras/stu322} {\bibfield  {journal}
  {\bibinfo  {journal} {\mnras}\ }\textbf {\bibinfo {volume} {440}},\ \bibinfo
  {pages} {555} (\bibinfo {year} {2014})},\ \Eprint
  {http://arxiv.org/abs/1309.0504} {arXiv:1309.0504 [astro-ph.CO]} \BibitemShut
  {NoStop}%
\bibitem [{\citenamefont {{Saito}}\ \emph {et~al.}(2014)\citenamefont
  {{Saito}}, \citenamefont {{Baldauf}}, \citenamefont {{Vlah}}, \citenamefont
  {{Seljak}}, \citenamefont {{Okumura}},\ and\ \citenamefont
  {{McDonald}}}]{SaiBalVla1405}%
  \BibitemOpen
  \bibfield  {author} {\bibinfo {author} {\bibfnamefont {S.}~\bibnamefont
  {{Saito}}}, \bibinfo {author} {\bibfnamefont {T.}~\bibnamefont {{Baldauf}}},
  \bibinfo {author} {\bibfnamefont {Z.}~\bibnamefont {{Vlah}}}, \bibinfo
  {author} {\bibfnamefont {U.}~\bibnamefont {{Seljak}}}, \bibinfo {author}
  {\bibfnamefont {T.}~\bibnamefont {{Okumura}}}, \ and\ \bibinfo {author}
  {\bibfnamefont {P.}~\bibnamefont {{McDonald}}},\ }\href {\doibase
  10.1103/PhysRevD.90.123522} {\bibfield  {journal} {\bibinfo  {journal}
  {\prd}\ }\textbf {\bibinfo {volume} {90}},\ \bibinfo {eid} {123522} (\bibinfo
  {year} {2014})},\ \Eprint {http://arxiv.org/abs/1405.1447} {arXiv:1405.1447
  [astro-ph.CO]} \BibitemShut {NoStop}%
\bibitem [{\citenamefont {{Biagetti}}\ \emph {et~al.}(2014)\citenamefont
  {{Biagetti}}, \citenamefont {{Desjacques}}, \citenamefont {{Kehagias}},\ and\
  \citenamefont {{Riotto}}}]{BiaDesKeh1408}%
  \BibitemOpen
  \bibfield  {author} {\bibinfo {author} {\bibfnamefont {M.}~\bibnamefont
  {{Biagetti}}}, \bibinfo {author} {\bibfnamefont {V.}~\bibnamefont
  {{Desjacques}}}, \bibinfo {author} {\bibfnamefont {A.}~\bibnamefont
  {{Kehagias}}}, \ and\ \bibinfo {author} {\bibfnamefont {A.}~\bibnamefont
  {{Riotto}}},\ }\href {\doibase 10.1103/PhysRevD.90.103529} {\bibfield
  {journal} {\bibinfo  {journal} {\prd}\ }\textbf {\bibinfo {volume} {90}},\
  \bibinfo {eid} {103529} (\bibinfo {year} {2014})},\ \Eprint
  {http://arxiv.org/abs/1408.0293} {arXiv:1408.0293 [astro-ph.CO]} \BibitemShut
  {NoStop}%
\bibitem [{\citenamefont {{Angulo}}\ \emph
  {et~al.}(2015{\natexlab{b}})\citenamefont {{Angulo}}, \citenamefont
  {{Fasiello}}, \citenamefont {{Senatore}},\ and\ \citenamefont
  {{Vlah}}}]{AngFasSen1503}%
  \BibitemOpen
  \bibfield  {author} {\bibinfo {author} {\bibfnamefont {R.}~\bibnamefont
  {{Angulo}}}, \bibinfo {author} {\bibfnamefont {M.}~\bibnamefont
  {{Fasiello}}}, \bibinfo {author} {\bibfnamefont {L.}~\bibnamefont
  {{Senatore}}}, \ and\ \bibinfo {author} {\bibfnamefont {Z.}~\bibnamefont
  {{Vlah}}},\ }\href {\doibase 10.1088/1475-7516/2015/09/029} {\bibfield
  {journal} {\bibinfo  {journal} {Journal of Cosmology and Astro-Particle
  Physics}\ }\textbf {\bibinfo {volume} {2015}},\ \bibinfo {eid} {029}
  (\bibinfo {year} {2015}{\natexlab{b}})},\ \Eprint
  {http://arxiv.org/abs/1503.08826} {arXiv:1503.08826 [astro-ph.CO]}
  \BibitemShut {NoStop}%
\bibitem [{\citenamefont {{Bel}}\ \emph {et~al.}(2015)\citenamefont {{Bel}},
  \citenamefont {{Hoffmann}},\ and\ \citenamefont
  {{Gazta{\~n}aga}}}]{BelHofGaz1510}%
  \BibitemOpen
  \bibfield  {author} {\bibinfo {author} {\bibfnamefont {J.}~\bibnamefont
  {{Bel}}}, \bibinfo {author} {\bibfnamefont {K.}~\bibnamefont {{Hoffmann}}}, \
  and\ \bibinfo {author} {\bibfnamefont {E.}~\bibnamefont {{Gazta{\~n}aga}}},\
  }\href {\doibase 10.1093/mnras/stv1600} {\bibfield  {journal} {\bibinfo
  {journal} {\mnras}\ }\textbf {\bibinfo {volume} {453}},\ \bibinfo {pages}
  {259} (\bibinfo {year} {2015})},\ \Eprint {http://arxiv.org/abs/1504.02074}
  {arXiv:1504.02074 [astro-ph.CO]} \BibitemShut {NoStop}%
\bibitem [{\citenamefont {{Lazeyras}}\ and\ \citenamefont
  {{Schmidt}}(2018)}]{LazSch1712}%
  \BibitemOpen
  \bibfield  {author} {\bibinfo {author} {\bibfnamefont {T.}~\bibnamefont
  {{Lazeyras}}}\ and\ \bibinfo {author} {\bibfnamefont {F.}~\bibnamefont
  {{Schmidt}}},\ }\href {\doibase 10.1088/1475-7516/2018/09/008} {\bibfield
  {journal} {\bibinfo  {journal} {Journal of Cosmology and Astro-Particle
  Physics}\ }\textbf {\bibinfo {volume} {2018}},\ \bibinfo {eid} {008}
  (\bibinfo {year} {2018})},\ \Eprint {http://arxiv.org/abs/1712.07531}
  {arXiv:1712.07531 [astro-ph.CO]} \BibitemShut {NoStop}%
\bibitem [{\citenamefont {{Abidi}}\ and\ \citenamefont
  {{Baldauf}}(2018)}]{AbiBal1807}%
  \BibitemOpen
  \bibfield  {author} {\bibinfo {author} {\bibfnamefont {M.~M.}\ \bibnamefont
  {{Abidi}}}\ and\ \bibinfo {author} {\bibfnamefont {T.}~\bibnamefont
  {{Baldauf}}},\ }\href {\doibase 10.1088/1475-7516/2018/07/029} {\bibfield
  {journal} {\bibinfo  {journal} {\jcap}\ }\textbf {\bibinfo {volume} {7}},\
  \bibinfo {eid} {029} (\bibinfo {year} {2018})},\ \Eprint
  {http://arxiv.org/abs/1802.07622} {arXiv:1802.07622} \BibitemShut {NoStop}%
\bibitem [{\citenamefont {{Oddo}}\ \emph {et~al.}(2020)\citenamefont {{Oddo}},
  \citenamefont {{Sefusatti}}, \citenamefont {{Porciani}}, \citenamefont
  {{Monaco}},\ and\ \citenamefont {{S{\'a}nchez}}}]{OddSefPor2003}%
  \BibitemOpen
  \bibfield  {author} {\bibinfo {author} {\bibfnamefont {A.}~\bibnamefont
  {{Oddo}}}, \bibinfo {author} {\bibfnamefont {E.}~\bibnamefont {{Sefusatti}}},
  \bibinfo {author} {\bibfnamefont {C.}~\bibnamefont {{Porciani}}}, \bibinfo
  {author} {\bibfnamefont {P.}~\bibnamefont {{Monaco}}}, \ and\ \bibinfo
  {author} {\bibfnamefont {A.~G.}\ \bibnamefont {{S{\'a}nchez}}},\ }\href
  {\doibase 10.1088/1475-7516/2020/03/056} {\bibfield  {journal} {\bibinfo
  {journal} {\jcap}\ }\textbf {\bibinfo {volume} {2020}},\ \bibinfo {eid} {056}
  (\bibinfo {year} {2020})},\ \Eprint {http://arxiv.org/abs/1908.01774}
  {arXiv:1908.01774 [astro-ph.CO]} \BibitemShut {NoStop}%
\bibitem [{\citenamefont {{Chiang}}\ \emph {et~al.}(2015)\citenamefont
  {{Chiang}}, \citenamefont {{Wagner}}, \citenamefont {{S{\'a}nchez}},
  \citenamefont {{Schmidt}},\ and\ \citenamefont {{Komatsu}}}]{ChiWagSan1509}%
  \BibitemOpen
  \bibfield  {author} {\bibinfo {author} {\bibfnamefont {C.-T.}\ \bibnamefont
  {{Chiang}}}, \bibinfo {author} {\bibfnamefont {C.}~\bibnamefont {{Wagner}}},
  \bibinfo {author} {\bibfnamefont {A.~G.}\ \bibnamefont {{S{\'a}nchez}}},
  \bibinfo {author} {\bibfnamefont {F.}~\bibnamefont {{Schmidt}}}, \ and\
  \bibinfo {author} {\bibfnamefont {E.}~\bibnamefont {{Komatsu}}},\ }\href
  {\doibase 10.1088/1475-7516/2015/09/028} {\bibfield  {journal} {\bibinfo
  {journal} {\jcap}\ }\textbf {\bibinfo {volume} {2015}},\ \bibinfo {eid} {028}
  (\bibinfo {year} {2015})},\ \Eprint {http://arxiv.org/abs/1504.03322}
  {arXiv:1504.03322 [astro-ph.CO]} \BibitemShut {NoStop}%
\bibitem [{\citenamefont {{Fry}}(1996)}]{Fry9604}%
  \BibitemOpen
  \bibfield  {author} {\bibinfo {author} {\bibfnamefont {J.~N.}\ \bibnamefont
  {{Fry}}},\ }\href {\doibase 10.1086/310006} {\bibfield  {journal} {\bibinfo
  {journal} {\apjl}\ }\textbf {\bibinfo {volume} {461}},\ \bibinfo {pages}
  {L65+} (\bibinfo {year} {1996})}\BibitemShut {NoStop}%
\bibitem [{\citenamefont {{S{\'a}nchez}}\ \emph {et~al.}(2017)\citenamefont
  {{S{\'a}nchez}}, \citenamefont {{Scoccimarro}}, \citenamefont {{Crocce}},
  \citenamefont {{Grieb}}, \citenamefont {{Salazar-Albornoz}}, \citenamefont
  {{Dalla Vecchia}}, \citenamefont {{Lippich}}, \citenamefont {{Beutler}},
  \citenamefont {{Brownstein}}, \citenamefont {{Chuang}}, \citenamefont
  {{Eisenstein}}, \citenamefont {{Kitaura}}, \citenamefont {{Olmstead}},
  \citenamefont {{Percival}}, \citenamefont {{Prada}}, \citenamefont
  {{Rodr{\'{\i}}guez-Torres}}, \citenamefont {{Ross}}, \citenamefont
  {{Samushia}}, \citenamefont {{Seo}}, \citenamefont {{Tinker}}, \citenamefont
  {{Tojeiro}}, \citenamefont {{Vargas-Maga{\~n}a}}, \citenamefont {{Wang}},\
  and\ \citenamefont {{Zhao}}}]{SanScoCro1701}%
  \BibitemOpen
  \bibfield  {author} {\bibinfo {author} {\bibfnamefont {A.~G.}\ \bibnamefont
  {{S{\'a}nchez}}}, \bibinfo {author} {\bibfnamefont {R.}~\bibnamefont
  {{Scoccimarro}}}, \bibinfo {author} {\bibfnamefont {M.}~\bibnamefont
  {{Crocce}}}, \bibinfo {author} {\bibfnamefont {J.~N.}\ \bibnamefont
  {{Grieb}}}, \bibinfo {author} {\bibfnamefont {S.}~\bibnamefont
  {{Salazar-Albornoz}}}, \bibinfo {author} {\bibfnamefont {C.}~\bibnamefont
  {{Dalla Vecchia}}}, \bibinfo {author} {\bibfnamefont {M.}~\bibnamefont
  {{Lippich}}}, \bibinfo {author} {\bibfnamefont {F.}~\bibnamefont
  {{Beutler}}}, \bibinfo {author} {\bibfnamefont {J.~R.}\ \bibnamefont
  {{Brownstein}}}, \bibinfo {author} {\bibfnamefont {C.-H.}\ \bibnamefont
  {{Chuang}}}, \bibinfo {author} {\bibfnamefont {D.~J.}\ \bibnamefont
  {{Eisenstein}}}, \bibinfo {author} {\bibfnamefont {F.-S.}\ \bibnamefont
  {{Kitaura}}}, \bibinfo {author} {\bibfnamefont {M.~D.}\ \bibnamefont
  {{Olmstead}}}, \bibinfo {author} {\bibfnamefont {W.~J.}\ \bibnamefont
  {{Percival}}}, \bibinfo {author} {\bibfnamefont {F.}~\bibnamefont {{Prada}}},
  \bibinfo {author} {\bibfnamefont {S.}~\bibnamefont
  {{Rodr{\'{\i}}guez-Torres}}}, \bibinfo {author} {\bibfnamefont {A.~J.}\
  \bibnamefont {{Ross}}}, \bibinfo {author} {\bibfnamefont {L.}~\bibnamefont
  {{Samushia}}}, \bibinfo {author} {\bibfnamefont {H.-J.}\ \bibnamefont
  {{Seo}}}, \bibinfo {author} {\bibfnamefont {J.}~\bibnamefont {{Tinker}}},
  \bibinfo {author} {\bibfnamefont {R.}~\bibnamefont {{Tojeiro}}}, \bibinfo
  {author} {\bibfnamefont {M.}~\bibnamefont {{Vargas-Maga{\~n}a}}}, \bibinfo
  {author} {\bibfnamefont {Y.}~\bibnamefont {{Wang}}}, \ and\ \bibinfo {author}
  {\bibfnamefont {G.-B.}\ \bibnamefont {{Zhao}}},\ }\href {\doibase
  10.1093/mnras/stw2443} {\bibfield  {journal} {\bibinfo  {journal} {\mnras}\
  }\textbf {\bibinfo {volume} {464}},\ \bibinfo {pages} {1640} (\bibinfo {year}
  {2017})},\ \Eprint {http://arxiv.org/abs/1607.03147} {arXiv:1607.03147}
  \BibitemShut {NoStop}%
\bibitem [{\citenamefont {{Gil-Mar{\'\i}n}}\ \emph
  {et~al.}(2017{\natexlab{b}})\citenamefont {{Gil-Mar{\'\i}n}}, \citenamefont
  {{Percival}}, \citenamefont {{Verde}}, \citenamefont {{Brownstein}},
  \citenamefont {{Chuang}}, \citenamefont {{Kitaura}}, \citenamefont
  {{Rodr{\'\i}guez-Torres}},\ and\ \citenamefont {{Olmstead}}}]{GilPerVer1606}%
  \BibitemOpen
  \bibfield  {author} {\bibinfo {author} {\bibfnamefont {H.}~\bibnamefont
  {{Gil-Mar{\'\i}n}}}, \bibinfo {author} {\bibfnamefont {W.~J.}\ \bibnamefont
  {{Percival}}}, \bibinfo {author} {\bibfnamefont {L.}~\bibnamefont {{Verde}}},
  \bibinfo {author} {\bibfnamefont {J.~R.}\ \bibnamefont {{Brownstein}}},
  \bibinfo {author} {\bibfnamefont {C.-H.}\ \bibnamefont {{Chuang}}}, \bibinfo
  {author} {\bibfnamefont {F.-S.}\ \bibnamefont {{Kitaura}}}, \bibinfo {author}
  {\bibfnamefont {S.~A.}\ \bibnamefont {{Rodr{\'\i}guez-Torres}}}, \ and\
  \bibinfo {author} {\bibfnamefont {M.~D.}\ \bibnamefont {{Olmstead}}},\ }\href
  {\doibase 10.1093/mnras/stw2679} {\bibfield  {journal} {\bibinfo  {journal}
  {\mnras}\ }\textbf {\bibinfo {volume} {465}},\ \bibinfo {pages} {1757}
  (\bibinfo {year} {2017}{\natexlab{b}})},\ \Eprint
  {http://arxiv.org/abs/1606.00439} {arXiv:1606.00439 [astro-ph.CO]}
  \BibitemShut {NoStop}%
\bibitem [{\citenamefont {{Beutler}}\ \emph {et~al.}(2017)\citenamefont
  {{Beutler}}, \citenamefont {{Seo}}, \citenamefont {{Saito}}, \citenamefont
  {{Chuang}}, \citenamefont {{Cuesta}}, \citenamefont {{Eisenstein}},
  \citenamefont {{Gil-Mar{\'\i}n}}, \citenamefont {{Grieb}}, \citenamefont
  {{Hand}}, \citenamefont {{Kitaura}}, \citenamefont {{Modi}}, \citenamefont
  {{Nichol}}, \citenamefont {{Olmstead}}, \citenamefont {{Percival}},
  \citenamefont {{Prada}}, \citenamefont {{S{\'a}nchez}}, \citenamefont
  {{Rodriguez-Torres}}, \citenamefont {{Ross}}, \citenamefont {{Ross}},
  \citenamefont {{Schneider}}, \citenamefont {{Tinker}}, \citenamefont
  {{Tojeiro}},\ and\ \citenamefont {{Vargas-Maga{\~n}a}}}]{BeuSeoSai1704}%
  \BibitemOpen
  \bibfield  {author} {\bibinfo {author} {\bibfnamefont {F.}~\bibnamefont
  {{Beutler}}}, \bibinfo {author} {\bibfnamefont {H.-J.}\ \bibnamefont
  {{Seo}}}, \bibinfo {author} {\bibfnamefont {S.}~\bibnamefont {{Saito}}},
  \bibinfo {author} {\bibfnamefont {C.-H.}\ \bibnamefont {{Chuang}}}, \bibinfo
  {author} {\bibfnamefont {A.~J.}\ \bibnamefont {{Cuesta}}}, \bibinfo {author}
  {\bibfnamefont {D.~J.}\ \bibnamefont {{Eisenstein}}}, \bibinfo {author}
  {\bibfnamefont {H.}~\bibnamefont {{Gil-Mar{\'\i}n}}}, \bibinfo {author}
  {\bibfnamefont {J.~N.}\ \bibnamefont {{Grieb}}}, \bibinfo {author}
  {\bibfnamefont {N.}~\bibnamefont {{Hand}}}, \bibinfo {author} {\bibfnamefont
  {F.-S.}\ \bibnamefont {{Kitaura}}}, \bibinfo {author} {\bibfnamefont
  {C.}~\bibnamefont {{Modi}}}, \bibinfo {author} {\bibfnamefont {R.~C.}\
  \bibnamefont {{Nichol}}}, \bibinfo {author} {\bibfnamefont {M.~D.}\
  \bibnamefont {{Olmstead}}}, \bibinfo {author} {\bibfnamefont {W.~J.}\
  \bibnamefont {{Percival}}}, \bibinfo {author} {\bibfnamefont
  {F.}~\bibnamefont {{Prada}}}, \bibinfo {author} {\bibfnamefont {A.~G.}\
  \bibnamefont {{S{\'a}nchez}}}, \bibinfo {author} {\bibfnamefont
  {S.}~\bibnamefont {{Rodriguez-Torres}}}, \bibinfo {author} {\bibfnamefont
  {A.~J.}\ \bibnamefont {{Ross}}}, \bibinfo {author} {\bibfnamefont {N.~P.}\
  \bibnamefont {{Ross}}}, \bibinfo {author} {\bibfnamefont {D.~P.}\
  \bibnamefont {{Schneider}}}, \bibinfo {author} {\bibfnamefont
  {J.}~\bibnamefont {{Tinker}}}, \bibinfo {author} {\bibfnamefont
  {R.}~\bibnamefont {{Tojeiro}}}, \ and\ \bibinfo {author} {\bibfnamefont
  {M.}~\bibnamefont {{Vargas-Maga{\~n}a}}},\ }\href {\doibase
  10.1093/mnras/stw3298} {\bibfield  {journal} {\bibinfo  {journal} {\mnras}\
  }\textbf {\bibinfo {volume} {466}},\ \bibinfo {pages} {2242} (\bibinfo {year}
  {2017})},\ \Eprint {http://arxiv.org/abs/1607.03150} {arXiv:1607.03150
  [astro-ph.CO]} \BibitemShut {NoStop}%
\bibitem [{\citenamefont {{Grieb}}\ \emph {et~al.}(2017)\citenamefont
  {{Grieb}}, \citenamefont {{S{\'a}nchez}}, \citenamefont {{Salazar-Albornoz}},
  \citenamefont {{Scoccimarro}}, \citenamefont {{Crocce}}, \citenamefont
  {{Dalla Vecchia}}, \citenamefont {{Montesano}}, \citenamefont
  {{Gil-Mar{\'{\i}}n}}, \citenamefont {{Ross}}, \citenamefont {{Beutler}},
  \citenamefont {{Rodr{\'{\i}}guez-Torres}}, \citenamefont {{Chuang}},
  \citenamefont {{Prada}}, \citenamefont {{Kitaura}}, \citenamefont {{Cuesta}},
  \citenamefont {{Eisenstein}}, \citenamefont {{Percival}}, \citenamefont
  {{Vargas-Maga{\~n}a}}, \citenamefont {{Tinker}}, \citenamefont {{Tojeiro}},
  \citenamefont {{Brownstein}}, \citenamefont {{Maraston}}, \citenamefont
  {{Nichol}}, \citenamefont {{Olmstead}}, \citenamefont {{Samushia}},
  \citenamefont {{Seo}}, \citenamefont {{Streblyanska}},\ and\ \citenamefont
  {{Zhao}}}]{GriSanSal1705}%
  \BibitemOpen
  \bibfield  {author} {\bibinfo {author} {\bibfnamefont {J.~N.}\ \bibnamefont
  {{Grieb}}}, \bibinfo {author} {\bibfnamefont {A.~G.}\ \bibnamefont
  {{S{\'a}nchez}}}, \bibinfo {author} {\bibfnamefont {S.}~\bibnamefont
  {{Salazar-Albornoz}}}, \bibinfo {author} {\bibfnamefont {R.}~\bibnamefont
  {{Scoccimarro}}}, \bibinfo {author} {\bibfnamefont {M.}~\bibnamefont
  {{Crocce}}}, \bibinfo {author} {\bibfnamefont {C.}~\bibnamefont {{Dalla
  Vecchia}}}, \bibinfo {author} {\bibfnamefont {F.}~\bibnamefont
  {{Montesano}}}, \bibinfo {author} {\bibfnamefont {H.}~\bibnamefont
  {{Gil-Mar{\'{\i}}n}}}, \bibinfo {author} {\bibfnamefont {A.~J.}\ \bibnamefont
  {{Ross}}}, \bibinfo {author} {\bibfnamefont {F.}~\bibnamefont {{Beutler}}},
  \bibinfo {author} {\bibfnamefont {S.}~\bibnamefont
  {{Rodr{\'{\i}}guez-Torres}}}, \bibinfo {author} {\bibfnamefont {C.-H.}\
  \bibnamefont {{Chuang}}}, \bibinfo {author} {\bibfnamefont {F.}~\bibnamefont
  {{Prada}}}, \bibinfo {author} {\bibfnamefont {F.-S.}\ \bibnamefont
  {{Kitaura}}}, \bibinfo {author} {\bibfnamefont {A.~J.}\ \bibnamefont
  {{Cuesta}}}, \bibinfo {author} {\bibfnamefont {D.~J.}\ \bibnamefont
  {{Eisenstein}}}, \bibinfo {author} {\bibfnamefont {W.~J.}\ \bibnamefont
  {{Percival}}}, \bibinfo {author} {\bibfnamefont {M.}~\bibnamefont
  {{Vargas-Maga{\~n}a}}}, \bibinfo {author} {\bibfnamefont {J.~L.}\
  \bibnamefont {{Tinker}}}, \bibinfo {author} {\bibfnamefont {R.}~\bibnamefont
  {{Tojeiro}}}, \bibinfo {author} {\bibfnamefont {J.~R.}\ \bibnamefont
  {{Brownstein}}}, \bibinfo {author} {\bibfnamefont {C.}~\bibnamefont
  {{Maraston}}}, \bibinfo {author} {\bibfnamefont {R.~C.}\ \bibnamefont
  {{Nichol}}}, \bibinfo {author} {\bibfnamefont {M.~D.}\ \bibnamefont
  {{Olmstead}}}, \bibinfo {author} {\bibfnamefont {L.}~\bibnamefont
  {{Samushia}}}, \bibinfo {author} {\bibfnamefont {H.-J.}\ \bibnamefont
  {{Seo}}}, \bibinfo {author} {\bibfnamefont {A.}~\bibnamefont
  {{Streblyanska}}}, \ and\ \bibinfo {author} {\bibfnamefont {G.-b.}\
  \bibnamefont {{Zhao}}},\ }\href {\doibase 10.1093/mnras/stw3384} {\bibfield
  {journal} {\bibinfo  {journal} {\mnras}\ }\textbf {\bibinfo {volume} {467}},\
  \bibinfo {pages} {2085} (\bibinfo {year} {2017})},\ \Eprint
  {http://arxiv.org/abs/1607.03143} {arXiv:1607.03143} \BibitemShut {NoStop}%
\bibitem [{\citenamefont {{Eggemeier}}\ \emph {et~al.}(2020)\citenamefont
  {{Eggemeier}}, \citenamefont {{Scoccimarro}}, \citenamefont {{Crocce}},
  \citenamefont {{Pezzotta}},\ and\ \citenamefont
  {{S{\'a}nchez}}}]{EggScoCro2006}%
  \BibitemOpen
  \bibfield  {author} {\bibinfo {author} {\bibfnamefont {A.}~\bibnamefont
  {{Eggemeier}}}, \bibinfo {author} {\bibfnamefont {R.}~\bibnamefont
  {{Scoccimarro}}}, \bibinfo {author} {\bibfnamefont {M.}~\bibnamefont
  {{Crocce}}}, \bibinfo {author} {\bibfnamefont {A.}~\bibnamefont
  {{Pezzotta}}}, \ and\ \bibinfo {author} {\bibfnamefont {A.~G.}\ \bibnamefont
  {{S{\'a}nchez}}},\ }\href@noop {} {\bibfield  {journal} {\bibinfo  {journal}
  {arXiv e-prints}\ ,\ \bibinfo {eid} {arXiv:2006.09729}} (\bibinfo {year}
  {2020})},\ \Eprint {http://arxiv.org/abs/2006.09729} {arXiv:2006.09729
  [astro-ph.CO]} \BibitemShut {NoStop}%
\bibitem [{\citenamefont {{Pezzotta et al.}}()}]{PezCroSco2101}%
  \BibitemOpen
  \bibfield  {author} {\bibinfo {author} {\bibfnamefont {A.}~\bibnamefont
  {{Pezzotta et al.}}},\ }\href@noop {} {\bibinfo  {journal} {in prep.}\
  }\BibitemShut {NoStop}%
\bibitem [{\citenamefont {{Senatore}}(2015)}]{Sen1406}%
  \BibitemOpen
\bibfield  {journal} {  }\bibfield  {author} {\bibinfo {author} {\bibfnamefont
  {L.}~\bibnamefont {{Senatore}}},\ }\href {\doibase
  10.1088/1475-7516/2015/11/007} {\bibfield  {journal} {\bibinfo  {journal}
  {Journal of Cosmology and Astro-Particle Physics}\ }\textbf {\bibinfo
  {volume} {2015}},\ \bibinfo {eid} {007} (\bibinfo {year} {2015})},\ \Eprint
  {http://arxiv.org/abs/1406.7843} {arXiv:1406.7843 [astro-ph.CO]} \BibitemShut
  {NoStop}%
\bibitem [{\citenamefont {{Mirbabayi}}\ \emph {et~al.}(2015)\citenamefont
  {{Mirbabayi}}, \citenamefont {{Schmidt}},\ and\ \citenamefont
  {{Zaldarriaga}}}]{Mirbabayi:2015}%
  \BibitemOpen
  \bibfield  {author} {\bibinfo {author} {\bibfnamefont {M.}~\bibnamefont
  {{Mirbabayi}}}, \bibinfo {author} {\bibfnamefont {F.}~\bibnamefont
  {{Schmidt}}}, \ and\ \bibinfo {author} {\bibfnamefont {M.}~\bibnamefont
  {{Zaldarriaga}}},\ }\href {\doibase 10.1088/1475-7516/2015/07/030} {\bibfield
   {journal} {\bibinfo  {journal} {\jcap}\ }\textbf {\bibinfo {volume} {7}},\
  \bibinfo {eid} {030} (\bibinfo {year} {2015})},\ \Eprint
  {http://arxiv.org/abs/1412.5169} {arXiv:1412.5169} \BibitemShut {NoStop}%
\bibitem [{\citenamefont {{Kofman}}\ and\ \citenamefont
  {{Pogosyan}}(1995)}]{KofPog9503}%
  \BibitemOpen
  \bibfield  {author} {\bibinfo {author} {\bibfnamefont {L.}~\bibnamefont
  {{Kofman}}}\ and\ \bibinfo {author} {\bibfnamefont {D.}~\bibnamefont
  {{Pogosyan}}},\ }\href {\doibase 10.1086/175419} {\bibfield  {journal}
  {\bibinfo  {journal} {\apj}\ }\textbf {\bibinfo {volume} {442}},\ \bibinfo
  {pages} {30} (\bibinfo {year} {1995})},\ \Eprint
  {http://arxiv.org/abs/astro-ph/9403029} {astro-ph/9403029} \BibitemShut
  {NoStop}%
\bibitem [{\citenamefont {{Bertschinger}}(1995)}]{Ber9501}%
  \BibitemOpen
  \bibfield  {author} {\bibinfo {author} {\bibfnamefont {E.}~\bibnamefont
  {{Bertschinger}}},\ }\href@noop {} {\enquote {\bibinfo {title} {{Cosmological
  dynamics}},}\ }\bibinfo {howpublished} {NASA STI/Recon Technical Report N}
  (\bibinfo {year} {1995}),\ \Eprint {http://arxiv.org/abs/astro-ph/9503125}
  {arXiv:astro-ph/9503125 [astro-ph]} \BibitemShut {NoStop}%
\bibitem [{\citenamefont {{McDonald}}(2006)}]{McDonald:2006}%
  \BibitemOpen
  \bibfield  {author} {\bibinfo {author} {\bibfnamefont {P.}~\bibnamefont
  {{McDonald}}},\ }\href {\doibase 10.1103/PhysRevD.74.103512} {\bibfield
  {journal} {\bibinfo  {journal} {\prd}\ }\textbf {\bibinfo {volume} {74}},\
  \bibinfo {eid} {103512} (\bibinfo {year} {2006})},\ \Eprint
  {http://arxiv.org/abs/astro-ph/0609413} {astro-ph/0609413} \BibitemShut
  {NoStop}%
\bibitem [{\citenamefont {{Crocce}}\ and\ \citenamefont
  {{Scoccimarro}}(2006{\natexlab{a}})}]{CroSco0603a}%
  \BibitemOpen
  \bibfield  {author} {\bibinfo {author} {\bibfnamefont {M.}~\bibnamefont
  {{Crocce}}}\ and\ \bibinfo {author} {\bibfnamefont {R.}~\bibnamefont
  {{Scoccimarro}}},\ }\href {\doibase 10.1103/PhysRevD.73.063519} {\bibfield
  {journal} {\bibinfo  {journal} {\prd}\ }\textbf {\bibinfo {volume} {73}},\
  \bibinfo {pages} {063519} (\bibinfo {year} {2006}{\natexlab{a}})},\ \Eprint
  {http://arxiv.org/abs/arXiv:astro-ph/0509418} {arXiv:astro-ph/0509418}
  \BibitemShut {NoStop}%
\bibitem [{\citenamefont {{Crocce}}\ and\ \citenamefont
  {{Scoccimarro}}(2006{\natexlab{b}})}]{CroSco0603b}%
  \BibitemOpen
  \bibfield  {author} {\bibinfo {author} {\bibfnamefont {M.}~\bibnamefont
  {{Crocce}}}\ and\ \bibinfo {author} {\bibfnamefont {R.}~\bibnamefont
  {{Scoccimarro}}},\ }\href {\doibase 10.1103/PhysRevD.73.063520} {\bibfield
  {journal} {\bibinfo  {journal} {\prd}\ }\textbf {\bibinfo {volume} {73}},\
  \bibinfo {pages} {063520} (\bibinfo {year} {2006}{\natexlab{b}})},\ \Eprint
  {http://arxiv.org/abs/arXiv:astro-ph/0509419} {arXiv:astro-ph/0509419}
  \BibitemShut {NoStop}%
\bibitem [{\citenamefont {{Mo}}\ \emph {et~al.}(1997)\citenamefont {{Mo}},
  \citenamefont {{Jing}},\ and\ \citenamefont {{White}}}]{MoJinWhi97}%
  \BibitemOpen
  \bibfield  {author} {\bibinfo {author} {\bibfnamefont {H.~J.}\ \bibnamefont
  {{Mo}}}, \bibinfo {author} {\bibfnamefont {Y.~P.}\ \bibnamefont {{Jing}}}, \
  and\ \bibinfo {author} {\bibfnamefont {S.~D.~M.}\ \bibnamefont {{White}}},\
  }\href@noop {} {\bibfield  {journal} {\bibinfo  {journal} {\mnras}\ }\textbf
  {\bibinfo {volume} {284}},\ \bibinfo {pages} {189} (\bibinfo {year}
  {1997})}\BibitemShut {NoStop}%
\bibitem [{\citenamefont {{Paranjape}}\ and\ \citenamefont
  {{Sheth}}(2012)}]{ParShe1211}%
  \BibitemOpen
  \bibfield  {author} {\bibinfo {author} {\bibfnamefont {A.}~\bibnamefont
  {{Paranjape}}}\ and\ \bibinfo {author} {\bibfnamefont {R.~K.}\ \bibnamefont
  {{Sheth}}},\ }\href {\doibase 10.1111/j.1365-2966.2012.21911.x} {\bibfield
  {journal} {\bibinfo  {journal} {\mnras}\ }\textbf {\bibinfo {volume} {426}},\
  \bibinfo {pages} {2789} (\bibinfo {year} {2012})},\ \Eprint
  {http://arxiv.org/abs/1206.3506} {arXiv:1206.3506 [astro-ph.CO]} \BibitemShut
  {NoStop}%
\bibitem [{\citenamefont {{Wagner}}\ \emph {et~al.}(2015)\citenamefont
  {{Wagner}}, \citenamefont {{Schmidt}}, \citenamefont {{Chiang}},\ and\
  \citenamefont {{Komatsu}}}]{WagSchChi1503}%
  \BibitemOpen
  \bibfield  {author} {\bibinfo {author} {\bibfnamefont {C.}~\bibnamefont
  {{Wagner}}}, \bibinfo {author} {\bibfnamefont {F.}~\bibnamefont {{Schmidt}}},
  \bibinfo {author} {\bibfnamefont {C.~T.}\ \bibnamefont {{Chiang}}}, \ and\
  \bibinfo {author} {\bibfnamefont {E.}~\bibnamefont {{Komatsu}}},\ }\href
  {\doibase 10.1093/mnrasl/slu187} {\bibfield  {journal} {\bibinfo  {journal}
  {\mnras}\ }\textbf {\bibinfo {volume} {448}},\ \bibinfo {pages} {L11}
  (\bibinfo {year} {2015})},\ \Eprint {http://arxiv.org/abs/1409.6294}
  {arXiv:1409.6294 [astro-ph.CO]} \BibitemShut {NoStop}%
\bibitem [{\citenamefont {{Lazeyras}}\ \emph {et~al.}(2016)\citenamefont
  {{Lazeyras}}, \citenamefont {{Wagner}}, \citenamefont {{Baldauf}},\ and\
  \citenamefont {{Schmidt}}}]{LazWagBal1602}%
  \BibitemOpen
  \bibfield  {author} {\bibinfo {author} {\bibfnamefont {T.}~\bibnamefont
  {{Lazeyras}}}, \bibinfo {author} {\bibfnamefont {C.}~\bibnamefont
  {{Wagner}}}, \bibinfo {author} {\bibfnamefont {T.}~\bibnamefont {{Baldauf}}},
  \ and\ \bibinfo {author} {\bibfnamefont {F.}~\bibnamefont {{Schmidt}}},\
  }\href {\doibase 10.1088/1475-7516/2016/02/018} {\bibfield  {journal}
  {\bibinfo  {journal} {Journal of Cosmology and Astro-Particle Physics}\
  }\textbf {\bibinfo {volume} {2016}},\ \bibinfo {eid} {018} (\bibinfo {year}
  {2016})},\ \Eprint {http://arxiv.org/abs/1511.01096} {arXiv:1511.01096
  [astro-ph.CO]} \BibitemShut {NoStop}%
\bibitem [{\citenamefont {{Matsubara}}(1995)}]{Matsubara:1995}%
  \BibitemOpen
  \bibfield  {author} {\bibinfo {author} {\bibfnamefont {T.}~\bibnamefont
  {{Matsubara}}},\ }\href {\doibase 10.1086/192231} {\bibfield  {journal}
  {\bibinfo  {journal} {\apjs}\ }\textbf {\bibinfo {volume} {101}},\ \bibinfo
  {pages} {1} (\bibinfo {year} {1995})},\ \Eprint
  {http://arxiv.org/abs/astro-ph/9501056} {astro-ph/9501056} \BibitemShut
  {NoStop}%
\bibitem [{\citenamefont {{Bernardeau}}\ \emph {et~al.}(2008)\citenamefont
  {{Bernardeau}}, \citenamefont {{Crocce}},\ and\ \citenamefont
  {{Scoccimarro}}}]{BerCroSco0811}%
  \BibitemOpen
  \bibfield  {author} {\bibinfo {author} {\bibfnamefont {F.}~\bibnamefont
  {{Bernardeau}}}, \bibinfo {author} {\bibfnamefont {M.}~\bibnamefont
  {{Crocce}}}, \ and\ \bibinfo {author} {\bibfnamefont {R.}~\bibnamefont
  {{Scoccimarro}}},\ }\href {\doibase 10.1103/PhysRevD.78.103521} {\bibfield
  {journal} {\bibinfo  {journal} {\prd}\ }\textbf {\bibinfo {volume} {78}},\
  \bibinfo {pages} {103521} (\bibinfo {year} {2008})},\ \Eprint
  {http://arxiv.org/abs/0806.2334} {arXiv:0806.2334} \BibitemShut {NoStop}%
\bibitem [{\citenamefont {{Scoccimarro}}(1997)}]{Sco97}%
  \BibitemOpen
  \bibfield  {author} {\bibinfo {author} {\bibfnamefont {R.}~\bibnamefont
  {{Scoccimarro}}},\ }\href@noop {} {\bibfield  {journal} {\bibinfo  {journal}
  {\apj}\ }\textbf {\bibinfo {volume} {487}},\ \bibinfo {pages} {1} (\bibinfo
  {year} {1997})}\BibitemShut {NoStop}%
\bibitem [{\citenamefont {{Desjacques}}(2008)}]{Des0811}%
  \BibitemOpen
  \bibfield  {author} {\bibinfo {author} {\bibfnamefont {V.}~\bibnamefont
  {{Desjacques}}},\ }\href {\doibase 10.1103/PhysRevD.78.103503} {\bibfield
  {journal} {\bibinfo  {journal} {\prd}\ }\textbf {\bibinfo {volume} {78}},\
  \bibinfo {eid} {103503} (\bibinfo {year} {2008})},\ \Eprint
  {http://arxiv.org/abs/0806.0007} {arXiv:0806.0007 [astro-ph]} \BibitemShut
  {NoStop}%
\bibitem [{\citenamefont {{Desjacques}}\ \emph {et~al.}(2010)\citenamefont
  {{Desjacques}}, \citenamefont {{Crocce}}, \citenamefont {{Scoccimarro}},\
  and\ \citenamefont {{Sheth}}}]{DesCroSco1011}%
  \BibitemOpen
  \bibfield  {author} {\bibinfo {author} {\bibfnamefont {V.}~\bibnamefont
  {{Desjacques}}}, \bibinfo {author} {\bibfnamefont {M.}~\bibnamefont
  {{Crocce}}}, \bibinfo {author} {\bibfnamefont {R.}~\bibnamefont
  {{Scoccimarro}}}, \ and\ \bibinfo {author} {\bibfnamefont {R.~K.}\
  \bibnamefont {{Sheth}}},\ }\href {\doibase 10.1103/PhysRevD.82.103529}
  {\bibfield  {journal} {\bibinfo  {journal} {\prd}\ }\textbf {\bibinfo
  {volume} {82}},\ \bibinfo {pages} {103529} (\bibinfo {year} {2010})},\
  \Eprint {http://arxiv.org/abs/1009.3449} {arXiv:1009.3449 [astro-ph.CO]}
  \BibitemShut {NoStop}%
\bibitem [{\citenamefont {{Bernardeau}}\ \emph {et~al.}(2002)\citenamefont
  {{Bernardeau}}, \citenamefont {{Colombi}}, \citenamefont {{Gazta{\~n}aga}},\
  and\ \citenamefont {{Scoccimarro}}}]{Bernardeau:2002}%
  \BibitemOpen
  \bibfield  {author} {\bibinfo {author} {\bibfnamefont {F.}~\bibnamefont
  {{Bernardeau}}}, \bibinfo {author} {\bibfnamefont {S.}~\bibnamefont
  {{Colombi}}}, \bibinfo {author} {\bibfnamefont {E.}~\bibnamefont
  {{Gazta{\~n}aga}}}, \ and\ \bibinfo {author} {\bibfnamefont {R.}~\bibnamefont
  {{Scoccimarro}}},\ }\href {\doibase 10.1016/S0370-1573(02)00135-7} {\bibfield
   {journal} {\bibinfo  {journal} {\physrep}\ }\textbf {\bibinfo {volume}
  {367}},\ \bibinfo {pages} {1} (\bibinfo {year} {2002})},\ \Eprint
  {http://arxiv.org/abs/astro-ph/0112551} {astro-ph/0112551} \BibitemShut
  {NoStop}%
\bibitem [{\citenamefont {{Lazeyras}}\ and\ \citenamefont
  {{Schmidt}}(2019)}]{LazSch1911}%
  \BibitemOpen
  \bibfield  {author} {\bibinfo {author} {\bibfnamefont {T.}~\bibnamefont
  {{Lazeyras}}}\ and\ \bibinfo {author} {\bibfnamefont {F.}~\bibnamefont
  {{Schmidt}}},\ }\href {\doibase 10.1088/1475-7516/2019/11/041} {\bibfield
  {journal} {\bibinfo  {journal} {\jcap}\ }\textbf {\bibinfo {volume} {2019}},\
  \bibinfo {eid} {041} (\bibinfo {year} {2019})},\ \Eprint
  {http://arxiv.org/abs/1904.11294} {arXiv:1904.11294 [astro-ph.CO]}
  \BibitemShut {NoStop}%
\bibitem [{\citenamefont {{Pueblas}}\ and\ \citenamefont
  {{Scoccimarro}}(2009)}]{PueSco0908}%
  \BibitemOpen
  \bibfield  {author} {\bibinfo {author} {\bibfnamefont {S.}~\bibnamefont
  {{Pueblas}}}\ and\ \bibinfo {author} {\bibfnamefont {R.}~\bibnamefont
  {{Scoccimarro}}},\ }\href {\doibase 10.1103/PhysRevD.80.043504} {\bibfield
  {journal} {\bibinfo  {journal} {\prd}\ }\textbf {\bibinfo {volume} {80}},\
  \bibinfo {pages} {043504} (\bibinfo {year} {2009})},\ \Eprint
  {http://arxiv.org/abs/0809.4606} {arXiv:0809.4606} \BibitemShut {NoStop}%
\bibitem [{\citenamefont {{Dekel}}\ and\ \citenamefont
  {{Lahav}}(1999)}]{Dekel:1999}%
  \BibitemOpen
  \bibfield  {author} {\bibinfo {author} {\bibfnamefont {A.}~\bibnamefont
  {{Dekel}}}\ and\ \bibinfo {author} {\bibfnamefont {O.}~\bibnamefont
  {{Lahav}}},\ }\href {\doibase 10.1086/307428} {\bibfield  {journal} {\bibinfo
   {journal} {\apj}\ }\textbf {\bibinfo {volume} {520}},\ \bibinfo {pages} {24}
  (\bibinfo {year} {1999})},\ \Eprint {http://arxiv.org/abs/astro-ph/9806193}
  {arXiv:astro-ph/9806193 [astro-ph]} \BibitemShut {NoStop}%
\bibitem [{\citenamefont {{Taruya}}\ and\ \citenamefont
  {{Soda}}(1999)}]{TarSod9909}%
  \BibitemOpen
  \bibfield  {author} {\bibinfo {author} {\bibfnamefont {A.}~\bibnamefont
  {{Taruya}}}\ and\ \bibinfo {author} {\bibfnamefont {J.}~\bibnamefont
  {{Soda}}},\ }\href {\doibase 10.1086/307612} {\bibfield  {journal} {\bibinfo
  {journal} {\apj}\ }\textbf {\bibinfo {volume} {522}},\ \bibinfo {pages} {46}
  (\bibinfo {year} {1999})},\ \Eprint {http://arxiv.org/abs/astro-ph/9809204}
  {arXiv:astro-ph/9809204 [astro-ph]} \BibitemShut {NoStop}%
\bibitem [{\citenamefont {{Matsubara}}(1999)}]{Mat9911}%
  \BibitemOpen
  \bibfield  {author} {\bibinfo {author} {\bibfnamefont {T.}~\bibnamefont
  {{Matsubara}}},\ }\href {\doibase 10.1086/307931} {\bibfield  {journal}
  {\bibinfo  {journal} {\apj}\ }\textbf {\bibinfo {volume} {525}},\ \bibinfo
  {pages} {543} (\bibinfo {year} {1999})},\ \Eprint
  {http://arxiv.org/abs/astro-ph/9906029} {arXiv:astro-ph/9906029 [astro-ph]}
  \BibitemShut {NoStop}%
\bibitem [{\citenamefont {{Mo}}\ and\ \citenamefont
  {{White}}(1996)}]{MoWhi9609}%
  \BibitemOpen
  \bibfield  {author} {\bibinfo {author} {\bibfnamefont {H.~J.}\ \bibnamefont
  {{Mo}}}\ and\ \bibinfo {author} {\bibfnamefont {S.~D.~M.}\ \bibnamefont
  {{White}}},\ }\href@noop {} {\bibfield  {journal} {\bibinfo  {journal}
  {\mnras}\ }\textbf {\bibinfo {volume} {282}},\ \bibinfo {pages} {347}
  (\bibinfo {year} {1996})},\ \Eprint
  {http://arxiv.org/abs/arXiv:astro-ph/9512127} {arXiv:astro-ph/9512127}
  \BibitemShut {NoStop}%
\bibitem [{\citenamefont {{Sheth}}\ and\ \citenamefont
  {{Lemson}}(1999)}]{SheLem99}%
  \BibitemOpen
  \bibfield  {author} {\bibinfo {author} {\bibfnamefont {R.}~\bibnamefont
  {{Sheth}}}\ and\ \bibinfo {author} {\bibfnamefont {G.}~\bibnamefont
  {{Lemson}}},\ }\href@noop {} {\bibfield  {journal} {\bibinfo  {journal}
  {\mnras}\ }\textbf {\bibinfo {volume} {304}},\ \bibinfo {pages} {767}
  (\bibinfo {year} {1999})}\BibitemShut {NoStop}%
\bibitem [{\citenamefont {{Smith}}\ \emph {et~al.}(2007)\citenamefont
  {{Smith}}, \citenamefont {{Scoccimarro}},\ and\ \citenamefont
  {{Sheth}}}]{SmiScoShe0703}%
  \BibitemOpen
  \bibfield  {author} {\bibinfo {author} {\bibfnamefont {R.~E.}\ \bibnamefont
  {{Smith}}}, \bibinfo {author} {\bibfnamefont {R.}~\bibnamefont
  {{Scoccimarro}}}, \ and\ \bibinfo {author} {\bibfnamefont {R.~K.}\
  \bibnamefont {{Sheth}}},\ }\href {\doibase 10.1103/PhysRevD.75.063512}
  {\bibfield  {journal} {\bibinfo  {journal} {\prd}\ }\textbf {\bibinfo
  {volume} {75}},\ \bibinfo {pages} {063512} (\bibinfo {year} {2007})},\
  \Eprint {http://arxiv.org/abs/arXiv:astro-ph/0609547}
  {arXiv:astro-ph/0609547} \BibitemShut {NoStop}%
\bibitem [{\citenamefont {{Baldauf}}\ \emph {et~al.}(2013)\citenamefont
  {{Baldauf}}, \citenamefont {{Seljak}}, \citenamefont {{Smith}}, \citenamefont
  {{Hamaus}},\ and\ \citenamefont {{Desjacques}}}]{BalSelSmi1310}%
  \BibitemOpen
  \bibfield  {author} {\bibinfo {author} {\bibfnamefont {T.}~\bibnamefont
  {{Baldauf}}}, \bibinfo {author} {\bibfnamefont {U.}~\bibnamefont {{Seljak}}},
  \bibinfo {author} {\bibfnamefont {R.~E.}\ \bibnamefont {{Smith}}}, \bibinfo
  {author} {\bibfnamefont {N.}~\bibnamefont {{Hamaus}}}, \ and\ \bibinfo
  {author} {\bibfnamefont {V.}~\bibnamefont {{Desjacques}}},\ }\href {\doibase
  10.1103/PhysRevD.88.083507} {\bibfield  {journal} {\bibinfo  {journal}
  {\prd}\ }\textbf {\bibinfo {volume} {88}},\ \bibinfo {eid} {083507} (\bibinfo
  {year} {2013})},\ \Eprint {http://arxiv.org/abs/1305.2917} {arXiv:1305.2917
  [astro-ph.CO]} \BibitemShut {NoStop}%
\bibitem [{\citenamefont {{Schmidt}}(2016)}]{Sch1603}%
  \BibitemOpen
  \bibfield  {author} {\bibinfo {author} {\bibfnamefont {F.}~\bibnamefont
  {{Schmidt}}},\ }\href {\doibase 10.1103/PhysRevD.93.063512} {\bibfield
  {journal} {\bibinfo  {journal} {\prd}\ }\textbf {\bibinfo {volume} {93}},\
  \bibinfo {eid} {063512} (\bibinfo {year} {2016})},\ \Eprint
  {http://arxiv.org/abs/1511.02231} {arXiv:1511.02231 [astro-ph.CO]}
  \BibitemShut {NoStop}%
\bibitem [{\citenamefont {{Gil-Mar{\'\i}n}}\ \emph
  {et~al.}(2015{\natexlab{b}})\citenamefont {{Gil-Mar{\'\i}n}}, \citenamefont
  {{Nore{\~n}a}}, \citenamefont {{Verde}}, \citenamefont {{Percival}},
  \citenamefont {{Wagner}}, \citenamefont {{Manera}},\ and\ \citenamefont
  {{Schneider}}}]{GilNorVer1407}%
  \BibitemOpen
  \bibfield  {author} {\bibinfo {author} {\bibfnamefont {H.}~\bibnamefont
  {{Gil-Mar{\'\i}n}}}, \bibinfo {author} {\bibfnamefont {J.}~\bibnamefont
  {{Nore{\~n}a}}}, \bibinfo {author} {\bibfnamefont {L.}~\bibnamefont
  {{Verde}}}, \bibinfo {author} {\bibfnamefont {W.~J.}\ \bibnamefont
  {{Percival}}}, \bibinfo {author} {\bibfnamefont {C.}~\bibnamefont
  {{Wagner}}}, \bibinfo {author} {\bibfnamefont {M.}~\bibnamefont {{Manera}}},
  \ and\ \bibinfo {author} {\bibfnamefont {D.~P.}\ \bibnamefont
  {{Schneider}}},\ }\href {\doibase 10.1093/mnras/stv961} {\bibfield  {journal}
  {\bibinfo  {journal} {\mnras}\ }\textbf {\bibinfo {volume} {451}},\ \bibinfo
  {pages} {539} (\bibinfo {year} {2015}{\natexlab{b}})},\ \Eprint
  {http://arxiv.org/abs/1407.5668} {arXiv:1407.5668 [astro-ph.CO]} \BibitemShut
  {NoStop}%
\bibitem [{\citenamefont {{Scoccimarro}}(2015)}]{Sco1510}%
  \BibitemOpen
  \bibfield  {author} {\bibinfo {author} {\bibfnamefont {R.}~\bibnamefont
  {{Scoccimarro}}},\ }\href {\doibase 10.1103/PhysRevD.92.083532} {\bibfield
  {journal} {\bibinfo  {journal} {\prd}\ }\textbf {\bibinfo {volume} {92}},\
  \bibinfo {eid} {083532} (\bibinfo {year} {2015})},\ \Eprint
  {http://arxiv.org/abs/1506.02729} {arXiv:1506.02729 [astro-ph.CO]}
  \BibitemShut {NoStop}%
\bibitem [{\citenamefont {{Sefusatti}}\ \emph {et~al.}(2016)\citenamefont
  {{Sefusatti}}, \citenamefont {{Crocce}}, \citenamefont {{Scoccimarro}},\ and\
  \citenamefont {{Couchman}}}]{SefCroSco1512}%
  \BibitemOpen
  \bibfield  {author} {\bibinfo {author} {\bibfnamefont {E.}~\bibnamefont
  {{Sefusatti}}}, \bibinfo {author} {\bibfnamefont {M.}~\bibnamefont
  {{Crocce}}}, \bibinfo {author} {\bibfnamefont {R.}~\bibnamefont
  {{Scoccimarro}}}, \ and\ \bibinfo {author} {\bibfnamefont {H.~M.~P.}\
  \bibnamefont {{Couchman}}},\ }\href {\doibase 10.1093/mnras/stw1229}
  {\bibfield  {journal} {\bibinfo  {journal} {\mnras}\ }\textbf {\bibinfo
  {volume} {460}},\ \bibinfo {pages} {3624} (\bibinfo {year} {2016})},\ \Eprint
  {http://arxiv.org/abs/1512.07295} {arXiv:1512.07295 [astro-ph.CO]}
  \BibitemShut {NoStop}%
\bibitem [{\citenamefont {{Peebles}}(1980)}]{Pee80}%
  \BibitemOpen
  \bibfield  {author} {\bibinfo {author} {\bibfnamefont {P.}~\bibnamefont
  {{Peebles}}},\ }\href@noop {} {\emph {\bibinfo {title} {{The large-scale
  structure of the universe}}}}\ (\bibinfo  {publisher} {Princeton University
  Press},\ \bibinfo {year} {1980})\BibitemShut {NoStop}%
\bibitem [{\citenamefont {{Scoccimarro}}\ \emph {et~al.}(1998)\citenamefont
  {{Scoccimarro}}, \citenamefont {{Colombi}}, \citenamefont {{Fry}},
  \citenamefont {{Frieman}}, \citenamefont {{Hivon}},\ and\ \citenamefont
  {{Melott}}}]{ScoColFry9803}%
  \BibitemOpen
  \bibfield  {author} {\bibinfo {author} {\bibfnamefont {R.}~\bibnamefont
  {{Scoccimarro}}}, \bibinfo {author} {\bibfnamefont {S.}~\bibnamefont
  {{Colombi}}}, \bibinfo {author} {\bibfnamefont {J.~N.}\ \bibnamefont
  {{Fry}}}, \bibinfo {author} {\bibfnamefont {J.~A.}\ \bibnamefont
  {{Frieman}}}, \bibinfo {author} {\bibfnamefont {E.}~\bibnamefont {{Hivon}}},
  \ and\ \bibinfo {author} {\bibfnamefont {A.}~\bibnamefont {{Melott}}},\
  }\href {\doibase 10.1086/305399} {\bibfield  {journal} {\bibinfo  {journal}
  {\apj}\ }\textbf {\bibinfo {volume} {496}},\ \bibinfo {pages} {586} (\bibinfo
  {year} {1998})},\ \Eprint {http://arxiv.org/abs/arXiv:astro-ph/9704075}
  {arXiv:astro-ph/9704075} \BibitemShut {NoStop}%
\bibitem [{\citenamefont {{Scoccimarro}}\ \emph {et~al.}(1999)\citenamefont
  {{Scoccimarro}}, \citenamefont {{Couchman}},\ and\ \citenamefont
  {{Frieman}}}]{ScoCouFri9906}%
  \BibitemOpen
  \bibfield  {author} {\bibinfo {author} {\bibfnamefont {R.}~\bibnamefont
  {{Scoccimarro}}}, \bibinfo {author} {\bibfnamefont {H.~M.~P.}\ \bibnamefont
  {{Couchman}}}, \ and\ \bibinfo {author} {\bibfnamefont {J.~A.}\ \bibnamefont
  {{Frieman}}},\ }\href {\doibase 10.1086/307220} {\bibfield  {journal}
  {\bibinfo  {journal} {\apj}\ }\textbf {\bibinfo {volume} {517}},\ \bibinfo
  {pages} {531} (\bibinfo {year} {1999})},\ \Eprint
  {http://arxiv.org/abs/astro-ph/9808305} {arXiv:astro-ph/9808305 [astro-ph]}
  \BibitemShut {NoStop}%
\bibitem [{\citenamefont {{Feldman}}\ \emph {et~al.}(1994)\citenamefont
  {{Feldman}}, \citenamefont {{Kaiser}},\ and\ \citenamefont
  {{Peacock}}}]{FelKaiPea9405}%
  \BibitemOpen
  \bibfield  {author} {\bibinfo {author} {\bibfnamefont {H.~A.}\ \bibnamefont
  {{Feldman}}}, \bibinfo {author} {\bibfnamefont {N.}~\bibnamefont {{Kaiser}}},
  \ and\ \bibinfo {author} {\bibfnamefont {J.~A.}\ \bibnamefont {{Peacock}}},\
  }\href {\doibase 10.1086/174036} {\bibfield  {journal} {\bibinfo  {journal}
  {\apj}\ }\textbf {\bibinfo {volume} {426}},\ \bibinfo {pages} {23} (\bibinfo
  {year} {1994})},\ \Eprint {http://arxiv.org/abs/arXiv:astro-ph/9304022}
  {arXiv:astro-ph/9304022} \BibitemShut {NoStop}%
\bibitem [{\citenamefont {{Tegmark}}(1997)}]{Teg9711}%
  \BibitemOpen
  \bibfield  {author} {\bibinfo {author} {\bibfnamefont {M.}~\bibnamefont
  {{Tegmark}}},\ }\href {\doibase 10.1103/PhysRevLett.79.3806} {\bibfield
  {journal} {\bibinfo  {journal} {\prl}\ }\textbf {\bibinfo {volume} {79}},\
  \bibinfo {pages} {3806} (\bibinfo {year} {1997})},\ \Eprint
  {http://arxiv.org/abs/astro-ph/9706198} {arXiv:astro-ph/9706198 [astro-ph]}
  \BibitemShut {NoStop}%
\bibitem [{\citenamefont {{Sefusatti}}\ \emph {et~al.}(2010)\citenamefont
  {{Sefusatti}}, \citenamefont {{Crocce}},\ and\ \citenamefont
  {{Desjacques}}}]{SefCroDes1008}%
  \BibitemOpen
  \bibfield  {author} {\bibinfo {author} {\bibfnamefont {E.}~\bibnamefont
  {{Sefusatti}}}, \bibinfo {author} {\bibfnamefont {M.}~\bibnamefont
  {{Crocce}}}, \ and\ \bibinfo {author} {\bibfnamefont {V.}~\bibnamefont
  {{Desjacques}}},\ }\href {\doibase 10.1111/j.1365-2966.2010.16723.x}
  {\bibfield  {journal} {\bibinfo  {journal} {\mnras}\ }\textbf {\bibinfo
  {volume} {406}},\ \bibinfo {pages} {1014} (\bibinfo {year} {2010})},\ \Eprint
  {http://arxiv.org/abs/1003.0007} {arXiv:1003.0007 [astro-ph.CO]} \BibitemShut
  {NoStop}%
\bibitem [{\citenamefont {{Gelman}}\ and\ \citenamefont
  {{Rubin}}(1992)}]{GelRub9201}%
  \BibitemOpen
  \bibfield  {author} {\bibinfo {author} {\bibfnamefont {A.}~\bibnamefont
  {{Gelman}}}\ and\ \bibinfo {author} {\bibfnamefont {D.~B.}\ \bibnamefont
  {{Rubin}}},\ }\href {\doibase 10.1214/ss/1177011136} {\bibfield  {journal}
  {\bibinfo  {journal} {Statistical Science}\ }\textbf {\bibinfo {volume}
  {7}},\ \bibinfo {pages} {457} (\bibinfo {year} {1992})}\BibitemShut {NoStop}%
\bibitem [{\citenamefont {Lewis}(2019)}]{Lew2019}%
  \BibitemOpen
  \bibfield  {author} {\bibinfo {author} {\bibfnamefont {A.}~\bibnamefont
  {Lewis}},\ }\href {https://getdist.readthedocs.io} {\  (\bibinfo {year}
  {2019})},\ \Eprint {http://arxiv.org/abs/1910.13970} {arXiv:1910.13970
  [astro-ph.IM]} \BibitemShut {NoStop}%
\bibitem [{\citenamefont {{Scoccimarro}}\ and\ \citenamefont
  {{Frieman}}(1999)}]{ScoFri9907}%
  \BibitemOpen
  \bibfield  {author} {\bibinfo {author} {\bibfnamefont {R.}~\bibnamefont
  {{Scoccimarro}}}\ and\ \bibinfo {author} {\bibfnamefont {J.~A.}\ \bibnamefont
  {{Frieman}}},\ }\href {\doibase 10.1086/307448} {\bibfield  {journal}
  {\bibinfo  {journal} {\apj}\ }\textbf {\bibinfo {volume} {520}},\ \bibinfo
  {pages} {35} (\bibinfo {year} {1999})},\ \Eprint
  {http://arxiv.org/abs/arXiv:astro-ph/9811184} {arXiv:astro-ph/9811184}
  \BibitemShut {NoStop}%
\bibitem [{\citenamefont {{Sefusatti}}\ and\ \citenamefont
  {{Scoccimarro}}(2005)}]{SefSco0503}%
  \BibitemOpen
  \bibfield  {author} {\bibinfo {author} {\bibfnamefont {E.}~\bibnamefont
  {{Sefusatti}}}\ and\ \bibinfo {author} {\bibfnamefont {R.}~\bibnamefont
  {{Scoccimarro}}},\ }\href {\doibase 10.1103/PhysRevD.71.063001} {\bibfield
  {journal} {\bibinfo  {journal} {\prd}\ }\textbf {\bibinfo {volume} {71}},\
  \bibinfo {eid} {063001} (\bibinfo {year} {2005})},\ \Eprint
  {http://arxiv.org/abs/astro-ph/0412626} {arXiv:astro-ph/0412626 [astro-ph]}
  \BibitemShut {NoStop}%
\bibitem [{\citenamefont {{Bernardeau}}\ \emph {et~al.}(2014)\citenamefont
  {{Bernardeau}}, \citenamefont {{Taruya}},\ and\ \citenamefont
  {{Nishimichi}}}]{BerTarNis1401}%
  \BibitemOpen
  \bibfield  {author} {\bibinfo {author} {\bibfnamefont {F.}~\bibnamefont
  {{Bernardeau}}}, \bibinfo {author} {\bibfnamefont {A.}~\bibnamefont
  {{Taruya}}}, \ and\ \bibinfo {author} {\bibfnamefont {T.}~\bibnamefont
  {{Nishimichi}}},\ }\href {\doibase 10.1103/PhysRevD.89.023502} {\bibfield
  {journal} {\bibinfo  {journal} {\prd}\ }\textbf {\bibinfo {volume} {89}},\
  \bibinfo {eid} {023502} (\bibinfo {year} {2014})}\BibitemShut {NoStop}%
\bibitem [{\citenamefont {{Nishimichi}}\ \emph {et~al.}(2016)\citenamefont
  {{Nishimichi}}, \citenamefont {{Bernardeau}},\ and\ \citenamefont
  {{Taruya}}}]{NisBerTar1611}%
  \BibitemOpen
  \bibfield  {author} {\bibinfo {author} {\bibfnamefont {T.}~\bibnamefont
  {{Nishimichi}}}, \bibinfo {author} {\bibfnamefont {F.}~\bibnamefont
  {{Bernardeau}}}, \ and\ \bibinfo {author} {\bibfnamefont {A.}~\bibnamefont
  {{Taruya}}},\ }\href {\doibase 10.1016/j.physletb.2016.09.035} {\bibfield
  {journal} {\bibinfo  {journal} {Physics Letters B}\ }\textbf {\bibinfo
  {volume} {762}},\ \bibinfo {pages} {247} (\bibinfo {year} {2016})},\ \Eprint
  {http://arxiv.org/abs/1411.2970} {arXiv:1411.2970 [astro-ph.CO]} \BibitemShut
  {NoStop}%
\bibitem [{\citenamefont {{Nishimichi}}\ \emph {et~al.}(2017)\citenamefont
  {{Nishimichi}}, \citenamefont {{Bernardeau}},\ and\ \citenamefont
  {{Taruya}}}]{NisBerTar1712}%
  \BibitemOpen
  \bibfield  {author} {\bibinfo {author} {\bibfnamefont {T.}~\bibnamefont
  {{Nishimichi}}}, \bibinfo {author} {\bibfnamefont {F.}~\bibnamefont
  {{Bernardeau}}}, \ and\ \bibinfo {author} {\bibfnamefont {A.}~\bibnamefont
  {{Taruya}}},\ }\href {\doibase 10.1103/PhysRevD.96.123515} {\bibfield
  {journal} {\bibinfo  {journal} {\prd}\ }\textbf {\bibinfo {volume} {96}},\
  \bibinfo {eid} {123515} (\bibinfo {year} {2017})},\ \Eprint
  {http://arxiv.org/abs/1708.08946} {arXiv:1708.08946 [astro-ph.CO]}
  \BibitemShut {NoStop}%
\bibitem [{\citenamefont {{Ivanov}}\ \emph {et~al.}(2019)\citenamefont
  {{Ivanov}}, \citenamefont {{Simonovi{\'c}}},\ and\ \citenamefont
  {{Zaldarriaga}}}]{IvaSimZal1909}%
  \BibitemOpen
  \bibfield  {author} {\bibinfo {author} {\bibfnamefont {M.~M.}\ \bibnamefont
  {{Ivanov}}}, \bibinfo {author} {\bibfnamefont {M.}~\bibnamefont
  {{Simonovi{\'c}}}}, \ and\ \bibinfo {author} {\bibfnamefont {M.}~\bibnamefont
  {{Zaldarriaga}}},\ }\href@noop {} {\bibfield  {journal} {\bibinfo  {journal}
  {arXiv e-prints}\ ,\ \bibinfo {eid} {arXiv:1909.05277}} (\bibinfo {year}
  {2019})},\ \Eprint {http://arxiv.org/abs/1909.05277} {arXiv:1909.05277
  [astro-ph.CO]} \BibitemShut {NoStop}%
\bibitem [{\citenamefont {{D'Amico}}\ \emph {et~al.}(2019)\citenamefont
  {{D'Amico}}, \citenamefont {{Gleyzes}}, \citenamefont {{Kokron}},
  \citenamefont {{Markovic}}, \citenamefont {{Senatore}}, \citenamefont
  {{Zhang}}, \citenamefont {{Beutler}},\ and\ \citenamefont
  {{Gil-Mar{\'\i}n}}}]{AmiGleKok1909}%
  \BibitemOpen
  \bibfield  {author} {\bibinfo {author} {\bibfnamefont {G.}~\bibnamefont
  {{D'Amico}}}, \bibinfo {author} {\bibfnamefont {J.}~\bibnamefont
  {{Gleyzes}}}, \bibinfo {author} {\bibfnamefont {N.}~\bibnamefont {{Kokron}}},
  \bibinfo {author} {\bibfnamefont {D.}~\bibnamefont {{Markovic}}}, \bibinfo
  {author} {\bibfnamefont {L.}~\bibnamefont {{Senatore}}}, \bibinfo {author}
  {\bibfnamefont {P.}~\bibnamefont {{Zhang}}}, \bibinfo {author} {\bibfnamefont
  {F.}~\bibnamefont {{Beutler}}}, \ and\ \bibinfo {author} {\bibfnamefont
  {H.}~\bibnamefont {{Gil-Mar{\'\i}n}}},\ }\href@noop {} {\bibfield  {journal}
  {\bibinfo  {journal} {arXiv e-prints}\ ,\ \bibinfo {eid} {arXiv:1909.05271}}
  (\bibinfo {year} {2019})},\ \Eprint {http://arxiv.org/abs/1909.05271}
  {arXiv:1909.05271 [astro-ph.CO]} \BibitemShut {NoStop}%
\bibitem [{\citenamefont {{Tr{\"o}ster}}\ \emph {et~al.}(2020)\citenamefont
  {{Tr{\"o}ster}}, \citenamefont {{S{\'a}nchez}}, \citenamefont {{Asgari}},
  \citenamefont {{Blake}}, \citenamefont {{Crocce}}, \citenamefont {{Heymans}},
  \citenamefont {{Hildebrandt}}, \citenamefont {{Joachimi}}, \citenamefont
  {{Joudaki}}, \citenamefont {{Kannawadi}}, \citenamefont {{Lin}},\ and\
  \citenamefont {{Wright}}}]{TroSanAsg2001}%
  \BibitemOpen
  \bibfield  {author} {\bibinfo {author} {\bibfnamefont {T.}~\bibnamefont
  {{Tr{\"o}ster}}}, \bibinfo {author} {\bibfnamefont {A.~G.}\ \bibnamefont
  {{S{\'a}nchez}}}, \bibinfo {author} {\bibfnamefont {M.}~\bibnamefont
  {{Asgari}}}, \bibinfo {author} {\bibfnamefont {C.}~\bibnamefont {{Blake}}},
  \bibinfo {author} {\bibfnamefont {M.}~\bibnamefont {{Crocce}}}, \bibinfo
  {author} {\bibfnamefont {C.}~\bibnamefont {{Heymans}}}, \bibinfo {author}
  {\bibfnamefont {H.}~\bibnamefont {{Hildebrandt}}}, \bibinfo {author}
  {\bibfnamefont {B.}~\bibnamefont {{Joachimi}}}, \bibinfo {author}
  {\bibfnamefont {S.}~\bibnamefont {{Joudaki}}}, \bibinfo {author}
  {\bibfnamefont {A.}~\bibnamefont {{Kannawadi}}}, \bibinfo {author}
  {\bibfnamefont {C.-A.}\ \bibnamefont {{Lin}}}, \ and\ \bibinfo {author}
  {\bibfnamefont {A.}~\bibnamefont {{Wright}}},\ }\href {\doibase
  10.1051/0004-6361/201936772} {\bibfield  {journal} {\bibinfo  {journal}
  {\aap}\ }\textbf {\bibinfo {volume} {633}},\ \bibinfo {eid} {L10} (\bibinfo
  {year} {2020})},\ \Eprint {http://arxiv.org/abs/1909.11006} {arXiv:1909.11006
  [astro-ph.CO]} \BibitemShut {NoStop}%
\bibitem [{\citenamefont {{Feldman}}\ \emph {et~al.}(2001)\citenamefont
  {{Feldman}}, \citenamefont {{Frieman}}, \citenamefont {{Fry}},\ and\
  \citenamefont {{Scoccimarro}}}]{FelFriFry0102}%
  \BibitemOpen
  \bibfield  {author} {\bibinfo {author} {\bibfnamefont {H.~A.}\ \bibnamefont
  {{Feldman}}}, \bibinfo {author} {\bibfnamefont {J.~A.}\ \bibnamefont
  {{Frieman}}}, \bibinfo {author} {\bibfnamefont {J.~N.}\ \bibnamefont
  {{Fry}}}, \ and\ \bibinfo {author} {\bibfnamefont {R.}~\bibnamefont
  {{Scoccimarro}}},\ }\href {\doibase 10.1103/PhysRevLett.86.1434} {\bibfield
  {journal} {\bibinfo  {journal} {Physical Review Letters}\ }\textbf {\bibinfo
  {volume} {86}},\ \bibinfo {pages} {1434} (\bibinfo {year} {2001})},\ \Eprint
  {http://arxiv.org/abs/arXiv:astro-ph/0010205} {arXiv:astro-ph/0010205}
  \BibitemShut {NoStop}%
\bibitem [{\citenamefont {{Sugiyama}}\ \emph {et~al.}(2021)\citenamefont
  {{Sugiyama}}, \citenamefont {{Saito}}, \citenamefont {{Beutler}},\ and\
  \citenamefont {{Seo}}}]{SugSaiBeu2102}%
  \BibitemOpen
  \bibfield  {author} {\bibinfo {author} {\bibfnamefont {N.~S.}\ \bibnamefont
  {{Sugiyama}}}, \bibinfo {author} {\bibfnamefont {S.}~\bibnamefont {{Saito}}},
  \bibinfo {author} {\bibfnamefont {F.}~\bibnamefont {{Beutler}}}, \ and\
  \bibinfo {author} {\bibfnamefont {H.-J.}\ \bibnamefont {{Seo}}},\ }\href
  {\doibase 10.1093/mnras/staa3725} {\bibfield  {journal} {\bibinfo  {journal}
  {\mnras}\ }\textbf {\bibinfo {volume} {501}},\ \bibinfo {pages} {2862}
  (\bibinfo {year} {2021})},\ \Eprint {http://arxiv.org/abs/2010.06179}
  {arXiv:2010.06179 [astro-ph.CO]} \BibitemShut {NoStop}%
\bibitem [{\citenamefont {Hunter}(2007)}]{Hunter:2007}%
  \BibitemOpen
  \bibfield  {author} {\bibinfo {author} {\bibfnamefont {J.~D.}\ \bibnamefont
  {Hunter}},\ }\href {\doibase 10.1109/MCSE.2007.55} {\bibfield  {journal}
  {\bibinfo  {journal} {Computing in Science \& Engineering}\ }\textbf
  {\bibinfo {volume} {9}},\ \bibinfo {pages} {90} (\bibinfo {year}
  {2007})}\BibitemShut {NoStop}%
\bibitem [{\citenamefont {{Scoccimarro}}(2001)}]{Sco0101}%
  \BibitemOpen
  \bibfield  {author} {\bibinfo {author} {\bibfnamefont {R.}~\bibnamefont
  {{Scoccimarro}}},\ }in\ \href {\doibase 10.1111/j.1749-6632.2001.tb05618.x}
  {\emph {\bibinfo {booktitle} {The Onset of Nonlinearity in Cosmology}}},\
  Vol.\ \bibinfo {volume} {927},\ \bibinfo {editor} {edited by\ \bibinfo
  {editor} {\bibfnamefont {J.~N.}\ \bibnamefont {{Fry}}}, \bibinfo {editor}
  {\bibfnamefont {J.~R.}\ \bibnamefont {{Buchler}}}, \ and\ \bibinfo {editor}
  {\bibfnamefont {H.}~\bibnamefont {{Kandrup}}}}\ (\bibinfo {year} {2001})\
  pp.\ \bibinfo {pages} {13--23},\ \Eprint
  {http://arxiv.org/abs/astro-ph/0008277} {arXiv:astro-ph/0008277 [astro-ph]}
  \BibitemShut {NoStop}%
\end{thebibliography}%

\end{document}